\documentclass[aps,prd,twocolumn,superscriptaddress,nofootinbib,eqsecnum,showpacs,letterpaper,11pt]{revtex4}

\usepackage{verbatim}
\usepackage[usenames,dvipsnames]{color}
\usepackage[pdftex]{graphicx}
\usepackage[english]{babel}
\usepackage{amsmath,amssymb}
\usepackage[colorlinks=true,citecolor=blue,linkcolor=magenta]{hyperref}
\usepackage{soul}
\begin{document}
\newcommand{\sttheta}{\mbox{\st{$\theta$}}}
\newcommand{\E}{\mathrm{E}}
\newcommand{\Var}{\mathrm{Var}}
\newcommand{\bra}[1]{\langle #1|}
\newcommand{\ket}[1]{|#1\rangle}
\newcommand{\braket}[2]{\langle #1|#2 \rangle}
\newcommand{\mean}[2]{\langle #1 #2 \rangle}
\newcommand{\be}{\begin{equation}}
\newcommand{\ee}{\end{equation}}
\newcommand{\ba}{\begin{eqnarray}}
\newcommand{\ea}{\end{eqnarray}}
\newcommand{\SD}[1]{{\color{magenta}#1}}
\newcommand{\rem}[1]{{\sout{#1}}}
\newcommand{\alert}[1]{\textbf{\color{red} \uwave{#1}}}
\newcommand{\Y}[1]{\textcolor{yellow}{#1}}
\newcommand{\R}[1]{\textcolor{red}{{\it[#1]}}}
\newcommand{\B}[1]{\textcolor{blue}{#1}}
\newcommand{\C}[1]{\textcolor{cyan}{#1}}
\newcommand{\db}{\color{darkblue}}
\newcommand{\aaron}[1]{\textcolor{blue}{#1}}
\newcommand{\intinfty}{\int_{-\infty}^{\infty}\!}
\newcommand{\Tr}{\mathop{\rm Tr}\nolimits}
\newcommand{\const}{\mathop{\rm const}\nolimits}
\makeatletter
\newcommand{\rmnum}[1]{\romannumeral #1}
\newcommand{\Rmnum}[1]{\expandafter\@slowromancap\romannumeral #1@}
\makeatother
\newcommand{\fan}[1]{\textcolor{black}{#1}}

\title{The scalar Green function of the Kerr spacetime}
\author{Huan Yang}
\affiliation{Theoretical Astrophysics 350-17, California Institute of Technology, Pasadena, CA 91125, USA}
\affiliation{Perimeter Institute for Theoretical Physics, Waterloo, Ontario N2L2Y5, Canada}
\affiliation{Institute for Quantum Computing, University of Waterloo, Waterloo, Ontario N2L3G1, Canada}
\author{Fan Zhang}
\affiliation{Theoretical Astrophysics 350-17, California Institute of Technology, Pasadena, CA 91125, USA}
\affiliation{\mbox{Department of Physics, West Virginia University, PO Box 6315, Morgantown, WV 26506, USA}}
\author{Aaron Zimmerman}
\affiliation{Theoretical Astrophysics 350-17, California Institute of Technology, Pasadena, CA 91125, USA}
\affiliation{Canadian Institute for Theoretical Astrophysics, 60 St. George Street, Toronto, ON, M5S 3H8, Canada}
\author{Yanbei Chen}
\affiliation{Theoretical Astrophysics 350-17, California Institute of Technology, Pasadena, CA 91125, USA}

\begin{abstract}
In this paper we study the scalar Green function in the Kerr spacetime using WKB methods. The Green function can be expressed by Fourier-transforming to its frequency-domain counterpart, and with the help of complex analysis it can be divided into parts: 1) the ``direct part" which propagates on the light cone and dominates at very early times; 2) the ``quasinormal-mode part" which represents the waves traveling around the photon sphere, and is important at early and intermediate times; 3) the ``tail part" which is due to scattering by the Coulomb-type potential and becomes more important at later times.  We focus on the ``quasinormal-mode part" of the Green function and derive an approximate analytical formula for it using WKB techniques. This approximate Green function diverges at points that are connected by null geodesics, and recovers the four-fold singular structure of Green functions that are seen in Schwarzschild and other spacetimes. It also carries unique signatures of the Kerr spacetime such as frame-dragging. Along the way, we also derive approximate quasinormal mode wavefunctions and expressions for the black hole excitation factors in the Kerr spacetime. We expect this work to benefit the understanding of both wave propagation and the problem of self-force in the Kerr spacetime. 
\end{abstract}

\pacs{4.30.-w, 4.25.Nx, 4.30.Nk}

\maketitle

\section{Introduction}
Supermassive Black Holes (SMBHs), sometimes also referred to as  Massive Black Holes (MBHs) are the black holes with masses higher than $10^5\,M_\odot$; they are believed to exist {at the centers of} almost all galaxies. The closest example, Sagittarius A* \cite{Melia}, is the one at the center of our own galaxy, and was discovered by radio observations. 
These SMBHs exist in the dense cores of galaxies, where there are a large number of stars and compact stellar remnants such as white dwarfs, neutron stars, and less-massive black holes. Though the vast majority of these stars are expected to be {at large distances compared to the} Schwarzschild radii of the SMBH, many mechanisms have been proposed to drive stars and compact objects closer. {For example, when  a galactic merger results in a binary system of SMBHs, Kozai-Lidov effects and additional scattering due to the companion SMBH can cause stars to be captured by one of the holes~\cite{Ivanov2005,chen}}.
Once a compact object moves into orbit close enough to the central SMBH, the radiation of gravitational waves dominates the evolution of the system, and the object will eventually merge with the SMBH due to radiation reaction. 
{Such a system is called an extreme mass ratio inspiral (EMRI), and these systems are a primary target for proposed, future space-based gravitational wave detectors, such as the eLISA concept~\cite{eLISA}. Such experiments require accurate modeling of the evolution of the EMRI up to merger.}

One way to compute the {evolution of an EMRI through the} effects of radiation reaction
was proposed by Mino, Sasaki and Tanaka \cite{Mino1996} (for recent reviews on this subject, see \cite{Barack2009, Poisson2011}). 
The basic idea is to express the metric perturbation as the convolution between the Kerr gravitational Green function and the less-massive object's stress energy tensor, and hence obtain the radiation reaction {``self-force'' on the less-massive object}. 
In this formalism, it is physically clear how the test object sources gravitational perturbations, which propagate in the curved spacetime and exert back-reaction onto the source. 
However, for realistic EMRI evolutions, it is highly non-trivial to obtain the Green function.

For the Schwarzschild spacetime, Dolan and Ottewill \cite{Dolan2011} used a spectral method to relate the scalar Green function to the quasinormal modes. 
By adopting a matched expansion technique, they managed to obtain an approximate analytical form of the Green function {in the high-frequency, eikonal limit.} 
Moreover, they showed that the Green function is singular on the lightcone, and it has a four-fold singular structure of alternating singularities: $\delta(\sigma), 1/\sigma, -\delta(\sigma), -1/\sigma$ (see Sec. \ref{secgf} for discussions), where $\sigma$ is the Synge's world function. This four-fold singular structure matches the earlier expectation by Casals {\it et al.} \cite{Casals2009a,Casals2009b}, which was proved by using the Hadamard ansatz for the direct part of the Green function. On the other hand, Zengino\u{g}lu and Galley \cite{Zenginoglu2012} used numerical methods to obtain the time-domain scalar Green function in Schwarzschild background. They also observed the four-fold singular structure as a result of caustic echoes. 

In this work, we study the scalar Green function for a Kerr background with generic spin. We use a spectral representation of the Green function {in eikonal limit, extending the results of~\cite{Dolan2011} to the Kerr spacetime. To do this,} we apply the WKB techniques developed in \cite{Yang2012a} to derive an approximate analytical formula for the quasinormal-mode (QNM) part of the Green function. Along the way, we present analytical Kerr QNM wavefunctions and black hole excitation factors in the eikonal limit, for the first time in the literature. The Green function is expected to be singular on the points connected by null geodesics, and we confirm this for our approximate Green function using numerical investigations. We also recover the four-fold singular structure of the Green function, as seen in other spacetimes~{\cite{Casals2009a,Harte:2012uw,Casals2012}, including Schwarzschild~\cite{Dolan2011,Zenginoglu2012}}. {This study makes progress towards solving the problem of the evolution of EMRI systems in the Kerr spacetime using the self-force approximation. With additional, future work on the ``direct part" and ``tail part" of the Kerr Green function, the method for calculating the self-force in Schwarzschild presented in~\cite{casals2013b} can be extended to Kerr.}

This paper is organized as follows. In Sec.~\ref{sec1} we discuss the spectral representation of the scalar Green function. In Sec.~\ref{sec2} we review the WKB approximation for QNM frequencies and present the details for computing the QNM wave functions. In Sec.~\ref{sec3} we explicitly obtain the so-called ``black hole excitation factors" from the WKB wave functions. Combining all the results in previous sections,  we derive an approximate formula for the scalar Green function in Sec.~\ref{sec5}. It is shown to posses the four-fold singular structure, and in Sec.~\ref{sec6} we verify numerically that it is singular between points that are connected by null geodesics.  We conclude ion Sec.~\ref{sec7}. {Throughout this paper, we use geometric units $G = c =1$, and unless otherwise specified we also take the black hole's mass to be unity, $M = 1$. This is in contrast to many works in the QNM literature, where authors often take $2 M = 1$.}

\section{Spectral Decomposition}
\label{sec1}

In this section we review the spectral decomposition of the scalar Green function in the Kerr spacetime. In a generic spacetime, the scalar Green function satisfies the equation
\begin{align}
\label{scalargreen}
\Box G_{\rm ret}(x,x')&=\frac{1}{\sqrt{-g}}\partial_{\mu}\left(\sqrt{-g}g^{\mu\nu}\partial_{\nu}G_{\rm ret}\right)
\notag \\ &
=-4\pi\delta^{(4)}(x-x')\,.
\end{align}
We only consider the retarded Green function $G_{\rm ret}(x,x')$, for which $x'$ lies on or within the future lightcone of $x$. In addition, we use Boyer-Lindquist coordinates, in which the line element for the Kerr spacetime is written as
\begin{align}
ds^2=&-\left(1-\frac{2Mr}{\Sigma}\right)dt^2-\frac{4aMr\sin^2{\theta}}{\Sigma}dt d\phi
\notag \\ &
+\frac{\Sigma}{\Delta}dr^2+\Sigma d\theta^2
\notag \\ & 
+\sin^2{\theta}\left(r^2+a^2+\frac{2Ma^2r\sin^2{\theta}}{\Sigma}\right)d\phi^2\,.
\end{align}
Here, $\Delta \equiv r^2-2Mr+a^2$, $\Sigma=r^2+a^2\cos^2{\theta}$, $M$ is the mass of the background black hole and $a$ is its spin parameter. 

The scalar wave equation~\eqref{scalargreen} in the Kerr spacetime is separable in the frequency domain, which allows us to write down the following spectral decomposition of the Green function:
\begin{align}
\label{eqsphdec}
G_{\rm ret}(x,x')=&\frac{2}{ \sqrt{r^2+a^2} \sqrt{r'^2+a^2}} \int d\omega e^{-i\omega(t-t')} \notag \\
&\times \sum_m e^{i m(\phi-\phi')} \sum_l S_{lm\omega}(\theta)S^*_{lm\omega}(\theta')\notag \\
&\times \tilde{G}_{lm\omega}(r,r')\,.
\end{align}
Here, $S_{lm\omega}(\theta)$ is the spheroidal harmonic function, which solves the scalar angular Teukolsky equation~\cite{Teukolsky1} (see also~\cite{Fackerell,casals}); its analytical approximation is given in~\cite{Yang2012a} for the case where $l \gg 1$, and we review this approximation in Sec.~\ref{sec2}. The function $\tilde{G}$ is the radial Green function, which satisfies
\begin{align}
\label{eqraw}
\frac{d^2 \tilde{G}}{dr_*^2}&+\left[ \frac{K^2-\Delta \lambda_0}{(r^2+a^2)^2}-H^2-\frac{dH}{dr_*}\right ]\tilde{G}
\nonumber \\ 
&=-\delta(r_*-r'_*)\,,
\end{align}
where 
\begin{align}
H &=r\Delta/(r^2+a^2)^2,\\
K &=ma-\omega(r^2+a^2), \\
\frac{dr_*}{dr} &=\frac{(r^2+a^2)}{\Delta}, \\ 
\lambda_0 & =A_{lm}+a^2\omega^2-2am \omega\,,\label{eqdrstar}
\end{align}
$A_{lm}$ is the eigenvalue of the angular Teukolsky equation~\eqref{AngTeuk}, and an in-going (out-going) boundary condition is used at the horizon (spatial infinity). 
For $a=0$, the expression~\eqref{eqsphdec} recovers the Green function in the Schwarzschild limit as given in \cite{Dolan2011}, {recalling that for $a=0$ the spheroidal harmonics $S_{lm\omega}(\theta)$ combined with the azimuthal wavefunctions $e^{im\phi}$ limit to the spherical harmonics $Y_{lm}(\theta,\phi)$, and that 
\begin{align}
\sum_m Y^*_{lm}(\theta',\phi') Y_{lm}(\theta,\phi) = \frac{2l+1}{4 \pi} P_l(\cos \gamma) \,,
\end{align}
where $P_l$ are the Legendre polynomials, and $\gamma$ is the angle between a point with angular coordinates $(\theta',\phi')$ and a point with coordinates $(\theta, \phi)$.}

The Green function is constructed using the two homogeneous solutions $u_{\rm in}$ and $u_{\rm out}$ of Eq.~\eqref{eqraw}. 
The in-going solution $u_{\rm in}$ satisfies Eq.~\eqref{eqraw} with an in-going boundary condition at the background black hole's horizon,
\begin{equation}
\label{equinasym}
u_{\rm in}(\omega,r) = \left\{
\begin{array}{c}
e^{-i\bar{\omega} r_*}\,, \qquad r_*\rightarrow-\infty\,, \\
\\
C^{-}_{lm\omega}e^{-i\omega r_*}+C^{+}_{lm\omega}e^{i\omega r_*}, \; r_* \rightarrow \infty\,,
\end{array}\right.
\end{equation}
where $\bar{\omega}=\omega-ma/(2Mr_+)$ and $r_+$ is the horizon radius.
Similarly, the out-going solution $u_{\rm out}$ satisfies an out-going boundary condition at spatial infinity,
\begin{equation}
\label{equoutasym}
u_{\rm out}(\omega,r) \equiv \left\{
\begin{array}{c}
D^{-}_{lm\omega}e^{-i\bar{\omega} r_*}+D^{+}_{lm\omega}e^{i\bar{\omega} r_*}\,,  \\
\qquad \qquad \qquad \qquad \qquad r_*\rightarrow-\infty\, \\
\\
e^{i\omega r_*}\,, \qquad \qquad  r_* \rightarrow \infty\,.
\end{array}\right.
\end{equation}
Using the in-going and out-going homogeneous solutions above, the radial Green function can be written as
\begin{equation}\label{eqradg}
\tilde{G}_{lm\omega}(r,r') = -\frac{u_{\rm in}(r_{<})u_{\rm out}(r_{>})}{W_{l\omega}}\,,
\end{equation}
with $r_{<}$ given by ${\rm min}(r,r')$ and $r_{>}$ given by ${\rm max}(r,r')$; the Wronskian $W_{lm}$ is a constant given by
\begin{equation}
W_{l\omega} = u_{\rm in}\frac{d u_{\rm out}}{d r_*}-u_{\rm out}\frac{d u_{\rm in}}{d r_*} = 2 i \omega C^-_{lm \omega}\,.
\end{equation}

At some particular complex-valued frequencies $\omega_{lmn}$, the in-going wave solution $u_{\rm in}$ also satisfies an out-going boundary condition at infinity: $C^{-}_{lm\omega}=0$. Given this boundary condition, the in-going solution must be a multiple of the out-going solution. In other words, $u_{\rm in}$ and $u_{\rm out}$ are degenerate at these frequencies. 
As a consequence, the out-going wave solution $u_{\rm out}$ must correspondingly satisfy the in-going wave condition at the horizon: $D^{+}_{lm\omega}=0$. These solutions are called the quasinormal modes (QNMs).
By construction, at these QNM frequencies $D^{-}_{l\omega_{lmn}}C^{+}_{l\omega_{lmn}}=1$ holds, which can be seen by combining Eqs.~\eqref{equinasym}, \eqref{equoutasym}, and the degeneracy condition. 

To evaluate the Green function, we insert Eq.~\eqref{eqradg} back into Eq.~\eqref{eqsphdec} to perform the integral over frequency $\omega$. As in the Schwarzschild case~\cite{leaver3}, this integral can be evaluated using the residue theorem and divided into three pieces. The first piece, called the ``direct part,'' is the integral on the high frequency arc, and it is expected to quickly approach zero after an initial pulse~\cite{leaver3}. The second piece is the integral on the branch cut on the imaginary frequency axis, which contributes to the power-law decay at late times and is also non-negligible at intermediate and early times~\cite{casals2012a, casals2012b, casals2013a, casals2013b}. The final piece comes from the residues of the poles whose frequencies correspond to those of the QNMs, and it is important at early and intermediate times. In this work we focus on the QNM contribution to the Green function. It has the following form (with $u_{\rm out}$ replaced by $u_{\rm in}$ for simplicity, as they are degenerate at the QNM frequencies): 

\begin{align}
\label{eqqnmgreen}
G_{\rm QNM}&(x,x')= \frac{8\pi}{ \sqrt{r^2+a^2} \sqrt{r'^2+a^2}}  \notag \\ &
\times {\rm Re}\Biggl[\sum_m e^{i m(\phi-\phi')}\sum_{l, n} S_{lm\omega}(\theta)S^*_{lm\omega}(\theta')  \notag \\ &
\qquad \times \mathcal{B}_{lmn}\tilde{u}_{\rm in}(r)\tilde{u}_{\rm in}(r')e^{-i\omega_{lmn} T}\Biggr].
\end{align}
As before, $\omega_{lmn}$ is the QNM frequency with spheroidal harmonic indices $l,m$ and overtone number $n$ (except for near extreme Kerr black holes, which may have two branches of QNMs~\cite{Yang2012b,Yang2013a}); we have defined $T = t -t ' - r_* - r_*'$; and we have used a normalized in-going wave solution $\tilde{u}_{\rm in}(r)$ (see also \cite{Dolan2011})
\begin{equation} 
\label{eqtildeu}
\tilde{u}_{\rm in}(r) \equiv u_{\rm in}(r) \times \left[ C^+_{lm\omega}e^{i\omega_{lmn}r_*}\right ]^{-1} \,,
\end{equation}
which asymptotes to $1$ when $r \rightarrow +\infty$.
The coefficient $\mathcal{B}_{lmn}$ is usually referred to as the black hole excitation factor, and it indicates the amount that the QNMs are excited by the $\delta$-function source in Eq.~\eqref{scalargreen}.  
It is given by
\begin{equation}\label{eqbhexc}
\mathcal{B}_{lmn} \equiv \left [\frac{C^+_{lm\omega}}{2\omega}\left(\frac{\partial C^-_{lm\omega}}{\partial \omega}\right)^{-1} \right ]_{\omega=\omega_{lmn}}\,.
\end{equation}
Although Eq.~\eqref{eqraw} is manifestly invariant under the coordinate freedom $r_* \rightarrow r_*+r_0$, $\mathcal{B}_{lmn}$ picks up a multiplicative factor of $e^{-2i\omega_{lmn}r_0}$ from the changes to the amplitudes $C_{lmn}^+$ and $C_{lmn}^-$ in the asymptotic form~\eqref{equinasym} of $u_{\rm in}$. This factor is cancelled by those coming from $C_{lmn}^+$ in Eq.~\eqref{eqtildeu}, leaving the Green function~\eqref{eqqnmgreen} invariant, as one would expect. 
We are therefore at liberty to choose $r_0$ such that 

\begin{align} \label{eq:TortR}
r_* =& r+\frac{2 r_+}{r_+-r_-}\log\left (\frac{r}{r_+}-1\right ) 
\notag \\ &
-\frac{2r_-}{r_+-r_-}\log\left (\frac{r}{r_-}-1\right )\, ,
\end{align}
where $r_{\pm}=1\pm\sqrt{1-a^2}$ are the outer- and inner-horizon radii. 
For $a=0$ expression~\eqref{eq:TortR} reduces to the commonly used relation $r_*=r+2\log(r/2-1)$ for Schwarzschild black holes. 

In the following sections, we use a WKB analysis and matched-expansion techniques to obtain approximate analytical forms of $S_{lm\omega}, \, \tilde{u}_{\rm in}$ and $\mathcal{B}_{lmn}$. 
After that we evaluate the summations in Eq.~\eqref{eqqnmgreen} for $G_{\rm QNM}(x,x')$ and discuss the resulting expression for the Green function.

\section{QNMs in the eikonal limit}
\label{sec2}

In order to evaluate the summation in Eq. (\ref{eqqnmgreen}) to obtain the QNM part of the Green function, we have to insert the frequencies and wavefunctions of the QNMs. While the exact frequencies and wavefunctions can only be obtained numerically, analytical approximations for them are available in the eikonal limit $l \gg 1$. {The study of QNMs in the eikonal limit has a long history, including the use of WKB techniques to solve for the QNMs~\cite{Schutz:1985km,Iyer,Dolan:2009nk,Dolan10}, and the use of a geometric-optics correspondence between the QNMs and the unstable null orbits of the spacetime, e.g.~\cite{Goebel1972,Ferrari,Cardoso:2008bp}}. In this section, we review the WKB analysis of the QNMs of the Kerr black hole. This review is based on the results of~\cite{Yang2012a}, which first extended the WKB analysis and the geometric correspondence with null orbits to Kerr black holes with arbitrary spins and azimuthal quantum numbers (cf.~\cite{Dolan10}).

For later convenience, let us first define a set of variables {to stand in for the angular quantum numbers $l$ and $m$ and the overtone number $n$,}
\begin{align}
L\equiv l+1/2, & &  N\equiv n+1/2, & & \mu\equiv m/L. 
\end{align}
In the eikonal limit, QNM frequencies are given by
\begin{align}
\omega_{lmn}&=\omega_R-i\omega_I \notag \\
& =L\Omega_R(\mu,a)-i N \Omega_I(\mu,a)+O(1/L)
\end{align} 
The functional forms of $\Omega_R(\mu,a)$ and $\Omega_I(\mu,a)$ was derived in~\cite{Yang2012a}, and they are reproduced in Appendix~\ref{sec:WKBFrequencies}. 
In the case of a Schwarzschild background with $a=0$, the two are equal, with $\Omega_R = \Omega_I= 1/\sqrt{27}$, and both are independent of $\mu$. 
In fact, for generic Kerr black holes, the function $\Omega_R(\mu,a)$ can be determined by the orbital and precession frequencies of a corresponding spherical photon orbit ($\mu$ can be viewed as parameterization of all spherical photon orbits), while $\Omega_I(\mu,a)$ is given by the Lyapunov exponent of the same orbit. 
This correspondence {provides a geometric understanding of the high-frequency QNMs} (more details are found in ~\cite{Yang2012a}).

\subsection{The radial wavefunction}

As shown by Teukolsky \cite{Teukolsky1}, the angular and radial dependencies of perturbations of Kerr can be separated in the frequency domain. The separated QNM wavefunctions can then be approximated analytically using a WKB analysis. The radial wavefunction $u_{\rm in}$ describes a scattering problem,
\begin{equation}
\label{eqteurad}
\frac{d^2 u_{\rm in}}{dr_*^2}+Q_{lm}(\omega,r)u_{\rm in}=\frac{d^2 u_{\rm in}}{dr_*^2}+\frac{K^2-\Delta \lambda_0}{(r^2+a^2)^2}u_{\rm in}=0\,.
\end{equation}
It is easy to see that $Q_{lm}(\omega,r)$ (in what follows, we suppress the $lm$ subscript for brevity) is of the order $L^2$, and lower order terms in $L$ have been dropped from the potential (here $\omega$ is a QNM frequency). It is natural to apply a WKB expansion, with the expansion parameter scaling as $1/L$:
 \begin{align}\label{eqradwkb}
u(r) &\sim e^{S_0+S_1+...} \,, \\
S_0&=\pm i\int^{r_*} \sqrt{Q(\omega,r)}dr_*, \\ 
S_1 &=-\frac{1}{4}\log[Q(\omega,r)]\,.
\end{align}
 
The leading term $S_0$ contributes mostly to the phase, while the next-to-leading term $S_1$ contributes mostly to the amplitude. Since $S_0$ scales as $L$ and $S_1$ scales as $\log L$, the phase varies much faster than the amplitude. In addition, because $\omega=\omega_R-i\omega_I$ is a complex number and $\omega_R \propto L^1,\,\omega_I \propto L^0$ ($\omega_R \gg \omega_I$), $Q(\omega_{lmn},r)$ is mostly real and $S_0$ also contains a relatively small real part, which contributes to the amplitude factor.  We can single it out by further expanding $Q(\omega_{lmn},r)$ as $Q(\omega_{R},r)-i\omega_I \partial Q/\partial \omega$, and correspondingly $S_0$ becomes
\begin{align}
S_0 \approx \pm i\int^{r_*}\sqrt{Q(\omega_R,r)}dr_* 
\pm  \int^{r_*} \mathcal{Q}_I\,dr_*\,,
\end{align}
where
\begin{align}
\mathcal{Q}_I \equiv  \frac{\partial_\omega Q|_{\omega_R}}{2\sqrt{Q(\omega_R,r)}} \omega_I,
\end{align}
and we only keep the leading order term for $S_1$,
\begin{align}
S_1 \approx -\frac{1}{4}\log[Q(\omega_R,r)]\,.
\end{align} 

\begin{figure}[t,b]
\includegraphics[width=0.48\columnwidth]{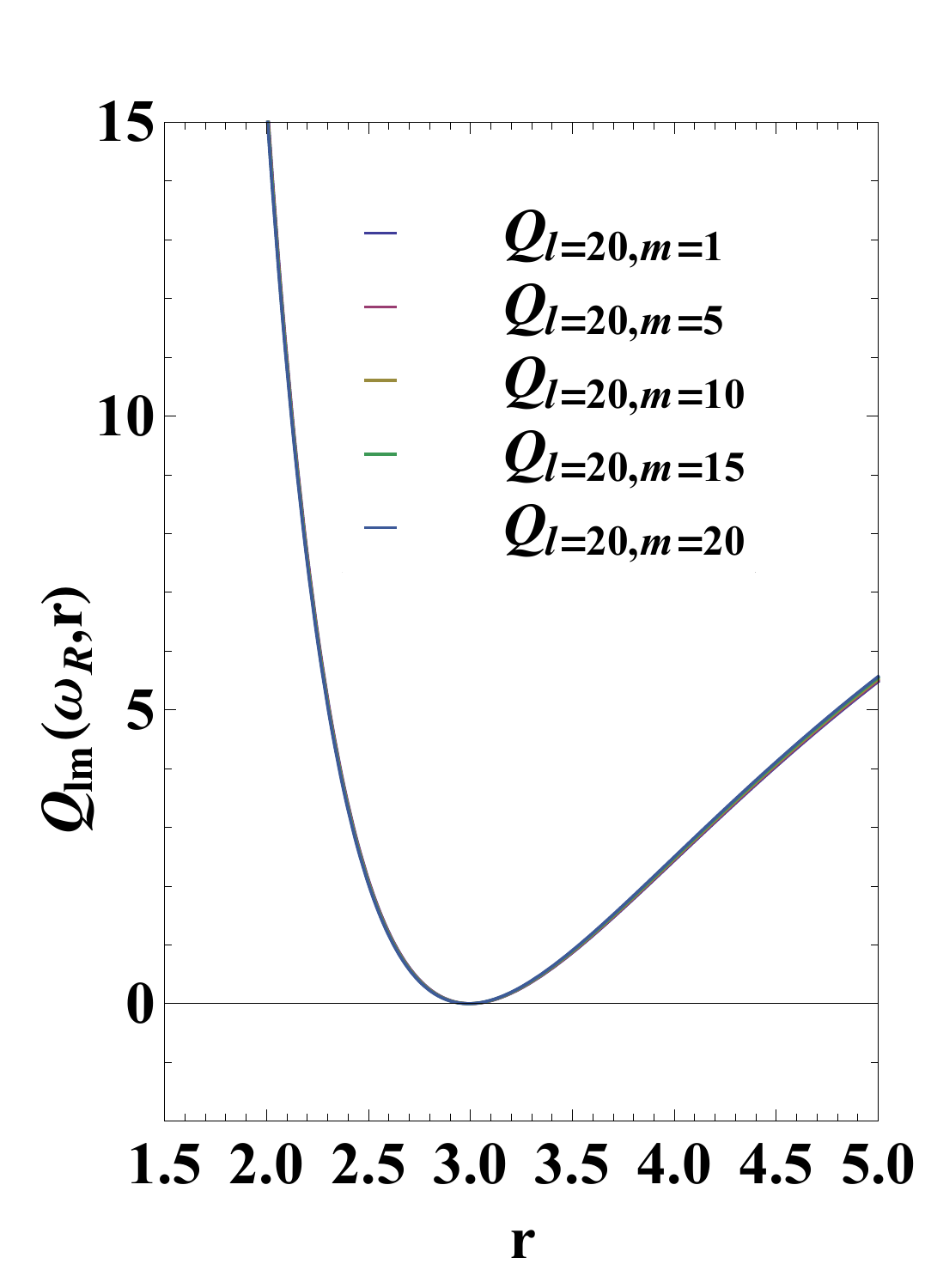}
\includegraphics[width=0.48\columnwidth]{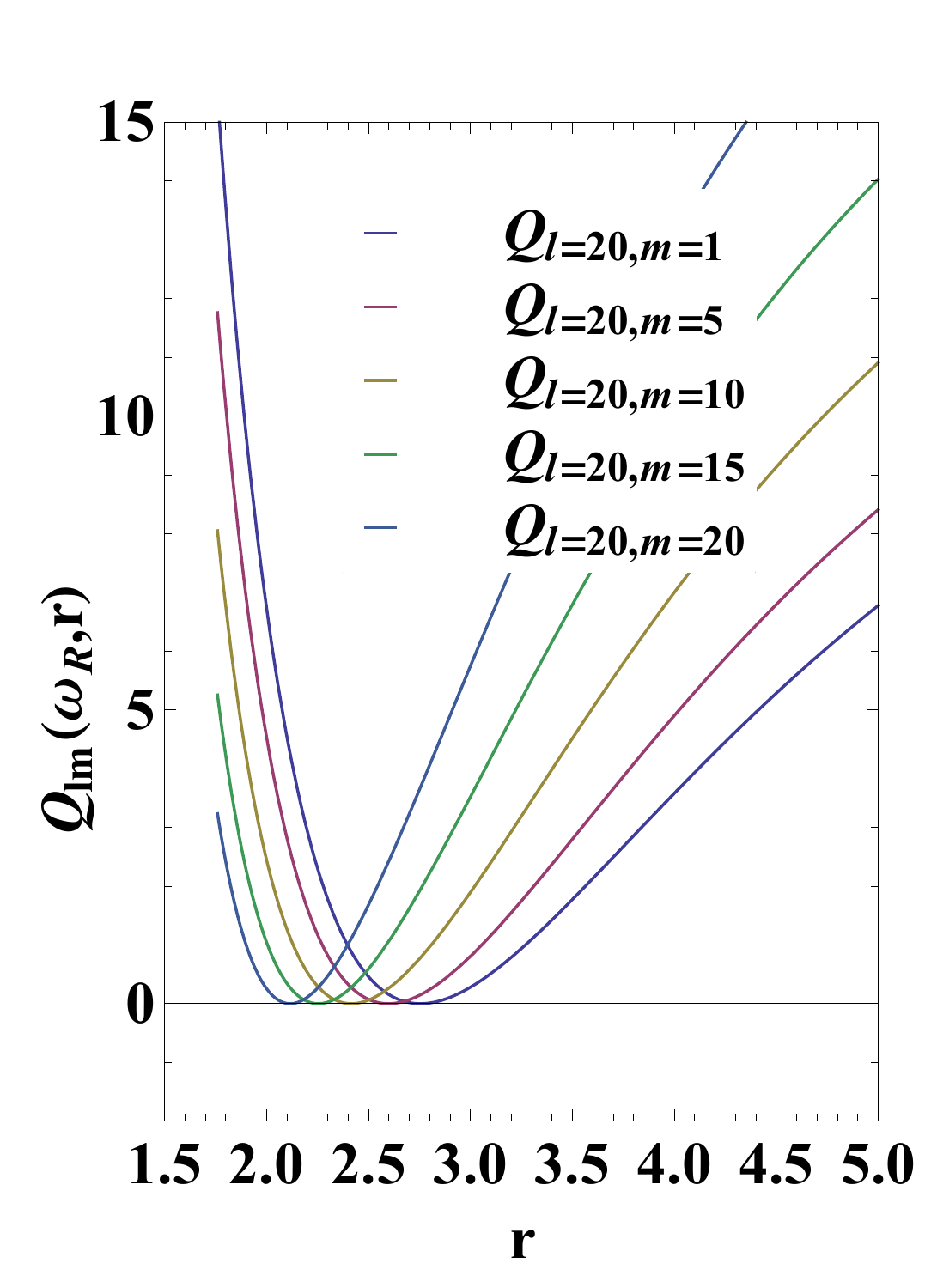}
\caption{ 
The radial potential $Q$ for different $(l,m)$ combinations. The figure on the left corresponds to $a=0.01$, and the figure on the right corresponds to $a=0.65$. On the right hand figure, the curves with extrema location from left to right correspond to $Q(\omega_R,r)$ for $(l=20,m=20)$, $(l=20,m=15)$, $(l=20,m=10)$, $(l=20,m=5)$ and $(l=20,m=1)$ modes. 
}
\label{fig:QvsR}
\end{figure}

The radial potential $Q(\Omega_R,r)$ is a positive function except when it is near its extrema, $r_p$, where in the eikonal limit it approaches zero~\cite{Iyer}. 
In Fig.~\ref{fig:QvsR} we plot $Q(\Omega_R,r)$ for various $m$ values and $l=20$ for two values of the spin parameter. 
We follow the convention of~\cite{Yang2012b,Yang2013a} in calling the extrema the ``peak'' of the potential, though it is a minimum of $Q(\Omega_R,r)$, because it is $\omega_R^2 - Q(\omega_R,r)$ that corresponds to the potential in the usual one dimensional, quantum mechanical scattering problem (see also~\cite{Iyer}). 

It is important to note that the WKB analysis breaks down near the peak of the potential. 
This means Eq.~\eqref{eqradwkb} only works in a region away from the position of the peak $r_p$, and we need a separate treatment for the wavefunction near the peak to connect to WKB solutions on either side of the peak. 
This matched-expansion procedure is carried out in Sec.~\ref{sec3}, where we work out the black hole excitation factors.

 \subsection{The angular wavefunction}
 
 The angular Teukolsky equation has the following form,
\begin{align}
\label{AngTeuk}
0= &\frac{d}{\sin{\theta}d\theta}\left(\sin{\theta}\frac{d u_{\theta}}{d \theta}
\right) + \biggl(a^2\omega^2\cos^2{\theta} 
\notag \\ 
& -\frac{m^2}{\sin^2{\theta}}+ {}_s A_{lm}
- 2 a\omega s \cos{\theta} 
 \nonumber \\
&
 - \frac{2 ms \cos{\theta}}{\sin^2{\theta}}
-s^2\cot^2{\theta}+s \biggr)u_{\theta}\,,
\end{align}
where $s$ is the spin index for the perturbation field: $s=0,\,-1,\,-2$ corresponds to scalar, electromagnetic and gravitational perturbations respectively. All the terms containing $s$ in the potential are sub-leading in $L$, and we neglect them from here on (the same is true for the radial equation, and we have already neglected the $s$-dependent terms in that analysis). 
The angular Teukolsky equation can also be written in a form suitable for WKB analysis. Defining $x \equiv \log[\tan(\theta/2)]$, the angular Teukolsky equation becomes
\begin{equation}
\label{eqbound}
\frac{d^2u_{\theta}}{d x^2}+V^\theta u_{\theta}=0 \,, 
\end{equation}
with
\begin{align}
V^\theta &= a^2\omega^2\cos^2{\theta}\sin^2{\theta}-m^2+A_{lm}\sin^2{\theta} \notag \\ 
&\equiv\Theta \sin^2\theta
\,.
\end{align}
This equation describes a bound state problem with $V^{\theta}$ serving as the potential well \cite{Yang2012a}. For $x \rightarrow \pm \infty$ ($\theta \rightarrow \pi, 0$), $V^{\theta}$ becomes negative and the wave solution tends to decay to zero: $u_{\theta} \rightarrow e^{-|mx|}$, so the wavefunction is trapped inside the potential well. By applying the WKB expansion to the second order we obtain
\begin{align}
u_{\theta} \sim \frac{1}{(\Theta \sin^2\theta)^{1/4}}e^{\pm i S_{\theta}}\,,
\end{align}
with 
\begin{align}
S_{\theta} =\int^{\theta}_{\pi/2}\sqrt{\Theta}d\theta'\,.
\end{align}
We have chosen $\theta = \pi/2$ as our origin for the integration because $\Theta$ is symmetric about that point; a different choice simply modifies the amplitude and phase of $u_\theta$.
As in the case of the radial potential $Q$, we expand $\Theta$ into real and imaginary parts as $\Theta = \Theta_R+i\Theta_I$. In the eikonal limit they are given by
\begin{align}
\Theta_R &=a^2\omega^2_R\cos^2\theta-\frac{m^2}{\sin^2\theta}+A_R \,, \\
\Theta_I &=A_I-2a^2\omega_R\omega_I\cos^2\theta \,,
\end{align}
where $A_{lm}=A_R+i A_I$. Here $A_R \propto L^2$ and $A_I \propto L$, and hence $\Theta_R \propto L^2$ and $\Theta_I \propto L$.

Using this expansion for $\Theta$, we can then separate the phase and amplitude contributions of $S_\theta$ to the wavefunction $u_{\theta}$, writing
\begin{align}
S_\theta \approx \int^{\theta}_{\pi/2} \sqrt{\Theta_R}d\theta' + i \int^{\theta}_{\pi/2}\frac{\Theta_I}{2\sqrt{\Theta_R}}d\theta',
\end{align}
where the first term gives the phase contribution, and the second term the amplitude.

We recall that the solutions for Eq.~\eqref{AngTeuk} are the spin-weighted spheroidal harmonics, or simply the spheroidal harmonics $S_{lm\omega}(\theta)$ since we have neglected the spin $s$. 
Noting that the spheroidal harmonics obey the identity ${}_s S_{lm\omega} ( \pi - \theta) = (-1)^{l+m}{}_{-s} S_{lm\omega} ( \theta)$, and that $\Theta(\pi - \theta) = \Theta(\theta)$, we see that we can construct $S_{lm\omega}(\theta)$ from a linear combination of the above two solutions $u_\theta$,
\begin{align}
\label{eqslm}
S_{lm\omega} =& \frac{C}{(\Theta_R \sin^2\theta)^{1/4}}\left[e^{ iS_\theta} 
+(-1)^{l+m}e^{- iS_\theta}\right]\,,
\end{align}
where $C$ is some constant which can be fixed by normalization condition $\int S_{lm\omega}S^*_{lm\omega}d\Omega=1$, or 
\begin{align}
\label{eqnnorm1}
1 & = 2\pi C^2 \int_{\theta_-}^{\theta_+} \frac{d \theta}{\sqrt{\Theta_R}} \left[e^{i (S_\theta -  S_\theta^*)} + e^{i (S^*_\theta -  S_\theta)} \right] 
\nonumber \\ 
& \qquad \qquad + (-1)^{l+m} \left[e^{i(S_\theta + S_\theta^*)} + e^{-i(S_\theta + S_\theta^*)} \right] 
\notag \\
&=2\pi C^2 \int_{\theta_-}^{\theta_+} \frac{d \theta}{\sqrt{\Theta_R}}\left[e^{ 2 N \upsilon(\theta)}+e^{ -2 N \upsilon(\theta)}\right]\,,
\end{align}
Here, $\theta_{\pm}$ are the angles at which $\Theta_R$ becomes zero. For future convenience, we have denoted contribution of $S_\theta$ to the amplitude as
\begin{align}
N \upsilon(\theta) = \int^{\theta}_{\pi/2}\frac{\Theta_I}{2\sqrt{\Theta_R}}d\theta'\,,
\end{align}
In going to the second line of Eq.~\eqref{eqnnorm1}, we have dropped the terms involving $S_\theta + S^*_\theta \propto L$, since those terms involve rapid oscillations of the phase in the exponential, which average to terms of relative $O(1/L)$ in the integration.

As shown in~\cite{Yang2012a}, the imaginary part of $A_{lm}$ arises solely through its functional dependence on the complex frequency $\omega_{lmn}$ at leading order in $L$. 
A practical method for computing $A_I$ is through the usual eigenvalue perturbation theory, keeping in mind that $A_R$ is the leading eigenvalue and using the WKB expressions for $S_{lm}(\omega_R, \theta)$ as the leading order eigenfunctions (so that there is no amplitude contribution to $S_\theta$). The perturbation to $V^\theta(\omega_R)$ from this perspective is given by $- 2 i a^2 \omega_R \omega_I \cos^2 \theta$, and the corresponding perturbation to the eigenvalue is $i A_I$. This allows us to write\footnote{Note that this equation has a sign difference compared to Eqs,~(2.24) and~(2.27) of \cite{Yang2012a}, which are in error.}
\begin{align}
A_I &= \int S^*_{lm\omega_R} S_{lm \omega_R}   \left( 2 a^2 \omega_R  \omega_I \cos^2 \theta\right) d\Omega \nonumber \\
&=   2 a^2 \omega_R  \omega_I \left(\int_{\theta_-}^{\theta_+}\frac{d\theta}{\sqrt{\Theta_R}}\right)^{-1}\int_{\theta_-}^{\theta_+} \frac{\cos^2 \theta}{\sqrt{\Theta_R}}  d \theta \nonumber \\
& \equiv 2 a^2 \omega_R \omega_I \langle \langle \cos^2 \theta\rangle \rangle \,,
\end{align}
where here the normalization for the spheroidal harmonics $S_{lm\omega_R}$ does not receive the contribution from the terms containing $\Theta_I$ which are present in Eq.~\eqref{eqnnorm1}. 
This expression for $A_I$ allows for the convenient simplification
\begin{align}
\Theta_I =  2 a^2 (L \Omega_R )(N \Omega_I) \left( \cos^2 \theta + \langle \langle \cos^2 \theta\rangle \rangle \right) \,,
\end{align}
which leads to a less compact but more explicit formula for $\upsilon(\theta)$,
\begin{align}
\label{eq:AngUpsilon}
\upsilon(\theta) =  a^2 L \Omega_R  \Omega_I \int^\theta_{\pi/2} \frac{  \cos^2 \theta' + \langle \langle \cos^2 \theta\rangle \rangle }{\sqrt{\Theta_R}}d\theta'\,.
\end{align}
It is important to recall that in both of the above equations, $\langle \langle \cos^2 \theta\rangle \rangle$ is a constant.

{We mention briefly that the geometric correspondence between QNMs and unstable null orbits gives the geometric interpretation of $\langle \langle \cos^2 \theta \rangle \rangle$ as the orbit-average of $\cos^2 \theta$ over one cycle of the orbit. From this perspective, the average is most conveniently expressed as an integral over the arclength of the orbit (rather than the affine parameter of the null orbit), also known as the ``Mino time'' $\xi$, and defined by $d\xi = d\theta/\sqrt{\Theta_R}$~\cite{Mino:2003yg}. The integral is performed over one cycle in the ``classical regime" of the orbit, where $\Theta_R \ge 0$, giving
\begin{align}
A_I &=2a^2\omega_R\omega_I\langle\langle\cos^2\theta\rangle\rangle \nonumber \\
&=2a^2\omega_R\omega_I \left(\oint d\xi \cos^2\theta \right) \left(\oint d\xi\right)^{-1} \nonumber \\
&=2a^2\omega_R\omega_I \xi_0^{-1}\oint d\xi \cos^2\theta\,.
\end{align}
For this integral, one requires the expression $\theta(\xi)$ for the null orbit. 
As a result, Eq.~\eqref{eqnnorm1} can be rewritten in a form that is related to geometric optics and spherical photon orbits,
\begin{align}
2\pi C^2 \oint & d\xi  \left[e^{2a^2\omega_R\omega_I\int^{\xi}_0(\cos^2\theta+\langle \langle \cos^2\theta \rangle\rangle)d\xi'} \right.
\notag \\ & \left.
+e^{-2a^2\omega_R\omega_I\int^{\xi}_0(\cos^2\theta+\langle \langle \cos^2\theta \rangle\rangle)d\xi'}\right]=1\,.
\end{align}

\begin{figure}[t,b]
\includegraphics[width=0.95\columnwidth]{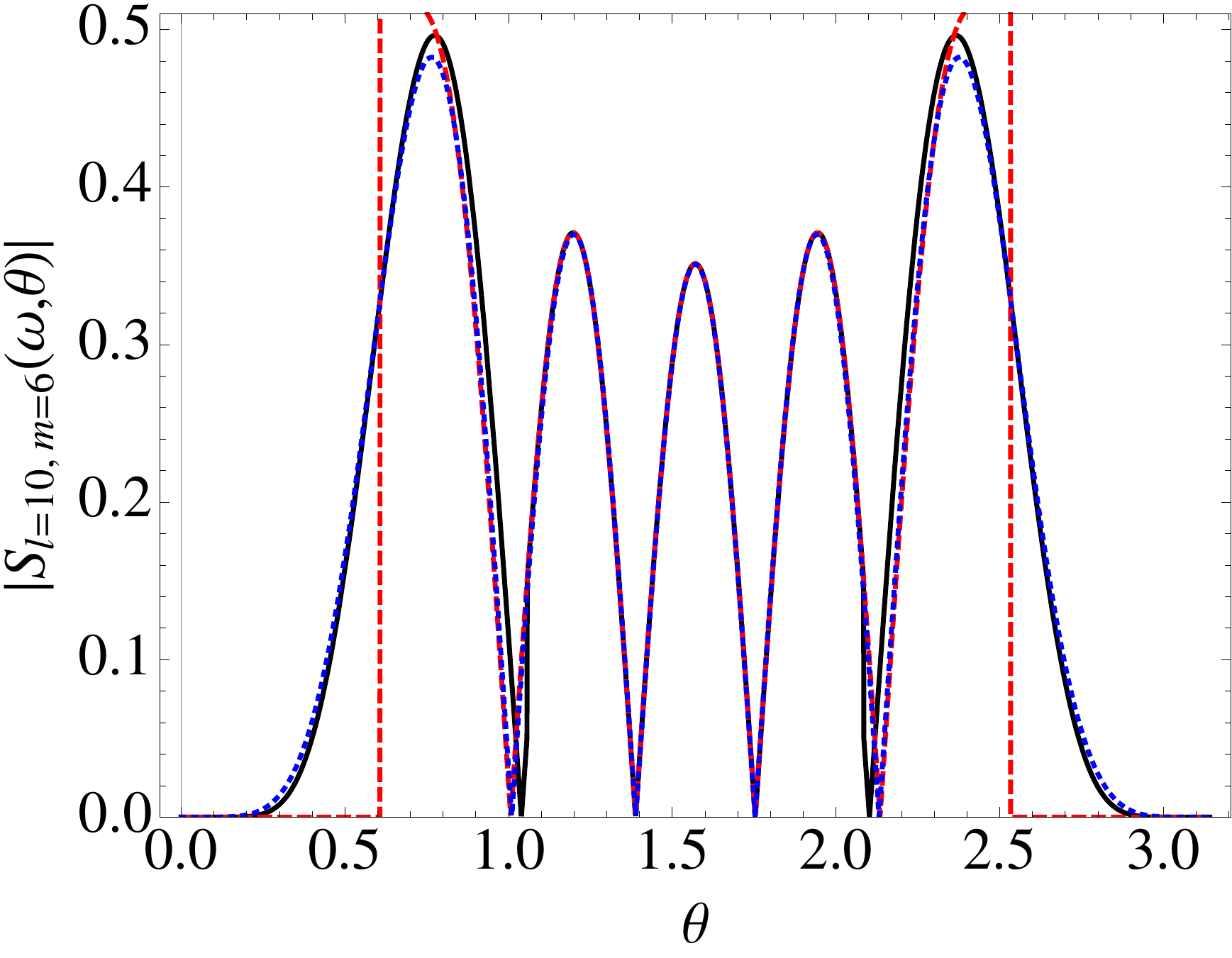}
\caption{(Color online.) WKB approximated wavefunction for $l=10, m=6$ spherical harmonics. The red dashed line corresponds to the WKB wavefunction described by Eq. (\ref{eqslm}), which becomes singular at the classical boundary; values outside the boundary are set to zero. The blue dotted line is the exact $Y_{lm}$ function. The solid black line corresponds to the solution obtained by matching the Airy function near the classical boundary with Eq.~\eqref{eqslm}. In order to generate this plot, we adopt the WKB function expression for $\sin\theta>\pi/L+m/L$ and the Airy function for the {rest of the domain. The wavefunction resulting from the matching is smooth enough that the matching point cannot be discerned by eye.}}
\label{fig:sph}
\end{figure}

The WKB approximation breaks down near the classical boundary $\Theta_R(\theta_{\pm})=0$, where a separate treatment is required to extend the wavefunction outside this classical regime.
In Fig. \ref{fig:sph}, we compare the wavefunction generated by Eq. (\ref{eqslm}) with exact spherical harmonics (for simplicity, we take $a=0$). The dashed red line is predicted by Eq. (\ref{eqslm}). It fits the exact spherical harmonic function (blue dotted line) well except near the classical boundary, where the WKB approximation breaks down and the WKB wavefunction blows up. 
In order to take care of the wavefunction near the classical boundary, we can expand $V^{\theta}$ as $V^\theta(x-x_{\pm}) \approx \partial_x V^\theta|_{x_{\pm}}(x-x_{\pm})$, {where $x_\pm$ are the positions of the boundary,} and then solve the equation
\begin{equation}
\frac{d^2 u_\theta}{dx^2}+(x-x_{\pm})V^\theta{}' u_\theta=0\,.
\end{equation}
The solution turns out to be an Airy function,
\begin{equation}
\label{eq:Airy}
u_\theta \sim {\rm Airy}\left [\left (2L^2\mu^2\sqrt{1-\mu^2} \right )^{1/3}(x-x_{\pm})\right ]
\end{equation}
which can be matched with Eq.~(\ref{eqslm}) in a buffer zone $1/L \ll |x_{\pm}|-|x| \ll1$, where both the WKB approximation and the linear expansion of $V_\theta$ are valid. In addition,  following asymptotic behavior of the Airy function as $ z \to - \infty$ is needed to complete the matching at, e.g. the right hand side of the classical regime,
 \begin{equation}
{\rm Airy}(-z) \sim \frac{\sin{(\frac{3}{2}z^{2/3}}-\pi/4)}{\sqrt{\pi}z^{1/4}}\,.
 \end{equation}
The explicit details of the matching are straightforward but not directly relevant for the results of this paper, and we omit them for brevity. When needed, the Airy function~\eqref{eq:Airy} can be used near the classical boundary in place of the WKB wavefunction~\eqref{eqslm} to obtain a better estimate for $S_{lm}$. The solid black line in Fig.~\ref{fig:sph} is generated using this method, and the result fits well with the spherical harmonic $Y_{lm}$ globally. Since one can show that the boundary treatment only contributes sub-leading WKB terms for Green function, in the remainder of the paper we use Eq.~\eqref{eqslm} to approximate the angular wavefunction.

\section{Matched expansions}\label{sec3}

Given the frequency and wave function of each QNM, the last quantity that needs to be computed is the black hole excitation factor $\mathcal{B}_{lmn}$ defined in Eq.~(\ref{eqbhexc}). This quantity determines the weight of each QNM's contribution to the Green function [cf. Eq.~\eqref{eqqnmgreen}]. Because the amplitude of the wave can be expressed as the convolution between the Green function and the source, this excitation factor also contributes to the weight of each QNM excitation due to a source distribution.

According to Eq.~\eqref{eqbhexc}, in order to compute $\mathcal{B}_{lmn}$, we have to obtain the frequency dependence of the reflection coefficients of both in-going and out-going wave solutions (i.e. $C^{-}_{lm\omega}$ and $C^{+}_{lm\omega}$). As we recall from the WKB analysis on the radial Teukolsky equations in Sec.~\ref{sec2}, the scattering potential $Q(\omega_R,r)$ is approximately zero near its peak, and that is where the WKB expansions fail. In fact, the WKB approximation works in two separate regions: one on each side of the scattering potential. In order to relate the boundary conditions at the horizon to those at spatial infinity we have to connect the WKB solutions on both sides of the potential. This can be done by writing down a separate solution near the peak of the potential and then matching it with WKB solutions on each side. This matched expansion procedure is illustrated in Fig~\ref{fig:potential}.

\begin{figure}[t,b]\centering
\includegraphics[width=1.0\columnwidth]{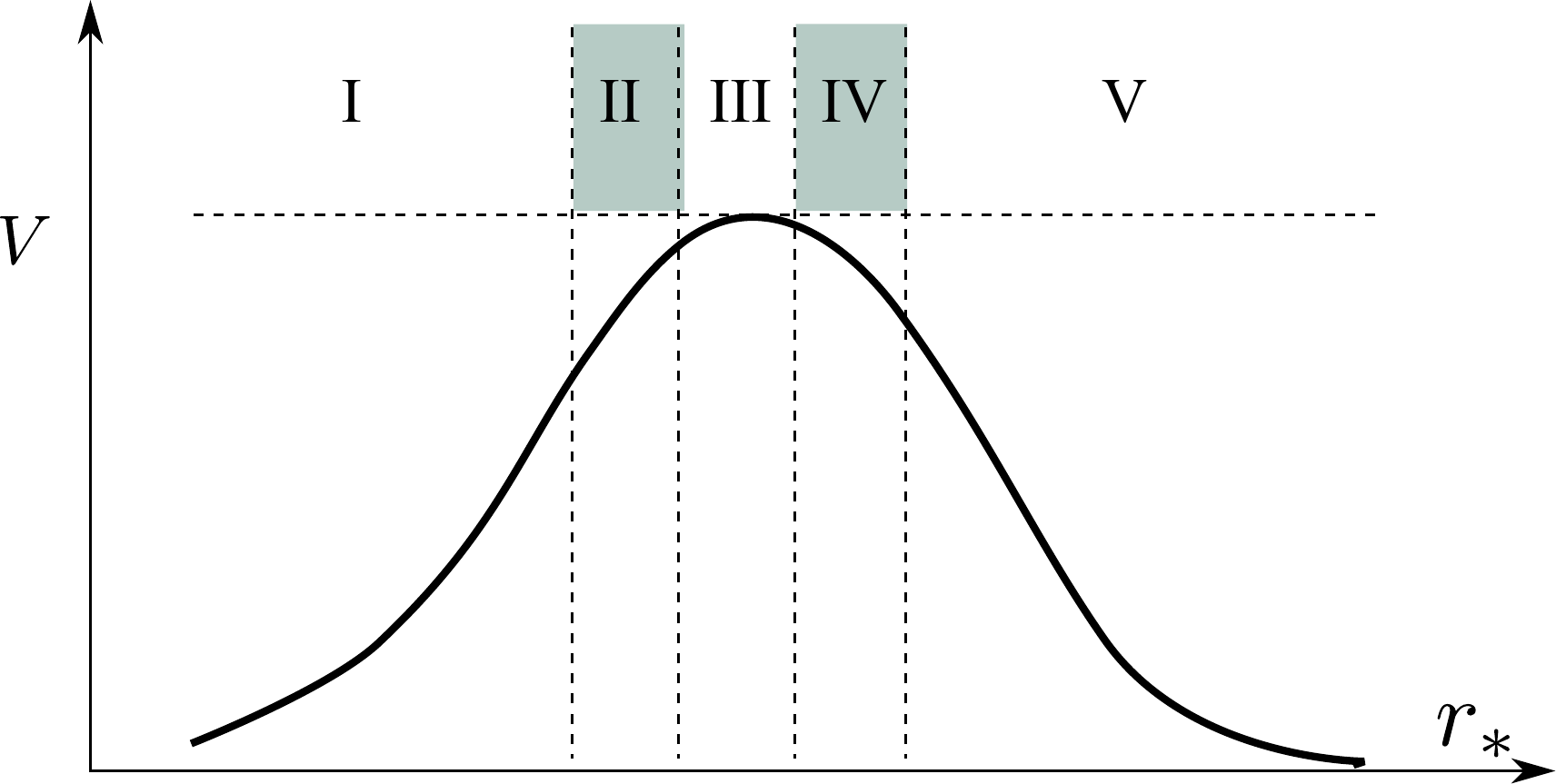}
\caption{(Color online.) Different regimes for the radial wavefunction. The WKB analysis is valid in regime I, II, IV and V. The near-peak regime is located at II, III and IV. Regime II and IV are the buffer zones to match near-peak and WKB solutions.}
\label{fig:potential}
\end{figure}

\subsection{The wavefunction near the peak of the scattering potential}
We start the matching procedure by solving the radial Teukolsky in the near-peak regime (regions II, III and IV in Fig. \ref{fig:potential}). Suppose the peak is located at $r=r_p$, or equivalently $r_*=r_{*p}$. By definition, we have
\begin{equation}
\left . \frac{\partial Q(\omega_R,r)}{\partial r}\right |_{r=r_p}=0\,.
\end{equation}
Combining this with the fact that the potential term also equals zero at its peak $Q(\omega_R, r_p)=0$~\cite{Iyer}\footnote{The geometric correspondence is that $Q=Q'=0$ is the condition for a massless particle to stay on the unstable spherical photon orbit~\cite{Yang2012a}.}, 
we can Taylor-expand $Q(\omega,r)$ around the peak of the potential (assuming $\omega = \omega_{R}-i\omega_I+\epsilon$), 
\begin{align}
\label{eq:QExpand}
Q(\omega,r) &= \frac{1}{2}Q_0''(r_*-r_{*p})^2+\left .\frac{\partial Q}{\partial \omega}\right |_{\omega_R,r_p}(\epsilon-i\omega_I) \nonumber \\
&= \frac{1}{2}Q''_0(r_*-r_{*p})^2 +N\sqrt{2Q''_0}\left(-i+\frac{\epsilon}{\omega_I}\right) \nonumber \\
& \equiv \sqrt{k}z^2+2N\sqrt{k}\left(-i+\frac{\epsilon}{\omega_I}\right)\,,
\end{align}
where $Q''_0 \equiv \partial^2_{r_*} Q |_{\omega = \omega_R,r=r_p}$ is positive near the peak of the potential, and
\begin{align}
z&=k^{1/4}(r_*-r_{*p}), \\
k&\equiv\frac{1}{2}Q''_0 \equiv \kappa L^2\,.
\end{align}

In arriving at the second line of Eq.~\eqref{eq:QExpand} we have also used the fact that the WKB analysis gives for the decay rate~\cite{Iyer}
\begin{align}
\label{eq:IyerDecayRate}
\omega_I = N \frac{\sqrt{2 Q_0''}}{\left. \partial Q/\partial \omega \right|_{r_p,\omega_R}}\,.
\end{align}
This expression for the decay rate is actually found by the matched expansions method outlined in this section, in the case where we set $\epsilon =0$. For now we consider Eq.~\eqref{eq:IyerDecayRate} to be an ansatz, and check its consistency at the end of the matching. The reason why we require a non-zero $\epsilon$ here is that we are interested in $\partial C^-_{lm\omega}/\partial \omega$ evaluated at the QNM frequency values, and so we need information about the wavefunctions in the vicinity of the QNM frequencies.

Since $k \propto L^2$, the rescaled radial parameter $z$ is proportional to $\sqrt{L}$. 
Therefore for small but finite $r_*-r_{*p}$ (where the leading order Taylor expansion is accurate) the corresponding $z$ ranges from $0$ to $\infty$ as we take the eikonal limit $L \rightarrow \infty$. 
In order to perform the matching, we need to set boundary conditions for the solution in region III. 
This can be done by taking $z \rightarrow \pm \infty$, which occurs in regions II and IV of Fig.~\ref{fig:potential}. 

We define $\psi \equiv u_r$ as the radial wavefunction in the near-peak regime, and with the new set of variables defined above, the radial Teukolsky equation can be rewritten in the more compact form
\begin{equation}
\frac{d^2 \psi}{d z^2}+\left [z^2+2N\left(-i+\frac{\epsilon}{\omega_I}\right) \right ]\psi=0\,.
\end{equation}
The solutions of the above equation can be expressed by parabolic cylinder functions~\cite{Dolan2011,Iyer,nist}. The two independent solutions are given by
\begin{align}
\psi_1&=D_{n+\eta}[z(-1+i)],\\
\psi_2&=D_{n+\eta}[z(1-i)]\,,
\end{align}
where $\eta=i\epsilon/\Omega_I$ and $n$ is the overtone number. In regime II and IV of Fig.~\ref{fig:potential}, the asymptotic behavior of $\psi_1$ is~\cite{nist}
 \begin{widetext}
\begin{align}\label{eqpsi1}
\psi_1 =\left\{
\begin{array}{cl}
2^{n/2}e^{-in\pi/4}|z|^ne^{iz^2/2}\,, & z\rightarrow-\infty\,, \\
\\
2^{n/2}e^{3in\pi/4}|z|^ne^{iz^2/2}+\eta\Gamma(n+1)(2\pi)^{1/2}e^{-3i\pi(n+1)/4}2^{-(n+1)/2}  |z|^{-(n+1)}e^{-iz^2/2}\,, &z \rightarrow \infty\,.
\end{array}\right.
\end{align}
For $z \to \infty$, the two terms in $\psi_1$ correspond to out-going and in-going waves, which we match to the WKB solutions in Sec.~\ref{sec:Matching}. Meanwhile, $\psi_2$ has similar asymptotic behavior, but with both out-going and in-going parts as $z \to - \infty$.
In order to satisfy the in-going boundary condition at the horizon, we have to pick $\psi_1$ as the physical solution.

\subsection{The WKB wavefunctions away from the peak of the potential} 

We must match $\psi_1$ with the WKB solution away from the peak of the potential. Let us recall the WKB solutions in Eq.~(\ref{eqradwkb}), and single out part of the $r$ dependence for later convenience:
\begin{equation}\label{eqwkbpm}
 u_{\pm} = \frac{1}{[Q(\omega_R,r)]^{1/4}}|r_*-r_{*p}|^{\pm N}e^{ \pm \bar{S}_0} \,,
\end{equation}
where $\bar{S}_0$ is given by
\begin{align} \label{eq:SpmWKB}
 \bar{S}_0 = \left\{
\begin{array}{cl}
i \displaystyle\int^{r_*}_{r_{*p}}\sqrt{Q(\omega_R,r)}dr_*
+  \displaystyle\int^{r_*}_{r_{*p}}\left [
\mathcal{Q}_I
-\frac{N}{|r_*-r_{*p}|}\right ]\,dr_*\,, & z>0\,, \\
\\
i \displaystyle\int^{r_{*p}}_{r_{*}}\sqrt{Q(\omega_R,r)}dr_*
+ \displaystyle \int^{r_{*p}}_{r_{*}}\left [
\mathcal{Q}_I
-\frac{N}{|r_*-r_{*p}|}\right ]\,dr_*\,, &z <0\,.
\end{array}\right.
\end{align}
{Note that these $u_\pm$ actually represent four distinct functions, two on each side of the peak. These WKB wavefunctions can be used to construct a single homogeneous solution of the radial Teukolsky equation through matching across the peak.} The $r \rightarrow r_p$ limit of these solutions is matched with the $z \rightarrow \pm \infty$ limit of $\psi_1$. As discussed in Sec.~\ref{sec2}, the first integral in the above expression~\eqref{eq:SpmWKB} for $\bar{S}_0$ corresponds to the phase and the second integral contributes to the amplitude. With the $|r_*-r_{*p}|^{\pm N}$ term factored out, the amplitude contribution in $\bar{S}_0$ (the second term) asymptotes zero in the $r \rightarrow r_p$ limit. Taking the limit that $r \rightarrow r_p$, but keeping $|z| \rightarrow \infty$
\footnote{For example, this can be achieved by requiring $r_*-r_{*p} \propto L^{-1/4}$. In the $L \rightarrow \infty$ limit, we then have $r \rightarrow r_p$ and $|z| \rightarrow \infty$.},  
i.e, in the buffer zone II and IV, the WKB solution in Eq.~(\ref{eqwkbpm}) can be greatly simplified using the rescaled radial position $z$:
\begin{equation}\label{eqpsi1}
u_{\pm} = k^{-1/4}\left (\frac{|z|}{k^{1/4}}\right )^{-1/2\pm N}e^{\pm iz^2/2}\,.
\end{equation}
It is worth pointing out that $u_+$ is the out-going solution when $r > r_p$ and the in-going solution when $r<r_p$ (and vice versa for $u_-$). This is because $\sqrt{Q(\omega_R,r)}\approx k^{1/4}z$ for $z>0$ and $-k^{1/4}z$ for $z<0$. For this reason, only $u_+$ is needed to construct the QNMs.

For generic frequencies, in order to compute the black hole excitation factor, we also have to know the asymptotic behavior of $u_{\pm}$ in the limit $r_* \rightarrow \pm \infty$. For later convenience we define the following phase factors:
\begin{align}
L\alpha_1(\mu,a) & \equiv \int^{\infty}_{r_{*p}}\left(\sqrt{Q(\omega_R,r)}-\omega_R \right)dr_*  -\omega_R r_{*p}\,,\nonumber \\
L\alpha_2(\mu,a) & \equiv  \int_{-\infty}^{r_{*p}}\left( \sqrt{Q(\omega_R,r)}-\bar{\omega}_R \right)dr_* +\bar{\omega}_R r_{*p} \,.
\end{align}
Here $ \bar{\omega}_R \equiv \omega_R-ma/(2Mr_+)$ is the radial frequency for QNMs near the horizon. The phases $\alpha_1$ and $\alpha_2$ are both finite numbers, which have the physical meaning of the accumulated phase factors at the position of the peak, if we were to extrapolate the $|r_*|\rightarrow \infty$ wavefunctions to the near zone. Similarly, we can define the accumulated amplitude factors as follows: 
\begin{align}
\label{eq:Gamma1}
&\gamma_1(\mu,a) \equiv \int ^{r_1}_{r_{*p}}\left [
\frac{\mathcal{Q}_I}{N}
-\frac{1}{r_*-r_{*p}}\right ]\,dr_*
+\log(r_1-r_{*p})+
\int _{r_1}^{\infty}\left [
\frac{\mathcal{Q}_I}{N}
-\Omega_I\right ]\,dr_* 
-\Omega_I r_1\,, \\
\label{eq:Gamma2}
&\gamma_2(\mu,a) \equiv \int _{r_2}^{r_{*p}}\left [
\frac{\mathcal{Q}_I}{N}
-\frac{1}{r_{*p}-r_{*}}\right ]\,dr_*
+\log(r_{*p}-r_2)+
\int ^{r_2}_{-\infty}\left [
\frac{\mathcal{Q}_I}{N}
-\Omega_I\right ]\,dr_*
+\Omega_I r_2\,,
\end{align}
\end{widetext}
where $r_1$ and $r_2$ are two constants satisfying $r_1>r_{*p}$ and $r_2<r_{*p}$. 
By taking the derivative of the above expressions with respect to $r_1$ and $r_2$, it is straightforward to show that $\gamma_1$ and $\gamma_2$ are independent of the choices of $r_1, r_2$. 
We introduce extra terms $(r_*-r_{*p})^{-1}$ and $\Omega_I$ into the integrands to ensure that the integrals are well defined in the $r \rightarrow r_p$ and $r_*\rightarrow \pm \infty$ limits. 
While all of the integrals are well defined, the first terms on the right hand sides of Eqs.~\eqref{eq:Gamma1} and~\eqref{eq:Gamma2} are sensitive to errors when evaluating them numerically. In practice we Taylor expand $\mathcal Q$ to second order, so that its first order piece exactly cancels with the $(r_* - r_{*p})^{-1}$ terms. 
We can then integrate the second order part, which is independent of $r_*$, keeping $\delta r_* \equiv r_1 - r_{*p}$ (or $r_{*p} - r_2$) small. 
Our expected fractional error on these integrals is then $\sim \delta r_*^2$, and we use $\delta r_* = 10^{-3}$.

We explore the dependence of $\alpha_1,\, \alpha_2,\, \gamma_1$, and $\gamma_2$ on $a$ and $\mu$ in Fig. \ref{fig:greenplot}. For Schwarzschild black holes with $a=0$, the corresponding $\alpha_1,\, \alpha_2,\, \gamma_1$, and $\gamma_2$ are all constants, because the Schwarzschild radial wavefunctions do not depend on $\mu$ (or equivalently on the azimuthal quantum number $m$). With non-zero spin parameter, these phase and amplitude factors develop a monotonic dependence on $\mu$.
{In addition, in the Schwarzschild limit we can explicitly compute the phase and amplitude factors. The results are
\begin{align}
\label{eq:SchAlpha1}
\alpha_1 = & \Omega_R [3 - 3\sqrt{3} + 4 \log 2 - 6 \log(2 + \sqrt{3}) ] \notag \\
 = & \Omega_R \zeta_{\rm DO} \,, \\
\alpha_2  =& \alpha_1 + \Omega_R(3 + 4 \log 2) \, \\
\label{eq:SchGamma1}
\gamma_1 =& 3 \sqrt{3} \Omega_I[ \log 2 + 3 \log 3 - \log( 2+\sqrt{3})]\notag \\
& + \Omega_I \zeta_{\rm DO} \,, \\
\gamma_2 =& \gamma_1 + 3 + 4 \log 2 - 3\sqrt{3} \log( 2+ \sqrt{3}) \,.
\end{align}
Here, $\zeta_{\rm DO} $ is one of the ``geometric constants'' defined by Dolan and Ottewill in~\cite{Dolan2011} and entering into their expressions for the black hole excitation factors.
These limits allow us to compare our black hole excitation factors to those computed by Dolan and Ottewill, recalling that for $a=0$, $\Omega_R = \Omega_I = 1/\sqrt{27}.$}

\begin{figure*}[t]
\centering
\includegraphics[width=1.0\columnwidth]{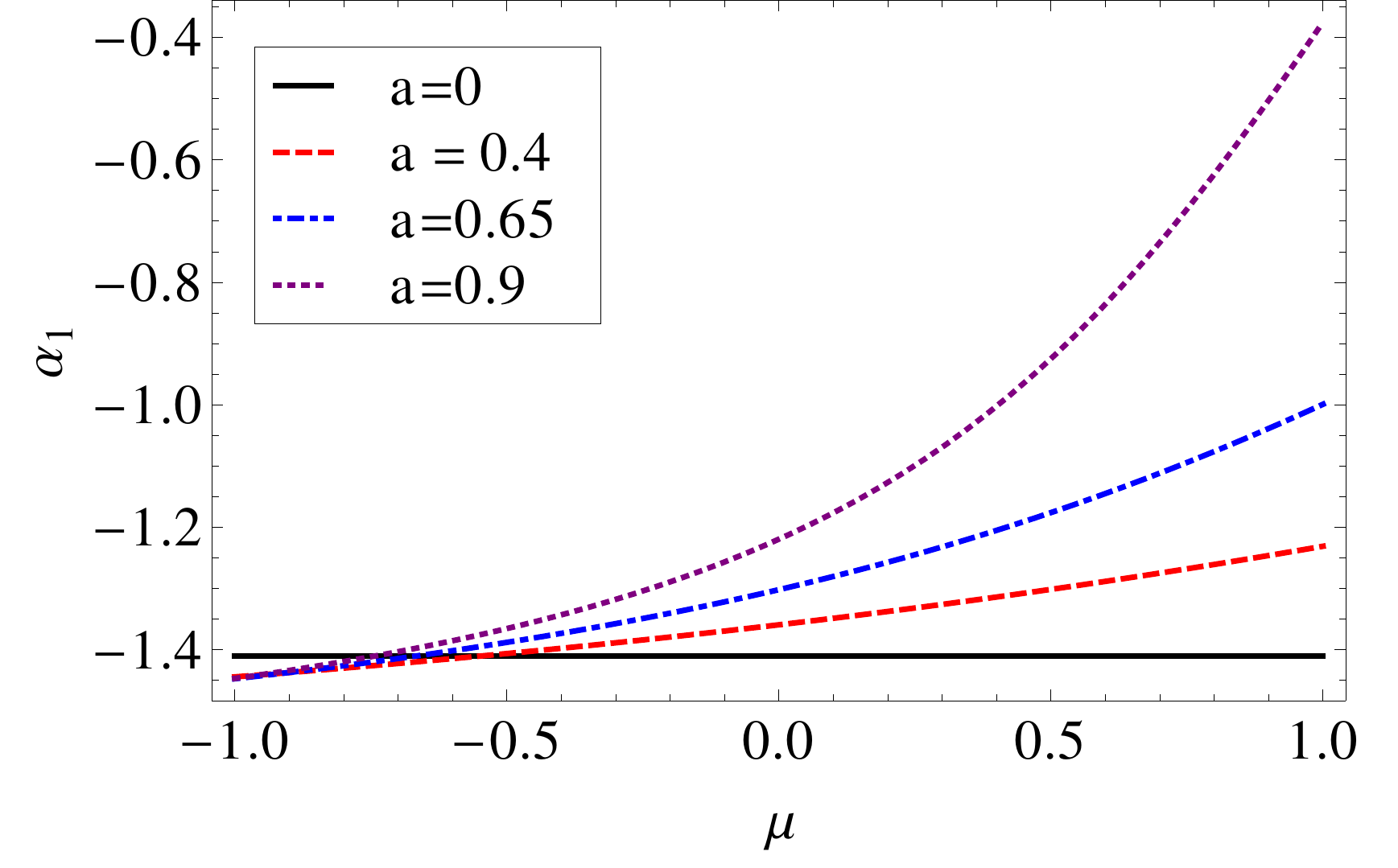}
\includegraphics[width=1.0\columnwidth]{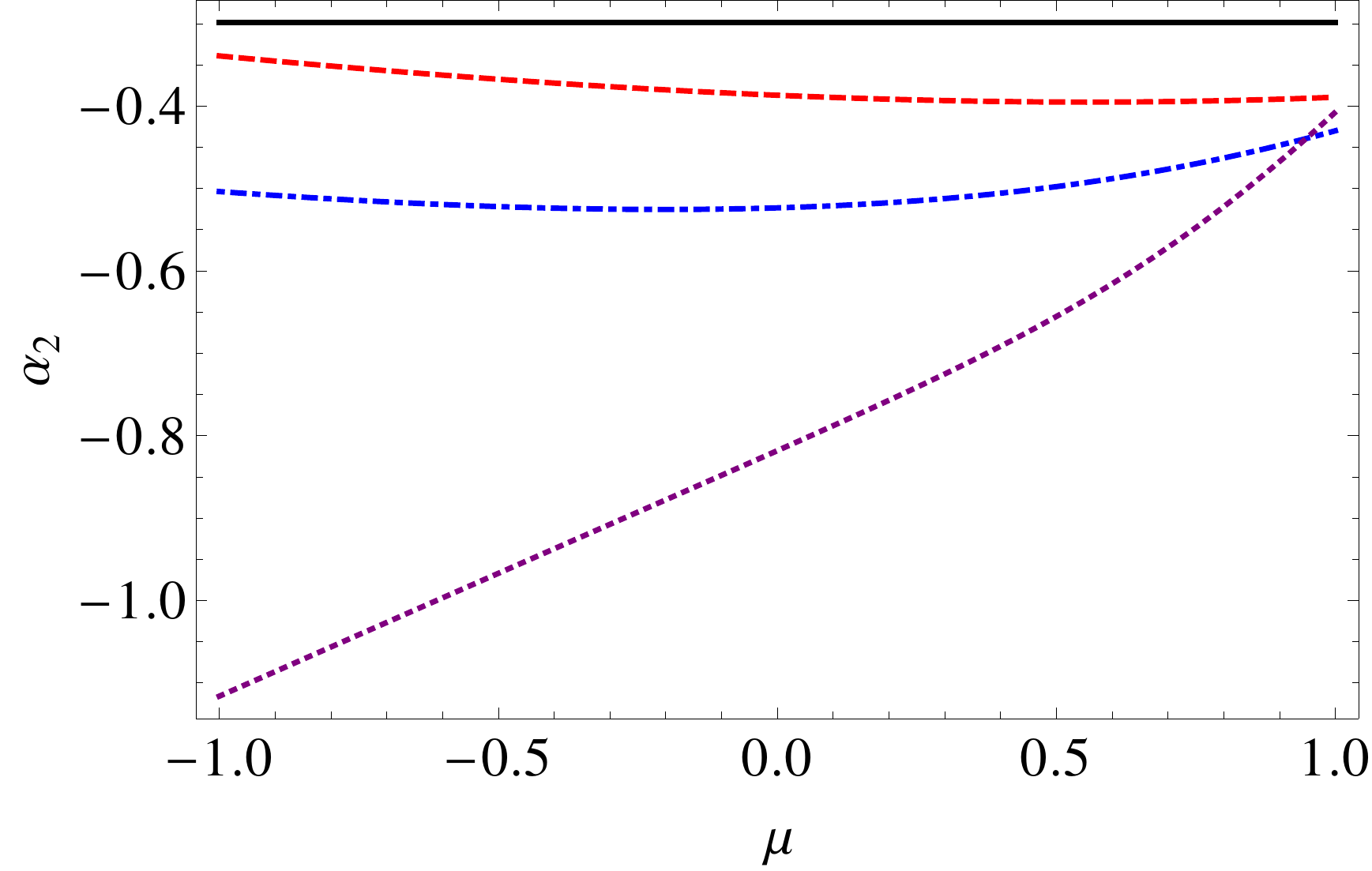} \\
\includegraphics[width=1.0\columnwidth]{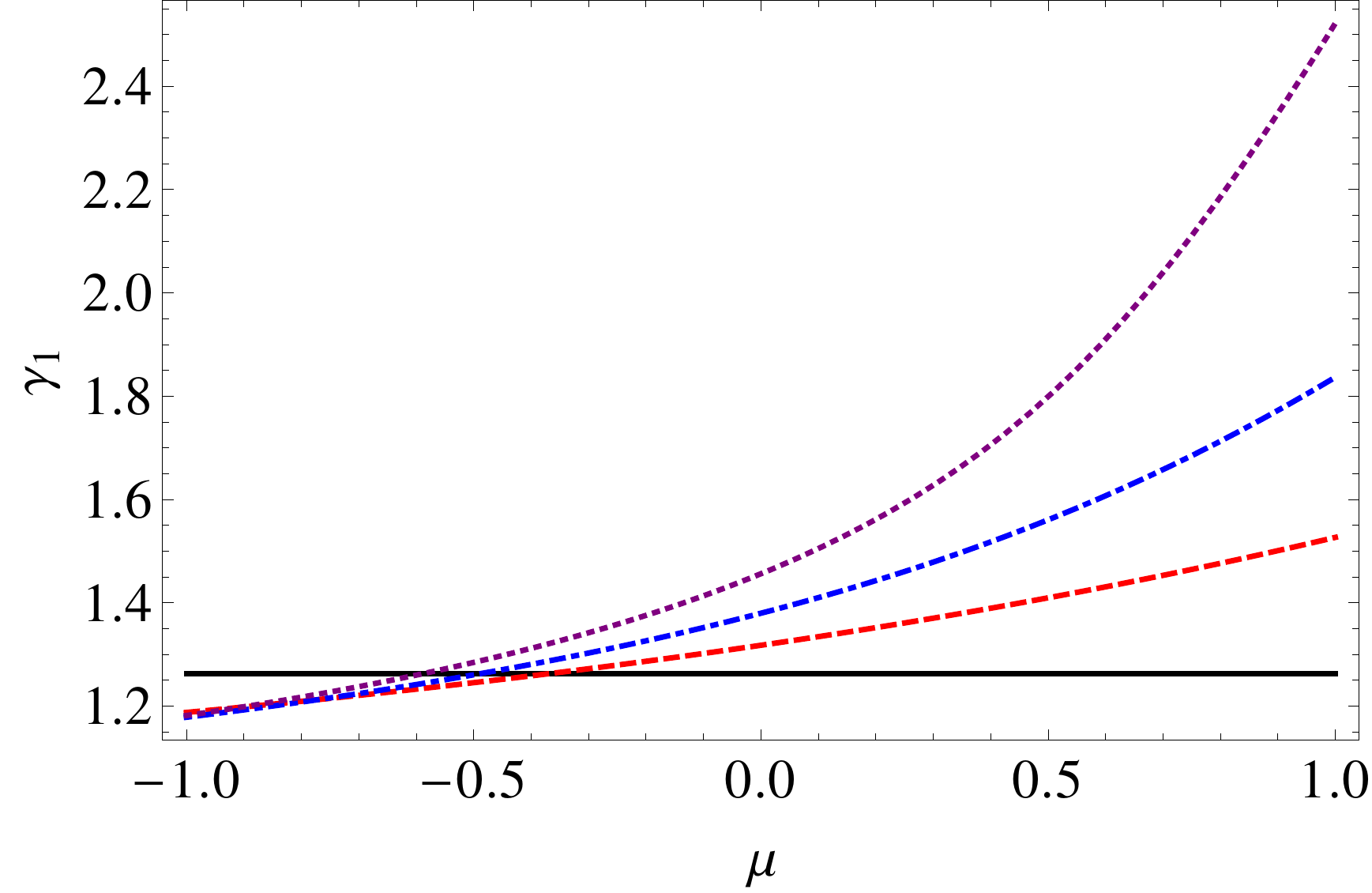}
\includegraphics[width=1.0\columnwidth]{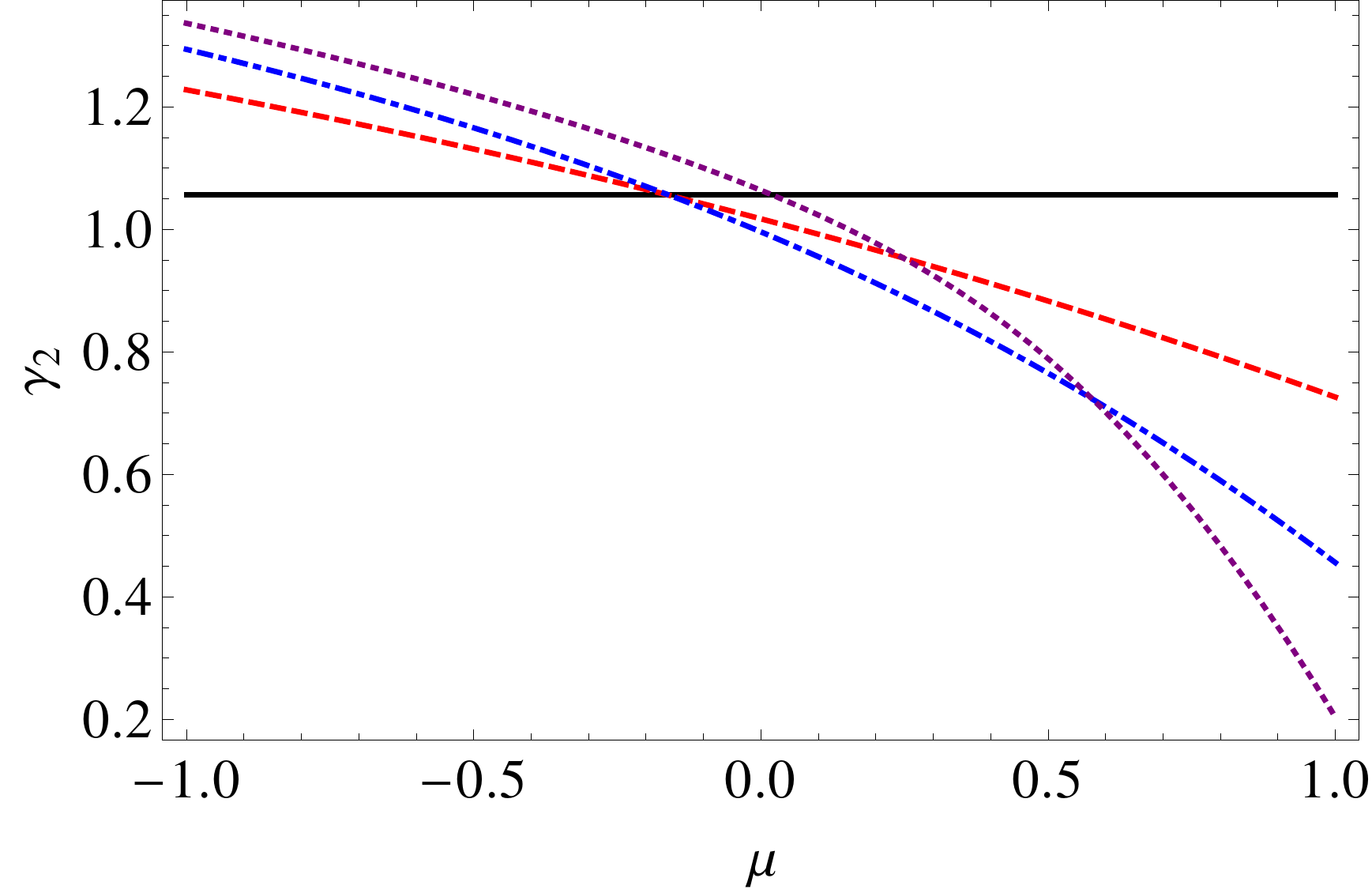}
\caption{Accumulated amplitude and phase factors as a function of $\mu$, plotted for the cases $a=0$ (black, solid lines), $a= 0.4$ (red, dashed lines), $a=0.65$ (blue, dot-dashed lines), and $a=0.9$ (purple, dotted lines). The Schwarzschild limit, $a=0$, does not depend on $\mu$ because of the spherical symmetry of the spacetime.}
\label{fig:greenplot}
\end{figure*}

\subsection{Matching solutions}
\label{sec:Matching}

The next step is to match the interior and exterior solutions in the buffer zone II and IV of Fig.~\ref{fig:potential}. The solution $\psi_1$ of region III has asymptotic solutions
\begin{equation}\label{eqpsi1bc}
\psi_1 \equiv \left\{
\begin{array}{cl}
C_{\rm in} u_{+}\,&z \rightarrow -\infty\,, \\
\\
B_{\rm in}u_- + B_{\rm out}u_+\,, &z \rightarrow \infty\,.
\end{array}\right.
\end{equation}
By comparing this with Eq.~\eqref{eqpsi1}, we can read off the coefficients $C_{\rm in},\, B_{\rm in},\,B_{\rm out}$,
\begin{align}
\label{eqcinbin}
C_{\rm in} &=k^{(n+1)/4} 2^{n/2}e^{-in\pi/4}, \\
 B_{\rm out} & =  (-1)^n  C_{\rm in}\,,\\
B_{\rm in} &=\eta \, n! (2\pi)^{1/2}e^{-3i\pi(n+1)/4} 2^{-(n+1)/2}  k^{-n/4}\,.
\end{align}

For QNMs, both $\eta$ and $\epsilon$ are zero. According to Eq.~\eqref{eqcinbin}, $B_{\rm in}$ is then zero and $u_+$ is the only surviving solution, as we expect. Now we can write down the asymptotic behavior of $u_+$ near the horizon or spatial infinity:
\begin{equation}\label{equplus}
u_+ = \left\{
\begin{array}{cl}
(\bar{\omega}_R)^{-1/2}e^{-i \bar{\omega}_{lmn}r_*+i L\alpha_2+N\gamma_2},&r_* \rightarrow -\infty\,, \\
\\
(\omega_R)^{-1/2}e^{i \omega_{lmn}r_*+i L\alpha_1+N\gamma_1}\,, &r_* \rightarrow +\infty\,,
\end{array}\right.
\end{equation}
and similarly for $u_-$ in the case $r_* \rightarrow +\infty$ ($u_-$ in the limit $r_* \rightarrow -\infty$ turns out not to be useful in our case),
\begin{align}
\label{equminus}
u_{-} =(\omega_R)^{-1/2} e^{-i \omega_{lmn}r_*-i L\alpha_1-N\gamma_1}, & & r_* \rightarrow +\infty\,.
\end{align}

By comparing Eq.~(\ref{equplus}) and Eq.~(\ref{equminus}) with Eq.~(\ref{equinasym}), Eq.~(\ref{eqpsi1bc}) and Eq.~(\ref{eqcinbin}), we can show that
\begin{align}
C^+_{lm\omega}& =\sqrt{\frac{\bar{\omega}_R}{\omega_R}} \frac{B_{\rm out}}{C_{\rm in}}e^{iL(\alpha_1-\alpha_2)+N(\gamma_1-\gamma_2)} \,,  \\
C^-_{lm\omega} & =\sqrt{\frac{\bar{\omega}_R}{\omega_R}}\frac{B_{\rm in}}{C_{\rm in}}e^{-iL(\alpha_1+ \alpha_2)- N(\gamma_1+\gamma_2)}\,.
\end{align}
Therefore the back hole excitation factors are given by
\begin{align}
\label{eqbhex}
\mathcal{B}_{lmn} &= \left [\frac{C^+_{lm\omega}}{2\omega}\left(\frac{\partial C^-_{lm\omega}}{\partial \omega}\right)^{-1} \right ]_{\omega=\omega_{lmn}} \nonumber \\
& = e^{2iL\alpha_1+2N\gamma_1}\frac{\Omega_I}{2\omega_{lmn}}2^nk^{n/2}\sqrt{\frac{i\sqrt{k}}{\pi}}\frac{(-i)^n}{n!} \nonumber \\
& \approx e^{2iL\alpha_1+2N\gamma_1}\frac{\Omega_I \sqrt{L}}{2\omega_R}\sqrt{\frac{i\sqrt{\kappa}}{\pi}}\frac{(-2i L \sqrt{\kappa})^n}{n!} \nonumber \\
&\equiv \sqrt{\frac{i}{L}} B e^{2iL\alpha_1}\frac{(-i L \xi)^n}{n!} \,,
\end{align}
with the constants $B,\,\xi$ given by (recall $\omega_R =L \Omega_R$)\,,
\begin{align}
B=\frac{e^{\gamma_1}\Omega_I}{2 \Omega_R} \frac{\kappa^{1/4}}{\sqrt{\pi}}, && \xi=2\sqrt{\kappa}e^{2\gamma_1}\,.
\end{align}
Since we neglect the sub-leading terms in $L$, the error of the expression for $\mathcal{B}_{lmn}$ scales as $L^{n-3/2}$ (a relative error of $1/L$). 
For $a=0$, it is straightforward to check that $\mathcal{B}_{lmn}$ recovers the Schwarzschild excitation factor derived in \cite{Dolan2011}, {using Eqs.~\eqref{eq:SchAlpha1} and~\eqref{eq:SchGamma1} for $\alpha_1$ and $\gamma_1$ in the Schwarzschild limit, together with the fact that $\kappa \to \Omega_R^4$ in this limit.} 

We check the black hole excitation factors derived using the WKB analysis against the recent results of~\cite{Zhang:2013ksa} (see also~\cite{Berti:2006wq}), where the excitation factors for Kerr were obtained numerically using Leaver's method and the MST formalism. These factors are available up to $l=7$ for all values of $s,\,m$ and several overtone numbers $n$ at~\cite{BertiSite}. The comparison is shown in Fig.~\ref{fig:bhexcitationplot} for $n=0$, three values of $a$, $l= 2,\,4,\,6$ and $7$, and all $m \in [-l,l]$. We make the same comparison for $n=1$ and a single value of spin in the bottom left panel of Fig.~\ref{fig:bhexcitationplot}. 
In this plot, the numerical values are given by the open circles, and the continuous curves are the WKB predictions for a given $l$ as a function of $\mu$. The crosses indicate the values of $\mathcal B_{lmn}$ from the WKB approximation at particular integer values of $m$. 

Although in~\cite{Zhang:2013ksa} the authors claim to use the same choice as we use for the origin of the coordinate $r_*$, we find that the results agree only after applying a simple shift $r_* \to r_* +r_0$ for each value of $a$, as discussed in Sec.~\ref{sec1}. We choose this shift by matching the phases of the numerical excitation factor and the WKB result at $l=7$, $m=7$, and $n=0$ for each value of $a$. 
This shift is then used for all other $l,\,m,\, n$ values, and is applied to the numerical values using the appropriate complex QNM frequency, which gives a leading order phase correction and a very small amplitude correction to the numerical values. We give $r_0$ in Fig.~\ref{fig:bhexcitationplot}. 
With this shift, the agreement is remarkable for all but the $l=2$ excitation factors. The discrepancy in the $l=2$ case for these generic values of $a$ is about the same as seen in the Schwarzschild case, cf.~\cite{Dolan2011}.

\begin{figure*}[t]
\centering
\includegraphics[width = 1.0 \columnwidth]{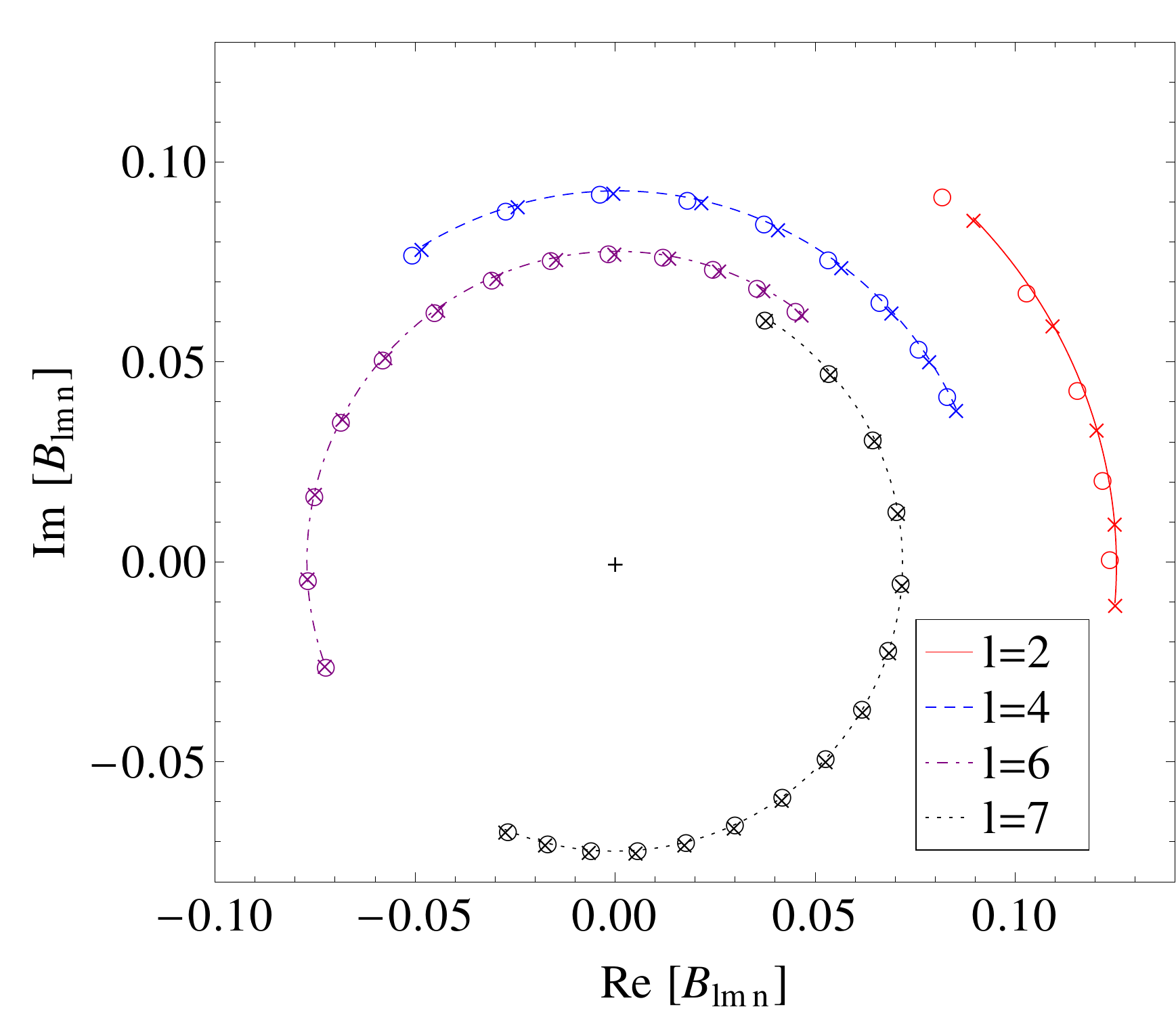}
\includegraphics[width = 1.0 \columnwidth]{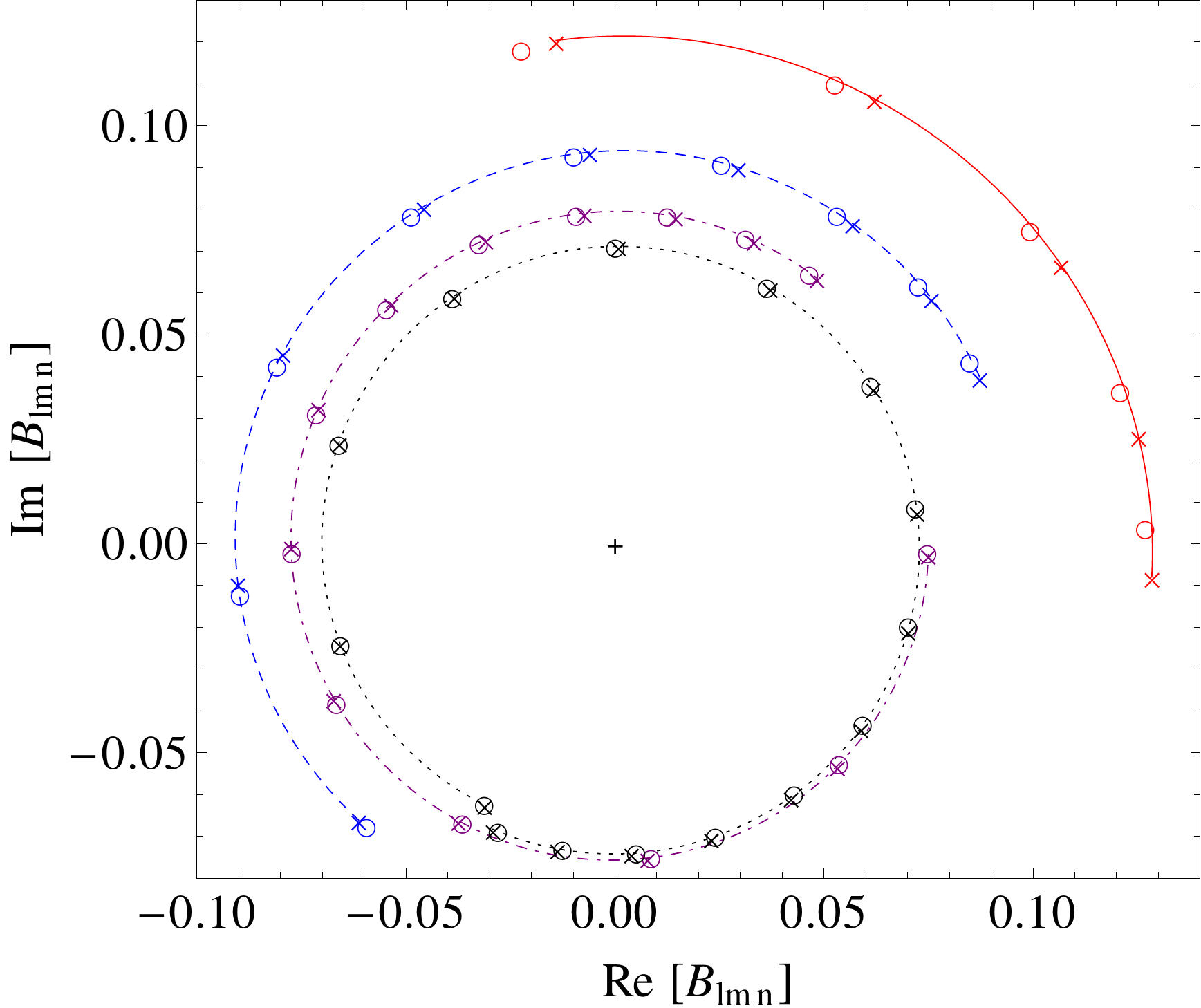}
\\
\includegraphics[width = 1.0 \columnwidth]{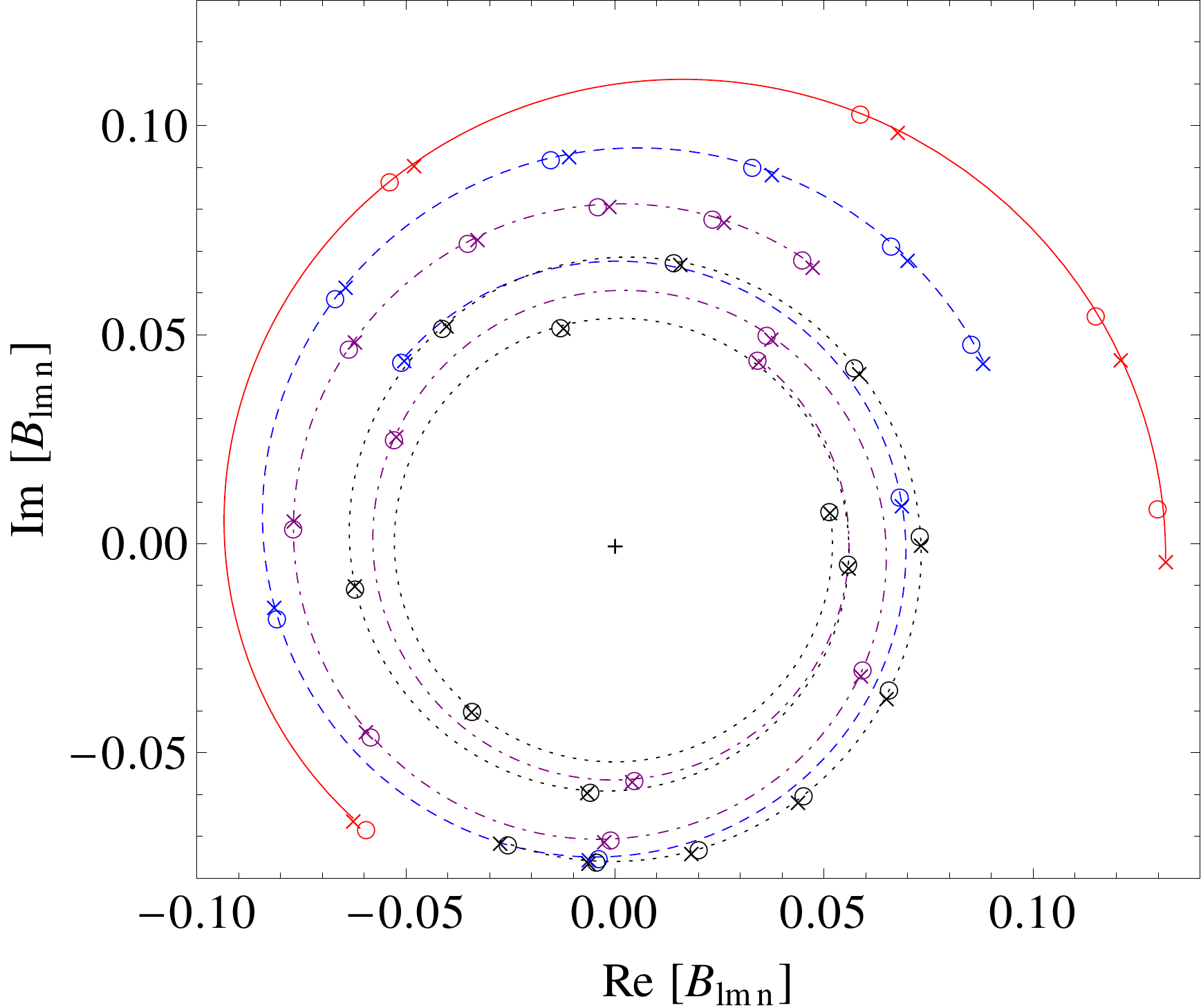}
\includegraphics[width = 1.0 \columnwidth]{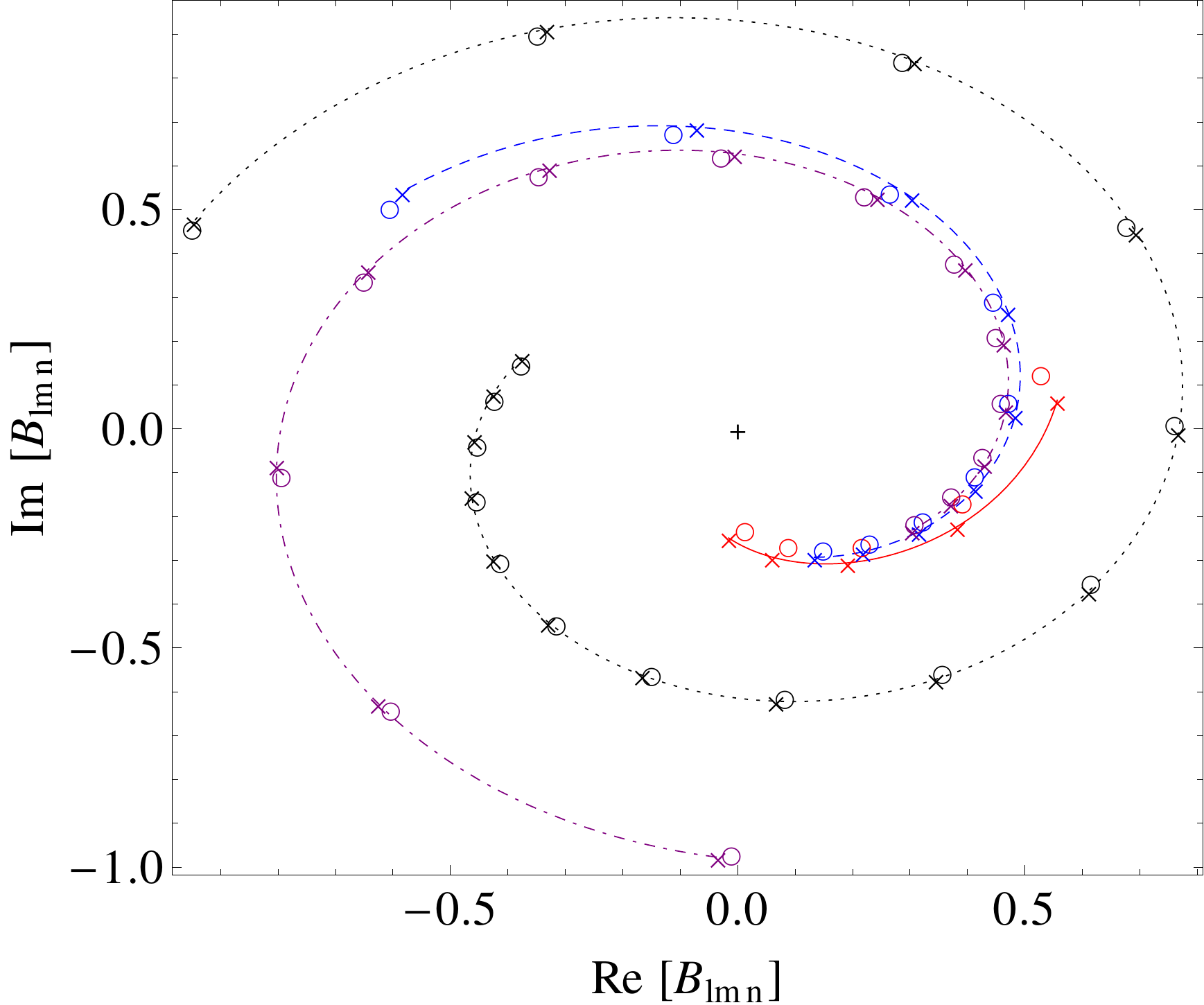}
\caption{Comparison of analytical (continuous curves and crosses) and numerical values (open circles) of the $n=0$ (top panels and bottom left panel) and $n=1$ (bottom right panel) black hole excitation factors for $l=2,\, 4,\, 6,$ and $7$. The higher $l$ curves lie inside the lower $l$ curves, and the values of $m$ increase as the curve is traversed counter-clockwise. Top left panel: Comparison for $a=0.4$, where $r_0 = -0.108$. Top right panel: Comparison for $a=0.65$, where $r_0 = -0.127$. Bottom left panel: Comparison for $a=0.9$, where $r_0 = 0.020$. Bottom right panel: Comparison for $a=0.65$ but for $n=1$, again using $r_0=-0.127$.}
\label{fig:bhexcitationplot}
\end{figure*}

\section{The Green function}
\label{sec5}
 
In this section, we combine all previous results on QNM frequencies, wave functions and the black hole excitation factors to obtain an approximate analytical expression for the Green function.
 
\subsection{Angular parts of the Green function}

We begin by investigating the angular components of the Green function~\eqref{eqqnmgreen}. 
First we focus on the spheroidal harmonic contributions,
\begin{align}
\label{eq:GFSlm}
S&_{lm\omega}^*(\theta') S_{lm\omega}(\theta)  \notag \\
&= \frac{C^2\left[e^{-i S_\theta^*(\theta')} + (-1)^{l+m}e^{i S_\theta^*(\theta')} \right]}{[\Theta_R(\theta')\sin^2\theta']^{1/4}} \notag \\  & \qquad \times 
\frac{\left[e^{i S_\theta(\theta)} + (-1)^{l+m}e^{-i S_\theta(\theta)} \right] }{ [\Theta_R(\theta)\sin^2\theta]^{1/4}} 
\notag \\
&=  \frac{C^2\left[e^{i \Phi(\theta, \theta') - N \Upsilon(\theta,\theta')} +e^{-i \Phi(\theta, \theta') + N \Upsilon(\theta,\theta')}   \right] }{[\Theta_R(\theta')\sin^2\theta' \, \Theta_R(\theta)\sin^2\theta]^{1/4}} \notag \\ &
 \qquad + (-1)^{l+m} \left(\theta' \to \pi - \theta '\right) \,,
\end{align}
where we have defined
\begin{align}
\Phi(\theta,\theta') &= \int^{\theta}_{\theta'} \sqrt{\Theta_R(\theta'')} d \theta'' \,, \\
\Upsilon(\theta,\theta') &= \upsilon(\theta) + \upsilon(\theta') \,,
\end{align}
and where the term $(\theta' \to \pi - \theta')$ indicates that the preceding expression is repeated with the given transformation. 
The fact that this is true follows from the symmetry of $S_{lm\omega} (\theta')$ under the given transform, and can be explicitly shown by recalling that $S_\theta(\pi - \theta) = - S_\theta(\theta)$, which separately requires that  $\upsilon(\pi - \theta) = - \upsilon(\theta)$ and that
\begin{align}
\int^{\pi - \theta}_{\pi/2} \sqrt{\Theta_R(\theta'')} d \theta'' 
=-\int^{\theta}_{\pi/2} \sqrt{\Theta_R(\theta'')} d \theta'' \,,
\end{align}
followed by expansion of the first equality in Eq.~\eqref{eq:GFSlm}.

In fact, if we restore the $\phi$ dependence $e^{i m \phi}$, the $(-1)^{m}$ factor can be generated by $\phi' \to \phi'+\pi$, together with the transformation $\theta' \to \pi - \theta'$, give the parity transformation $\hat{P}: (\theta',\phi')\to(\pi-\theta',\phi'+\pi)$. 
The additional $(-1)^l$ arises from the symmetry property of spheroidal function under the parity transformation:   $(-1)^l S_{lm}(\theta,\phi) = S_{lm}(\pi-\theta,\phi+\pi)$. In what follows, we split the analysis of the Green function into {multiple parts. 
First we evaluate the summations in the Green function contribution involving the first term following the second equality in Eq.~\eqref{eq:GFSlm}. 
The evaluation of the second set of terms, those indicated by $(\theta' \to \pi - \theta')$, follows in the same manner. A label ``P'' is used for this second set of terms, which enforce a parity symmetry for the $(\theta', \phi')$ coordinates.}

 \subsection{Summation over all QNM contributions}
 \label{sec:SumOverQNM}
Now we are ready to evaluate the summation in Eq.~\eqref{eqqnmgreen} to obtain the Green function in the eikonal limit. 
{We focus in this study} on the case where both $r>r_p$ and $r'>r_p$, where the normalized in-going radial function $\tilde{u}_{\rm in}(r)$ can be expressed as
 \begin{equation}
 \tilde{u}_{\rm in}(r)=U(r)[\rho(r)]^n \exp\left[-i L\tilde{\alpha}_1(r)\right]\,,
 \end{equation}
with $U(r)$, $\rho(r)$ and $\tilde{\alpha}_1(r)$ given by
\begin{align}
\tilde{\alpha}_1(r) & \equiv \int^{\infty}_{r_{*}}\left (\frac{\sqrt{Q(\omega_R,r)}}{L}-\Omega_R\right)dr_* \,,\\
\log[\rho(r)] & \equiv -\int _{r_{*}}^{\infty} \Omega_I \left [\frac{\partial_\omega Q|_{\omega_R}}{2\sqrt{Q(\omega_R,r)}} -1\right ]\,dr_* \,,\\
U(r)& \equiv \left( \frac{\omega_R \, \rho(r)}{\sqrt{Q(\omega_R,r)}}\right)^{1/2}\,.
\end{align}
 Note that $\rho(r) \propto (r_*-r_{*p})$ and $Q(\omega_R,r) \propto k (r_*-r_{p*})^2$ when $r \rightarrow r_{p}$, so $U(r)$ limits to a constant when $r \rightarrow r_{p}$. {Our analysis generalizes in a straightforward way to the other cases where either of $r$ or $r'$ are inside of $r_p$, by the use of the appropriate expressions for $\tilde u_{\rm in}$ inside the peak}.

Equation~\eqref{eqqnmgreen} contains a summation over all the indices $n,m,l$, and we begin by evaluating the summation over overtone number $n$. We split the Green function into two parts with two distinct terms each, as determined by contribution of the angular functions, Eq.~\eqref{eq:GFSlm}. The two terms in the first part differ only in the sign of the argument of the exponential, and so we can treat them at the same time as $\pm$ cases of $\Phi(\theta,\theta')$ and $\mp$ cases of $\Upsilon(\theta,\theta')$. The relevant parts of Eq.~\eqref{eqqnmgreen} are
\begin{widetext}
\begin{align}
&\sum_n \mathcal{B}_{lmn} e^{\mp N \Upsilon(\theta,\theta')}\tilde{u}_{\rm in}(r) \tilde{u}_{\rm in}(r')e^{-N \Omega_I T} \nonumber \\
& = B\sqrt{\frac{i}{L}}e^{iL[2\alpha_1-\tilde{\alpha}_1(r)-\tilde{\alpha}(r')]}e^{[-\Omega_I T\mp\Upsilon(\theta,\theta')]/2} U(r)U(r') \sum_n \frac{[-i\xi\rho(r)\rho(r')L]^n e^{-n\Omega_I T\mp n\Upsilon(\theta,\theta')}+O(L^{n-1})}{n!} \nonumber \\
& \approx  B\sqrt{\frac{i}{L}} e^{[-\Omega_I T\mp\Upsilon(\theta,\theta')]/2}U(r)U(r') 
\exp\left(iL\left [\bar{\alpha}_1(r)+\bar{\alpha}_1(r')-\xi\rho(r)\rho(r')e^{-\Omega_I T \mp \Upsilon(\theta,\theta')}\right ]\right) ,
\end{align}
with
\begin{align}
\bar{\alpha}_1(r) &  \equiv \int_{r_{*p}}^{r_{*}}\frac{\sqrt{Q(\omega_R,r)}}{L} dr_* -\Omega_R r_* \,.
\end{align}

After dealing with the summation over overtone $n$, we compute the summation over $m$. Now the relevant terms are
\begin{align}
&\sum_m e^{-i\omega_R T+im(\phi-\phi')\pm i \Phi(\theta,\theta')} \exp\left( iL\left[\bar{\alpha}_1(r)+\bar{\alpha}_1(r')-\xi\rho(r)\rho(r')e^{-\Omega_I T\mp \Upsilon(\theta,\theta')}\right] \right)  \nonumber \\
& \qquad \times B \sqrt{\frac{i}{L}}e^{[-\Omega_I T \mp\Upsilon(\theta,\theta')]/2}U(r)U(r')\frac{C^2}{[\Theta_R(\theta)\sin^2\theta \,\Theta_R(\theta')\sin^2\theta']^{1/4}}\,.
\end{align}
\end{widetext}
All the terms in the first line rapidly change in phase as $m$ varies, because the arguments of the exponentials are all proportional to $L$. 
The functions in the second line also depend on $m$, but they change slowly in amplitude with $m$. To compute the sum, we recall that $\mu=m/L$ and we apply the approximation that $\sum_m \rightarrow L\int d\mu$, as $L \gg 1$. 
We can then use the stationary phase approximation, also called the method of steepest descent, to evaluate the integral (see e.g.~\cite{ArfkenWeber}). The approximation states that in the limit $L\gg1$
\begin{align}
\label{eqsteepdec}
\int d\mu f(\mu) e^{i L g(\mu)} \approx & \sqrt{\frac{2 i \pi }{L |g''(\mu_0)|}}f(\mu_0)e^{i L g(\mu_0)}
\notag \\ &
\times  e^{-i \arg g''(\mu_0)/2}\,,
\end{align}
where $\mu_0$ is the extremum of $g(\mu)$: $\left .g'(\mu)\right |_{\mu_0}=0$, here a prime indicates a derivative with respect to $\mu$, and $f(\mu)$ is a function which varies slowly with $\mu$. In fact, one can view the above integral as a simplified version of path integral, and in the classical limit $L \gg 1$, we only pick the paths near the classical trajectory $g'(\mu)=0$, where $g(\mu)$ here is the ``geometric phase'' explicitly given by 
\begin{align}
\label{eqact}
g_{\pm}(\mu) =& -\Omega_R(\mu,a) T + \mu (\phi-\phi') 
\notag \\ &
\pm \frac{\Phi(\theta,\theta')}{L} +\bar{\alpha}_1(r)+\bar{\alpha}_1(r') \notag \\ &
-\xi\rho(r)\rho(r')e^{-\Omega_I(\mu,a) T\mp \Upsilon(\theta,\theta')}\,.
\end{align}
Note that it is possible for $g_\pm(\mu)$ to have multiple extrema, in which case Eq.~\eqref{eqsteepdec} must be extended to include contributions from all the stationary points. Physically this occurs when more than one null geodesic connects the points $x$ and $x'$, which corresponds to a caustic. 

In addition to the two terms involving $g_{\pm}(\mu)$, we recall the other family of terms in Eq.~\eqref{eq:GFSlm}, with $\theta' \rightarrow \pi-\theta'$ and an extra factor $(-1)^{l+m}$. The geometric phases associated with these terms are
\begin{align}
\label{eqact2}
g^{\rm P}_{\pm}(\mu) =& -\Omega_R(\mu,a) T \mp \pi+ \mu (\phi+W^{\rm P} \pi-\phi')  \notag \\ &
\pm \frac{\Phi(\theta,\pi-\theta')}{L} +\bar{\alpha}_1(r)+\bar{\alpha}_1(r') \notag \\ &
-\xi\rho(r)\rho(r')e^{-\Omega_I(\mu,a) T\mp \Upsilon(\theta,\pi-\theta')}\,.
\end{align}
{There are two contributions to the phases $g^{\rm P}_\pm$ that require discussion. The first is the factor involving} $W^{\rm P} \equiv 2 w +1$ for $w \in \mathbb{Z}$, which arises from the $(-1)^m$ term by noticing that $(-1)^m=(-1)^{(2w+1)m}$. This multiplicity in $W^{\rm P}$ is related to the multi-valued nature of $\phi$ and $\phi'$. When we fix our convention so that $(\phi, \phi') \in [0,2\pi)$ and vary $w$, there is at most one particular $w$ which allows for one (or both) of the phases $g^{\rm P}_{\pm}$ to have an extremum. We refer $w$ as the ``winding number " of the geometric phase. We can see the need for such a winding number from the following argument. Consider the case where both $x$ and $x'$ lie on a null geodesic orbiting the black hole on the photon sphere. When we follow the geodesic $w$ times around the black hole, the term $- \Omega_R T$ accumulates a phase factor $- w \Omega_{\rm R} T_{\rm period} = - 2\pi w(1-|\mu|)$ \cite{Yang2012a}. The $2\pi w |\mu|$ factor must be absorbed by the $\mu W^{\rm P}$ term, or else we cannot find an extremum for either of $g^{\rm P}_\pm$.

The second contribution to discuss is the factor $\mp \pi$. This term comes from the fact that $(-1)^l = (-1)^{-l}$ and can contribute a factor of $\pi$ with either sign when we convert the integer $l$ into the continuous variable $L$. While it might seem that we would require a second winding number to account for this ambiguity, our numerical investigations indicate that the choice of $\mp \pi$ is sufficient to guarantee the continuity of the Green function when we fix the source coordinates $x'$ and the spatial coordinates $(r,\theta,\phi$) of an observer, but allowing $t$ to grow. It may be that for some choices of $x, x'$ additional phases are required to keep the phase of the Green function continuous for an observer. Note that while this extra ambiguity does not change the position of the extrema of the phases $g^{\rm P}_\pm$, we must also keep track of the sign of the prefactor $\pm i$ which arises from $(-1)^{\mp l} = \pm i \exp( \mp i \pi L)$.

Following the same logic as above, we also need a ``winding number" term in $g(\mu)$\footnote{It can be mathematically introduced by multiplying $(-1)^{2wm}$ and extracting the associate phase change.}:
\begin{align}
\label{eqact3}
g_{\pm}(\mu) =& -\Omega_R(\mu,a) T + \mu (\pi W+\phi-\phi') 
\notag \\ &
\pm \frac{\Phi(\theta,\theta')}{L} +\bar{\alpha}_1(r)+\bar{\alpha}_1(r') \notag \\ &
-\xi\rho(r)\rho(r')e^{-\Omega_I(\mu,a) T\mp \Upsilon(\theta,\theta')}\,,
\end{align}
where $W=2w$ and $w \in \mathbb{Z}$, and at any given moment, there is at most one particular ``winding number" which gives an extremum of $g(\mu)$ for $\mu \in (-1,1)$.

We observe numerically that for reasonable $(x,x')$
\footnote{For example, $T$ has to larger than zero for the contour integral to converge. },
at most one pair of $g_{\pm}(\mu)$ and $g^{\rm P}_{\pm}(\mu)$ have extrema in $\mu \in  (-1,1)$. It could be true that in certain cases no extremum can be found. For each pair of phases with an extremum, one has $g_\pm''$ or $g_\pm^{\rm P''}$ greater than zero at the extremum, and the other has $g_\pm''$ or $g_\pm^{\rm P''}$ less than zero at the other extremum. The sign of $g_\pm''$ or $g_\pm^{\rm P''}$ determines whether there is an extra factor of $-i$ in Eq.~\eqref{eqsteepdec}, which determines how the related term contributes to the Green function (see Sec.~\ref{secgf}). In practice, we numerically compute all four functions $g_{\pm},\,g^{\rm P}_{\pm}$, and search for possible extrema.

{When searching for the extrema $\mu_0$ for a fixed $\theta, \, \theta'$ it is useful to note that the function $\Phi$ is only real in some range $\mu_{\rm min} \leq \mu \leq \mu_{\max}$, and the search must be carried out in this range, which is determined by whichever of the angles $\theta$ or $\theta'$ is closer to the boundary values $0$ and $\pi$. 
Next, we consider the behavior of the geometric phases as we fix $x'$ and $(r,\theta,\phi$) and allow $t$ to increase. 
We find that the particular pair of geometric phases which achieve an extremum change over time. Each time that a particular phase has its extremum leave the range $[\mu_{\rm min}, \mu_{\rm max}]$, another phase function takes its place. 
Matching of the phases at these points tells us how to fix the ambiguity $\mp \pi$, and over time we must iterate the winding number $w$ by one so that the phases which lose their extrema can again contribute to the stationary phase integral.}

As argued in \cite{Dolan2011}, the Schwarzschild Green function becomes singular when all terms in the summation over $l$ are resonant in phase. 
We can see that in our case, this can be translated into the condition that one of the geometric phases obeys $g(\mu_0)=2\pi j$ where $j$ is an integer. 
{When this occurs, the summation over $l$ does not converge, and the Green function is singular.} 
On the other hand, the Green function should be singular along any null geodesic connecting $x'$ to $x$~(see e.g. \cite{Kay:1996hj,Casals2009a,Casals2012}). 
This observation gives a consistency check of our method, since it requires that when one of the phases obeys $g(\mu_0)=2\pi j$, the points $x$ and $x'$ are connected by a geodesic. 
{In Sec.~\ref{sec6} we carry out a numerical study to test this relation between the singular points of the Green function and null geodesics. 
We also discuss the geometric phases from the perspective of the geometric correspondence between QNMs and unstable null orbits.}
 
{Before proceeding to the final evaluation of the Green function, which reveals the singular structure discussed above explicitly, we verify that integration over $\mu$ using the stationary phase approximation successfully recovers the results for Schwarzschild presented in~\cite{Dolan2011}.}
 
\subsection{Recovering the Green function for Schwarzschild}
\label{sec:GFschw}

The Green function simplifies greatly in the Schwarzschild limit. With $a = 0$, we have that $\Upsilon = 0$, and
\begin{align}
\Theta_R  &= L^2\left( 1 - \mu^2 \csc^2\theta\right)\,, \\
\frac{\Phi(\theta,\theta')}{L} &= \int_{\theta'}^{\theta} \sqrt{1 - \mu^2 \csc^2 \theta''} d\theta''\,,
\end{align}
The other terms are similarly reduced to their Schwarzschild limits. Aside from $\Phi$, the only explicit $\mu$-dependence in $g_\pm$ and $g^{\rm P}_\pm$ are terms like $\mu (\phi - \phi' +2 \pi W)$.

Next, we use the rotational and reflection symmetries of the spacetime and set $\phi = \phi'$, and $\pi \geq \theta > \theta' \geq 0$. Then the stationary phase condition on $g^\pm$ reduces to
\begin{align}
0 = g'_\pm & = \pm \int_{\theta'}^\theta \frac{- \mu \csc \theta}{\sqrt{\sin^2 \theta - \mu^2} } d \theta + \pi W \notag \\
& = \pm \tan^{-1} \left[ \frac{\mu \cos\theta}{\sqrt{\sin^2 \theta - \mu^2}} \right]_{\theta'}^\theta + \pi W \,, 
\end{align}
which is solved by setting $\mu_0^\pm = 0$ and requiring $W = 0$ for both cases. There is no need for the inclusion of the winding number in this case, because we have no $\mu$-dependent term which monotonically grows with time, and because of our rotation to a fixed azimuthal plane. With this stationary point, $g_\pm'' (\mu_0^\pm)$ is
\begin{align}
g_\pm {}''(\mu = 0) &= \mp \int_{\theta'}^\theta   \csc^2 \theta d \theta = \pm [\cot(\theta) - \cot(\theta')] \notag \\
& = \mp \frac{\sin(\theta - \theta')}{\sin \theta \sin \theta'} \,.
\end{align}
Further, $\Phi (\mu = 0)/ L = \theta - \theta' = \gamma$, where again $\gamma$ is the angle between the position of the source and that of the receiver. In addition, $\Theta_R(\mu = 0) = 1$, $\arg g_+{}''(0) = \pi$, $\arg g_-{}'' = 0$, and we can rewrite the relevant terms resulting from the stationary phase integrals involving $g_\pm$ as
\begin{align}
\label{eq:PoloidalGF}
L \sqrt{\frac{2 \pi}{L \sin \gamma}} \left(e^{-i \pi/4} e^{-i L \Psi^-} + e^{i \pi/4} e^{- i L \Psi^+} \right) \,,
\end{align}
where $\Psi^{\pm}$ is the expression for the geometric phase derived by Dolan and Ottewill, Eq.~(44) in~\cite{Dolan2011} and reproduced here:
\begin{align} 
\label{eq:PsiDO}
\Psi^{\pm}  \equiv& \frac{T}{\sqrt{27}} \pm \gamma - \frac{[\mathcal{R}_{\text{Sch}}(r)+\mathcal{R}_{\text{Sch}}(r') + 2\zeta_{\rm DO}]}{\sqrt{27}} \notag \\
& + \xi_{\text{Sch}} \rho_{\text{Sch}}(r)\rho_{\text{Sch}}(r')e^{-T/\sqrt{27}} \,.
\end{align} 
This fully recovers the results of Dolan and Ottewill, up to the final summation over the angular quantum number $l$, and the remainder of their results follow directly. 
This means that we expect $g^{\rm P}_\pm$ to make no contribution.

To verify this, we consider the stationary phase condition for $g^{\rm P}_\pm$,
\begin{align}
g^{\rm P}_\pm{}'
& = \pm \tan^{-1} \left[ \frac{\mu \cos\theta}{\sqrt{\sin^2 \theta - \mu^2}} \right]_{\pi - \theta'}^\theta \mp \pi W^{\rm P} \notag \\
& = 0\,.
\end{align}
Recalling that $-\pi/2 \leq \tan^{-1} x \leq \pi/2$, the condition can only be met if we take the argument of $\tan^{-1} x $ to $\pm \infty$. This is possible if we take $\theta' = \theta = 0$ or $\pi$, along with $\mu = 0$ and $W^{\rm P} = 1$. In this case, the physical picture is the evaluation of the Green function at a caustic after some number of full revolutions about the black hole. We make no attempt in this study to evaluate the Green function near a caustic, where multiple geodesics connect $x$ and $x'$, and additional care is required.

\subsection{The Green function}
\label{secgf}

{Following the stationary phase integral, the four parts of the QNM Green function involving the four phases $g_\pm$ and $g^{\rm P}_\pm$ reduce to two surviving parts, which we denote ``even'' and ``odd'' for reasons which become clear momentarily. We label those parts with subscripts e for even and o for odd.} In order to compute the Green function (and not only understand its singular structure), we need to evaluate the summation over $l$. To perform the sum over $l$, we apply the technique in~\cite{Dolan2011,Casals2009a}, using the Poisson summation formula to convert the summation over $l$ to an integral over $L$. In the eikonal approximation the QNM contribution becomes
\begin{widetext}
\begin{align}\label{eqintel}
G_{\rm eik} &= {\rm Re} \sum_l  \left[ \chi_{\rm e}(x,x') e^{ i L g_{\rm e}(\mu_{\rm e})} + i \chi_{\rm o}(x,x')e^{ i L g_{\rm o}(\mu_{\rm o})} \right] \nonumber \\
& = \sum_{s=-\infty}^{s=+\infty} (-1)^s {\rm Re} \int^{\infty}_0 d L e^{2\pi i L s } \left[ \chi_{\rm e} e^{ i L g_{\rm e}(\mu_{\rm e})}+i \chi_{\rm o}e^{i L g_{\rm o}(\mu_{\rm o})}\right] \,,
\end{align}
where $\chi_{\rm e}(x,x'),\, \chi_{\rm o}(x,x')$ are positive definite functions defined below. 
As we can see from the above equation, the criteria to separate even and odd contributions to the Green function is to check whether there is a prefactor of $i$ in the otherwise real amplitude. 
The overall prefactor receives contributions from the $i^{1/2}$ factor in Eq.~\eqref{eqbhex}, the $(-1)^{\mp l}=(-1)^{\mp(L-1/2)}$ factor in Eq.~\eqref{eq:GFSlm}, the $i^{1/2}$ factor and the possible $e^{-i\arg g''(\mu_0)/2}$ factor in Eq.~\eqref{eqsteepdec}. 
This means that after searching for the extrema of the phases $g_{\pm}$ and $g^{\rm P}_{\pm}$, we have to check whether there is an overall prefactor of $i$ in that term, before assigning it an even or odd label. 
Once we do this, we label the phase evaluated at its extremum $g_{\rm e}(\mu_{\rm e})$ or $g_{\rm o}(\mu_{\rm o})$ as appropriate. The functions $\chi$  are given by
\begin{align}
\chi_{\rm e}(x,x') & =  \frac{8\pi}{\sqrt{r^2+a^2}\sqrt{r'^2+a^2}}\sqrt{\frac{2\pi}{  |g_{\rm e}''(\mu_{\rm e})|}} 
\left [B\frac{C^2 U(r)U(r')  e^{-\Omega_I T/2 \mp \Upsilon(\theta, \theta')/2}}{[\Theta_R(\theta)\sin^2\theta \,\Theta_R(\theta')\sin^2\theta']^{1/4}}\right ]_{\mu_{\rm e}(x,x')}\,, \\
\chi_{\rm o} (x, x') & =  \chi_{\rm e} (g_{\rm e}'' (\mu_{\rm e}) \to g_{\rm o}''(\mu_{\rm o}), \, \mu_{\rm e} \to \mu_{\rm o}) \,,
\end{align}
where the sign in front of $\Upsilon$ depends on which $g_\pm$ or $g^{\rm P}_\pm$ functions they originate from. Since $C \propto L^{1/2}$ and $\Theta_R \propto L^2$, $\chi$ and $\chi_{\rm P}$ are independent of $L$. We manipulate the sum over $s$ in order to rewrite Eq.~\eqref{eqintel} in the form
\begin{align}
G_{\rm eik} & = \sum_{k=0}^{k=\infty} I_{k} \,,
\end{align}
where 
\begin{equation}\label{eqifact1}
I_{k} = \left\{
\begin{array}{cl}
\displaystyle {\rm Re} \int_0^{\infty}dL (-1)^{k/2}\left ( \chi_{\rm e} e^{i L[ \pi k  + g_{\rm e}(\mu_{\rm e}) ]} + i \chi_{\rm o}e^{-iL[\pi k - g_{\rm o}(\mu_{\rm o})]} \right ) ,&k\quad{\rm even}\,, \\
\\
\displaystyle {\rm Re} \int_0^{\infty} d L (-1)^{(k+1)/2}\left (i\chi_{\rm o}e^{iL[\pi(k+1) + g_{\rm o}(\mu_{\rm 0})]} +  \chi_{\rm e} e^{-i L [\pi(k+1) -g_{\rm e}(\mu_{\rm e})]} \right )\,, &k \quad{\rm odd}\,.
\end{array}\right.
\end{equation} 
\end{widetext} 
The integral can be evaluated using the identity
\begin{equation}
\int^{\infty}_0 d L e^{i L (q+i\epsilon^+)}=\frac{i}{q}+\pi\delta(q)\,,
\end{equation}
where $\epsilon^+$ is a positive infinitesimal running constant used to regulate the integral. The Green function is singular when the phase factor $q$ becomes zero. 
Here we are only interested in the case where $T>0$ (which is required for the QNM sum to converge), and those parts of the Green function which become singular. 
With $T>0$, the Green function can only become singular when the factors $g_{\rm e}$ or $g_{\rm o}$, which are negative and decreasing with increasing $T$, can cancel with the positive terms $\pi k$ or $\pi (k+1)$. 
Therefore we drop the second term in each line of Eq.~\eqref{eqifact1}, and it becomes 
\begin{equation}
\label{eqifact2}
I_{k} = \left\{
\begin{array}{cl}
\displaystyle (-1)^{k/2} \pi \, \chi_{\rm e} \, \delta[\pi k+g_{\rm e}(\mu_{\rm e})],&k\ {\rm even}\,, \\
\\ 
\displaystyle (-1)^{(k+1)/2} \frac{\chi_{\rm o}}{-\pi(k+1)-g_{\rm o}(\mu_{\rm o})}\,, &k\ {\rm odd}\,,
\end{array}\right.
\end{equation}
which {explains our use of the labels even and odd for the various contributions to the Green function. This expression} recovers the four-fold singular structure seen in the Schwarzschild case \cite{Dolan2011,Zenginoglu2012}, and in other spacetimes
~\cite{Casals2009a,Harte:2012uw,Casals2012}. 
In addition, it confirms our earlier argument that the Green function becomes singular when one of the phases obeys $g(\mu_0)=2 \pi j$, where $j$ is an integer.

In reality, as suggested by~\cite{Dolan2011} in the Schwarzschild case and more generally by~\cite{Zenginoglu2012}, the $g_{\rm o}$ related terms are turned on not only after $T>0$, but also after the previous $g_{\rm e}$ pulse. 
We notice that this feature is not captured by our method.
Adopting this understanding, the $I_{k}$ factor of the Green function should be 
\begin{equation}
\label{eqifact3}
I_{k} = \left\{
\begin{array}{cl}
\displaystyle (-1)^{k/2} \pi \, \chi_{\rm e} \, \delta[\pi k+g_{\rm e}(\mu_{\rm e})],&k\ {\rm even}\,, \\
\\ 
\displaystyle (-1)^{(k+1)/2} \frac{\chi_{\rm o} \, H [-\pi k-g_{\rm e}(\mu_{\rm e})]}{-\pi(k+1)-g_{\rm o}(\mu_{\rm o})}\,, &k\ {\rm odd}\,,
\end{array}\right.
\end{equation}
where the Heaviside function satisfies $H(x) =0$ for any $x<0$ and $H(x)=1$ for $x \ge 0$.

We illustrate a generic Green function at early times in Fig.~\ref{fig:gfplot}, in the case of $(r,\theta,\phi)=(8M, 5\pi/6, \pi/6)$, and $(r',\theta',\phi')=(8M, \pi/2, 0)$. The spin of the black hole is $a = 0.65$. 
The initial pulse (direct piece of the Green function) happens at $t-t'<10M$, and is not shown in the plot. 
Following the initial pulse, the QNM part of the Green function can be described by $I_1$ in Eq.~(\ref{eqifact3}), which diverges at $t'-t=36.06M$, as depicted by the vertical solid blue line. 
At $t-t=47.34M$, there is the $\delta$-function piece from $I_2$, after which the contribution from $I_3$ also turns on.

We emphasize that the WKB approximation fails at the places where wavefronts intersect each other, i.e., at the caustics. 
This fact has been discussed previously for the Schwarzschild spacetime \cite{Zenginoglu2012} and other spacetimes~\cite{Casals2009a,Casals2012}, and we expect a similar breakdown for Kerr black holes. The Green function near the caustics requires separate treatment, and we shall leave this analysis for future investigation. 

\begin{figure}[t,b]\centering
\includegraphics[width=1.0\columnwidth]{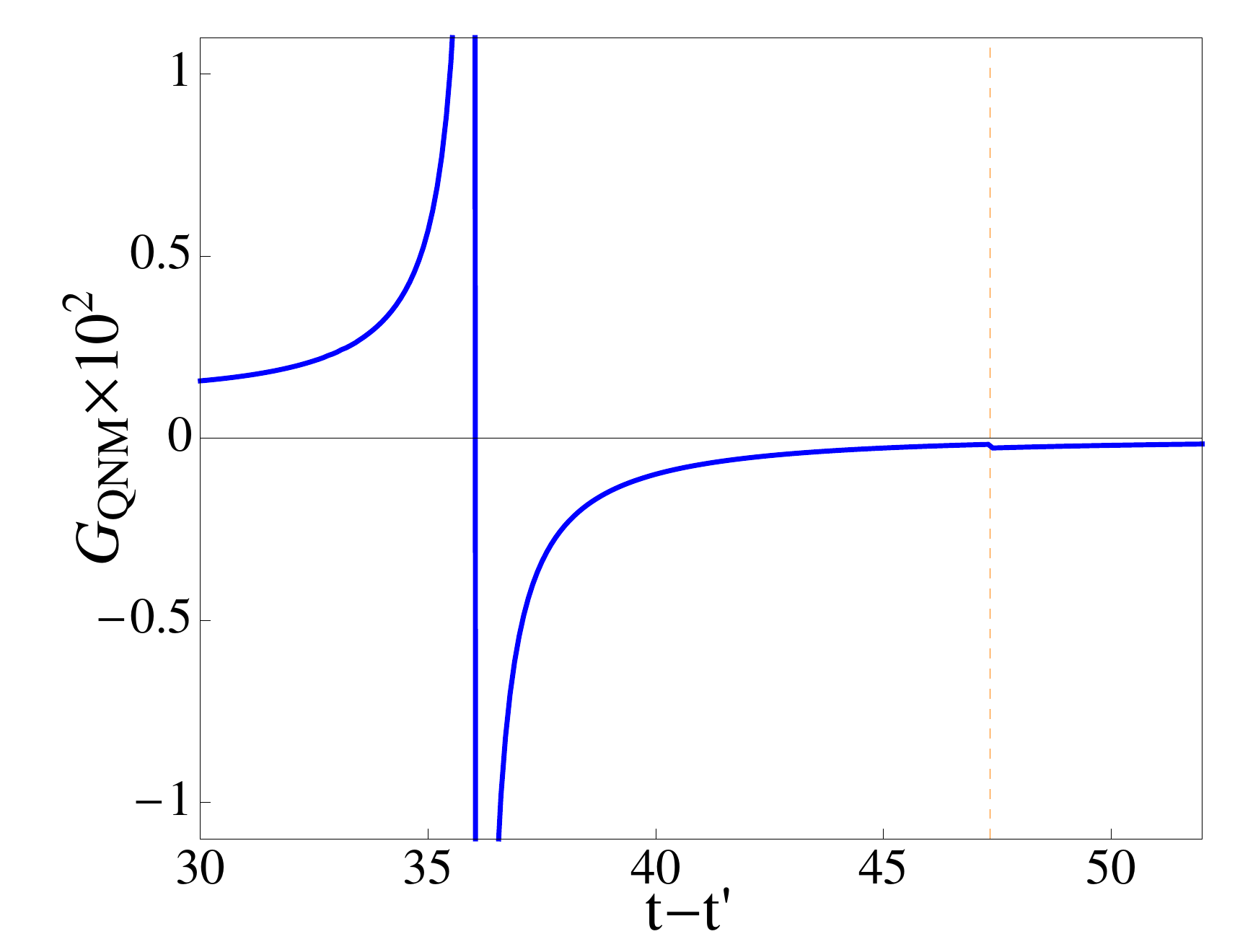}
\caption{(Color online.) QNM part of the Green function for black holes with $a=0.65M$. The solid blue line is described by the $I_1$ term in Eq.~(\ref{eqifact3}), which diverges at $t-t'=36.06M$ (the vertical solid blue line). The $\delta$-function contribution from $I_2$ is shown with the orange dashed line, at $t-t'=47.36M$. After that time the Green function is contributed to by both $I_1$ and $I_3$.}
\label{fig:gfplot}
\end{figure}

\section{Numerical examination of the geometric phases}
\label{sec6}

In this section, we numerically study the relationship between the singular structure of the Green function and null geodesics. 
We first {review the geometric correspondence between QNMs and the geodesic equations in Kerr, and we} note that there is an ``inclination measure'' $\tilde{\mu}_0 = \mathcal{L}_z/\mathcal{L}$ which we can associate with any null geodesic in the Kerr spacetime, where $\mathcal{L}$ is a conserved quantity analogous to angular momentum. 
It is important to note that while this inclination measure is equal to the cosine of the inclination angle of a photon orbit in Schwarzschild, there is no such interpretation in Kerr. 
Nevertheless, the inclination measure proves to be useful even in the Kerr spacetime, as we show below. 
We examine the relationship between $\tilde{\mu}_0$ and the extremum $\mu_0$ obtained from the stationary phase condition $g'(\mu_0)=0$ in Sec.~\ref{sec:MuvsMu}. Later, in Sec.~\ref{sec:Residual}, we show through examples that the correspondence between the singularities of the Green function and the null geodesics discussed in \cite{Dolan2011} for Schwarzschild black holes generalizes to the Kerr case. In this section, we will generally indicate the mass $M$ of the black hole explicitly, to facilitate comparison to previous work.

\subsection{The geometric correspondence and the principle function}
\label{sec:PrincipleFunction}

Here we briefly recount the details of the geometric correspondence between null geodesics and the QNMs. We also present the geodesic equations in Kerr for reference throughout this section.

A null geodesic in Kerr has three associated conserved quantities, the energy $\mathcal E$, the angular momentum about the spin axis of the black hole $\mathcal L_z$, and the Carter constant $\mathcal Q$.
A family of null geodesics in Kerr orbit the black hole at a constant radius on unstable orbits. We call the sphere these orbits are confined to the photon sphere. In the eikonal limit, the conserved quantities of these unstable null orbits correspond to the QNM parameters: $\omega_R$ to $\mathcal E$, $m$ to $\mathcal L_z$, and $A_R$ to $\mathcal Q + \mathcal L_z^2$~\cite{Yang2012a}.

We can also associate an inclination measure $\tilde{\mu}_0 = \mathcal{L}_z/\mathcal{L}$ to the corresponding photon orbit, {where $\mathcal L$ is implicitly defined through
\begin{align}
\label{eq:AngularMom}
\mathcal L^2 = \mathcal Q + \mathcal L_z^2 + \frac{a^2 \mathcal E^2}{2} \left( 1 - \frac{\mathcal L_z^2}{\mathcal L^2}\right)\,.
\end{align}
The analogue of angular momentum $\mathcal L$ is therefore also conserved. In the Kerr spacetime, the quantity $\mathcal L$ does not have a geometric correspondence, or indeed any physically relevant meaning, unless we use the additional supplemental approximation for $A_R$ in the eikonal limit~\cite{Yang2012a}
\begin{align}
\label{eq:AlmApp}
A_R \approx L^2 \left[ 1 - \frac{a^2\Omega_R^2}{2} \left( 1 - \mu^2 \right) \right] \,.
\end{align}
Using this supplemental approximation, the geometric correspondence of $\mathcal L$ is to the parameter $L$ of the corresponding QNM.} 

Also associated with a null geodesic is the Hamilton Jacobi principle function~\cite{MTW,Carter:1968rr},
\begin{align}
S & = S_t + S_r + S_\theta + S_\phi \notag \\
& = - \mathcal E t + \int \frac{\mathcal R}{\Delta} dr + \int \sqrt{\Theta} d \theta + \mathcal L_z \phi\,,
\end{align}
where
\begin{align}
\mathcal{R}(r) =& \left[ \mathcal{E} (r^2+a^2) - \mathcal{L}_z a\right]^2 \notag \\
&-\Delta \left[(\mathcal L_z - a\mathcal{E})^2 +\mathcal{Q} \right], \\
\Theta(\theta) =& \mathcal{Q} - \cos^2\theta \left( \frac{\mathcal{L}_z^2}{\sin^2\theta} - a^2 \mathcal{E}^2\right).
\end{align}
The resemblance between the principle function $S$ and the geometric phases $g_\pm$ and $g^{\rm P}_\pm$ indicates that these two are in correspondence when a null geodesic connects $x$ and $x'$, with the additional terms in the phase involving $\Omega_I$ and $\Upsilon$ correcting for the fact that the spacetime points $x$ and $x'$ are not necessarily on the photon sphere. 
In particular, one can show that if the null geodesic touches the photon sphere, the principal function $S$ and the phase function $g_{\pm}$ or $g^{\rm P}_{\pm}$ are exactly the same; if not, but assuming that the classical turning point of the radial motion is close to the photon sphere radius, the null geodesic can still be mapped to a ``parallel" geodesic on the photon sphere with appropriate $\mu_0$, and the terms in the phase functions involving $\Omega_I$ and $\Upsilon$ effectively account for the correction of the additional phase in the vicinity of the photon sphere. 
It is also reasonable to expect that $g_{\pm}$ and $g^{\rm P}_{\pm}$ will not be able to accurately describe the phase of a null geodesic if the geodesic is nowhere close to the photon sphere.
We build some evidence for this understanding in subsequent sections.

{Finally, by extremizing $S$ with respect to the conserved quantities, we can derive the equations of motion for null geodesics in Kerr,}
\begin{align}
\frac{dt}{d\xi} =& \frac{r^2+a^2}{\Delta} \left[\mathcal{E}(r^2+a^2)-\mathcal{L}_z a \right] \notag \\
& - a \left( a\mathcal{E} \sin^2\theta - \mathcal{L}_z\right), \\
\frac{d\phi}{d\xi} =& -\left( a\mathcal{E} - \frac{\mathcal{L}_z}{\sin^2\theta}\right) \notag \\
&+ \frac{a}{\Delta} \left[\mathcal{E}(r^2+a^2)-\mathcal{L}_z a \right], \\
\frac{dr}{d\xi} &= \pm \sqrt{\mathcal{R}}, \quad \frac{d\theta}{d\lambda} =\pm \sqrt{\Theta},
\end{align}
where $\xi$ is {the Mino time} along the geodesic, {and note that $\xi$ has units of inverse length}.

{We now turn to a discussion of the relationship between the stationary phase extrema $\mu^\pm_0$ and the inclination measure $\tilde \mu_0$ of null orbits.}

\subsection{Extremum $\mu_0$ versus the geodesic inclination measure}
\label{sec:MuvsMu}
The $\mu_0$ parameter originates from the stationary phase integral, so its physical significance is that quasinormal modes whose $m/L$ ratio match this value contribute the most to the (QNM part of the) Green function. 
Given the correspondence between QNMs and the spherical photon orbits, if our $x$ and $x'$ points are connected by a spherical photon orbit we would therefore na\"{i}vely expect the QNMs corresponding to that orbit to dominate the Green function, and that the stationary phase $\mu_0$ would be equal to inclination measure $\tilde{\mu}_0$ of the connecting geodesic.

We can see this is indeed true for the Schwarzschild case by noting that the spherical photon orbit condition $r=r'=r_p$ implies $\bar{\alpha}_1(r)=-\Omega_R r_{p*}$, $\bar{\alpha}_1(r') = -\Omega_R r_{p*}'$, and $\rho(r)=0=\rho(r')$. Consequently 
\begin{equation}
L g(\mu) =-\omega_R (t-t') + S_{\phi} \pm S_{\theta},
\end{equation}
{where here $S_\phi$ and $S_\theta$ are the $\phi$- and $\theta$-dependent parts of the principle function $S$, using the geometric correspondence.} In this case, the requirement that $\left .g_\pm'(\mu)\right |_{\mu_0}=0$ for stationary-phase trajectories is the same as the requirement that $S$ be maximized for geodesics, and this} ensures that the trajectories are the spherical photon geodesics. The case with a spinning black hole is more complicated, because $r_p$ has a $\mu$-dependence, so the $r$-dependent terms appear in the $g'(\mu)$ expression. This $\mu$-independence of $r_p$ for Schwarzschild black holes (a result of spherical symmetry) in fact ensures $\mu_0 = \tilde{\mu}_0$ even when $x$ and $x'$ are moved off of the spherical geodesics. Namely, when $a=0$, only 
\begin{equation}
m(\phi-\phi')\pm \Phi(\theta,\theta') =S_{\phi}\pm S_{\theta}
\end{equation}
in Eq.~\eqref{eqact} depend on $m$, so as far as the derivatives $g'_\pm$ are concerned, we have the same situation as the spherical photon geodesic case, and $\mu_0$ should be equal to the $\tilde{\mu}_0$ of the connecting geodesic, which is determined by the same two terms in the principal function. 

When the spherical symmetry is broken by a non-vanishing spin, more terms in the phases $g_\pm$ and $g^{\rm P}_\pm$ pick up $\mu$-dependence, so the situation is more complicated. Nevertheless the similarity between the phases and the principle function indicate that they are still approximated by the principal function provided the geodesic gets close to $r_p$ at some stage.
We verify this numerically for both the slowly-spinning limit of Kerr, with $a = 0.01$ and for a generic spin of $a= 0.65$, which we call the rapidly-spinning case\footnote{We use $a=0.01$ instead of exactly $a=0$, in order to avoid having to treat indeterminacy in expressions such as Eq.~\eqref{eq:TortR} with special codes. This allows us to test the effectiveness of the same numerical implementation as is used in the faster-spinning Kerr case.}. 
A spherical photon orbit has only one free parameter, which can be taken to be the photon sphere radius $r$, which lies between the co-rotating and counter-rotating equatorial radii, from
\begin{equation*}
2M(1+\cos[(2/3)\text{arccos}(-|a|/M)]) 
\end{equation*}
to 
\begin{equation*}
2M(1+\cos[(2/3)\text{arccos}(|a|/M)]).
\end{equation*}
For simplicity, we choose $x'$ on the equatorial plane, $\theta' =\pi/2$, in which case the constants of motion are related to $r$ by 
\begin{align}
\frac{\mathcal{L}_z}{\mathcal{E}} &= -\frac{(r^3 -3Mr^2 +a^2 r+a^2M)}{a(r-M)}, \\
\frac{\mathcal{Q}}{\mathcal{E}^2} &= -\frac{r^3(r^3-6Mr^2 + 9M^2 r -4a^2M)}{a^2(r-M)^2}, 
\end{align}
the geodesic initially has $dr/d\lambda=0$, and 
\begin{equation} 
\label{eq:SphereOrbitInitDir}
\frac{d\theta}{d\phi} = \frac{\sqrt{\mathcal{Q}/\mathcal{E}^2} \Delta}{-(a-\mathcal{L}_z/\mathcal{E})+a(r^2+a^2-a \mathcal{L}_z/\mathcal{E})}.
\end{equation}
We launch geodesics in this direction for various choices of $r$ in the $a=0.01$ and $a=0.65$, and allow the geodesics to evolve for a Boyer-Lindquist time of $3M$ in order to acquire our $x$, which then supplies us with the function $g(\mu)|_{x,x'}$ and the value $\mu_0$ as its extremum. 

To obtain an approximation for $\tilde{\mu}_0$ on the other hand, we note that in the Schwarzschild limit, $\sqrt{\mathcal{Q}}$ is the projection of the total angular momentum into the plane orthogonal to $z$, so $\mathcal{L}^2 = \mathcal{L}^2_z +\mathcal{Q}$ {[this can also be seen by setting $a =0$ in Eq.~\eqref{eq:AngularMom}]}, and we have
\begin{equation}
\tilde{\mu}_0 = \frac{\mathcal{L}_z/\mathcal{E}}{\sqrt{\mathcal{L}_z^2/\mathcal{E}^2 + \mathcal{Q}/\mathcal{E}^2 } }\,.
\end{equation}
Furthermore, for equatorial $x'$ the geodesic equations give us $d\phi/d\theta = \pm \mathcal{L}_z/\sqrt{\mathcal Q} = \pm \tilde{\mu}_0/\sqrt{1-\tilde{\mu}_0^2}$. This means
$\tilde{\mu}_0$ is just $\cos \beta$, with $\beta$ being the angle between the projection of the initial velocity into the $(\partial_{\theta},\partial_{\phi})$ plane and the $\partial_{\phi}$ direction. For non-negligible spin $a$, the calculation of $\mathcal{L}$ is less straight-forward, but we can nevertheless be assured that $\tilde{\mu}_0 =0$ if we set $\mathcal{L}_z =0$. For $a=0.65$, this corresponds to a radius $r=2.79473 M$. 

\begin{figure}[tb] 
\includegraphics[width=0.95\columnwidth]{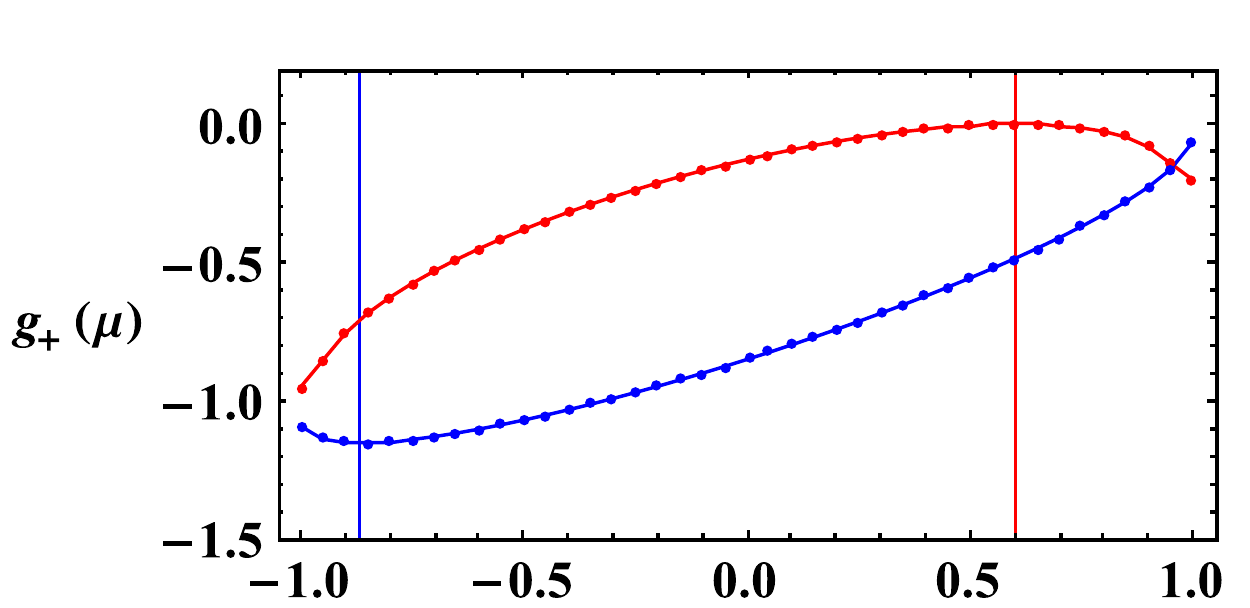}
\includegraphics[width=0.95\columnwidth]{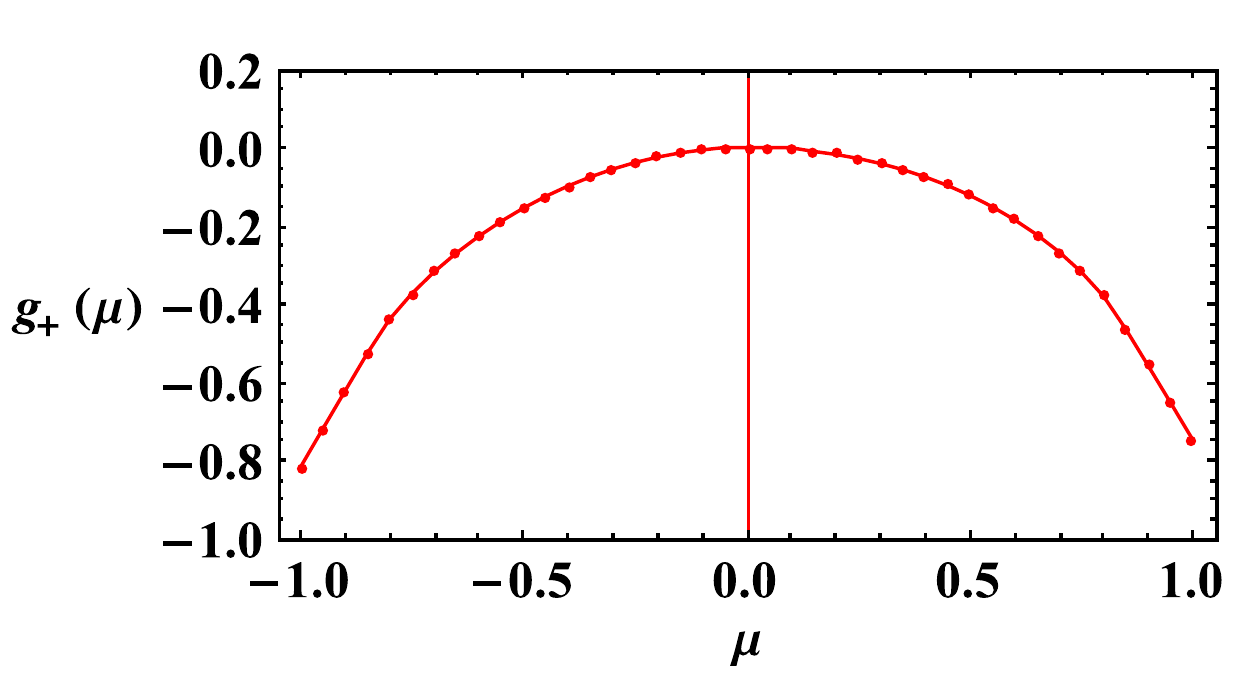}
\caption{ 
Top and bottom panels show $g_+(\mu)$ for certain spherical photon orbits for the cases $a=0.01$ and $a=0.65$ respectively, and the vertical lines are $\tilde{\mu}_0$. The figures suggest that $\mu_0 = \tilde{\mu}_0$ for both cases.    
}
\label{fig:MuVsMu}
\end{figure}

We plot in Fig.~\ref{fig:MuVsMu} the function $g_+(\mu)$ for both $a=0.01$ and $a=0.65$ (top and bottom panels, respectively). We have chosen two generic inclinations in the case where $a=0.01$ and used the Schwarzschild expressions to calculate their $\tilde{\mu}_0$. For the case $a=0.65$, we pick $\tilde{\mu}_0 = 0$. In the figures, the predicted values of $\tilde{\mu}_0$ are shown as vertical lines, and they agree with $\mu_0$ for both the slowly-spinning and rapidly-spinning cases.  

Next, we examine the case when $x$ is moved off the photon sphere, while $x'$ is left on it. Such a scenario still satisfies the condition that the geodesic be close to $r_p$ at some stage of its history, so we expect $\mu_0 = \tilde{\mu}_0$ to remain a valid prediction for $\mu_0$. 
To numerically study this case, we need to give the geodesics a non-vanishing radial velocity. 

For the slowly-spinning case, this entails nothing more than {adding} 
in a radial component while keeping the transverse spatial components unchanged (with the temporal component of the four-velocity tuned accordingly so the geodesic remains null). This way the angle $\beta$ is unchanged, as is $\tilde{\mu}_0$.

For the rapidly-spinning case, we need to adjust the radial and angular components in synch in order to maintain $\tilde{\mu}_0 = 0$. 
When $\mathcal L_z =0$ the geodesics form a one parameter family, and we choose to use $\mathcal{E}/\sqrt{\mathcal Q}$ as our parameter. Explicitly, the geodesic equations are given by
\begin{align}
\label{eq:OneParam1}
\frac{1}{\sqrt{\mathcal Q}}\frac{d\phi}{d\xi} &= a\left( \frac{2Mr}{\Delta} \right) \frac{\mathcal{E}}{\sqrt{\mathcal Q}}\,, \\
\frac{1}{\sqrt{\mathcal Q}}\frac{dr}{d\xi} &= \pm \sqrt{ \left(r^4+a^2r^2+2a^2Mr \right) \frac{\mathcal{E}^2}{\mathcal Q} -\Delta} \,, \\
\label{eq:OneParam3}
\frac{1}{\sqrt{\mathcal Q} }\frac{d\theta}{d\xi} &= \pm 1\,.
\end{align}
The parameter $\mathcal{E}/\sqrt{\mathcal Q}$ essentially determines the angle that the initial velocity makes with the radial direction, and in solving these equations we can absorb $\sqrt{\mathcal Q}$ into the definition of $\xi$.\footnote{In practice we actually use the angular momentum $\mathcal L$ in our codes in place of $\mathcal Q$ in these equations, because of its correspondence to $L$ and our eikonal equations. 
When $\mathcal L_z = 0$, Eq.~\eqref{eq:AngularMom} guarantees that these methods are equivalent.} If we restrict to the special angle that gives the spherical photon orbits, $\mathcal{E}$ reduces to $\omega_R$, and the geodesic's initial velocity as given by these expressions reduces to Eq.~\eqref{eq:SphereOrbitInitDir}. 
We in fact choose a value of $\mathcal{E}$ slightly larger than $\omega_R$, and pick the sign choices appropriately, so that the geodesic moves outwards away from the photon sphere. 

\begin{figure}[tb]
\includegraphics[width=0.95\columnwidth]{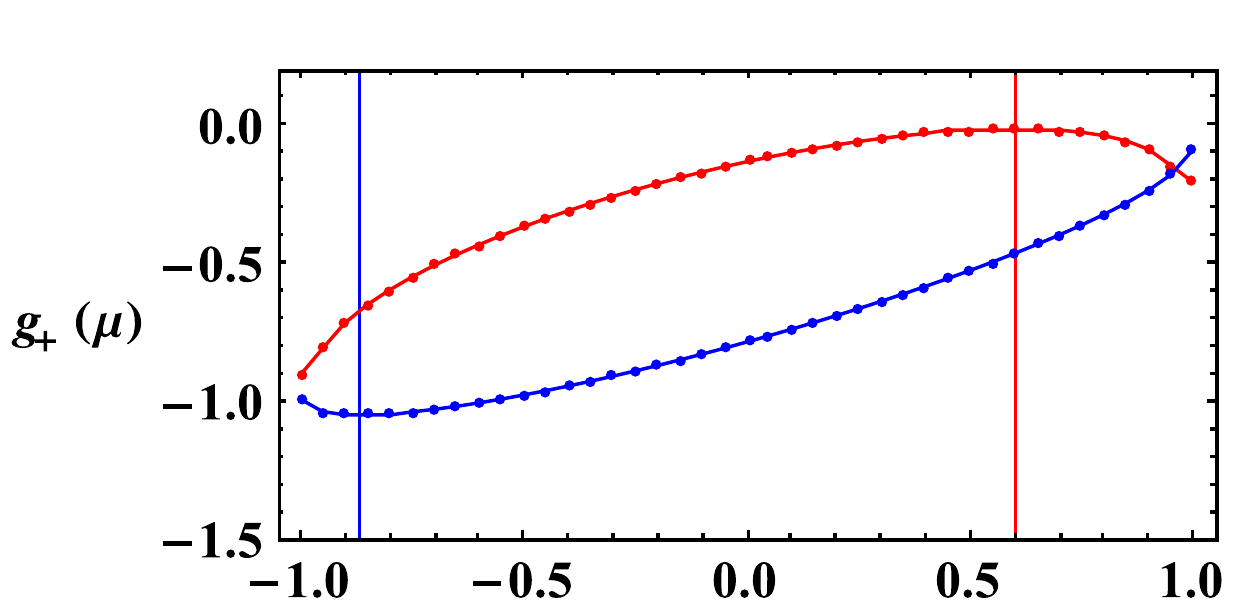}
\includegraphics[width=0.95\columnwidth]{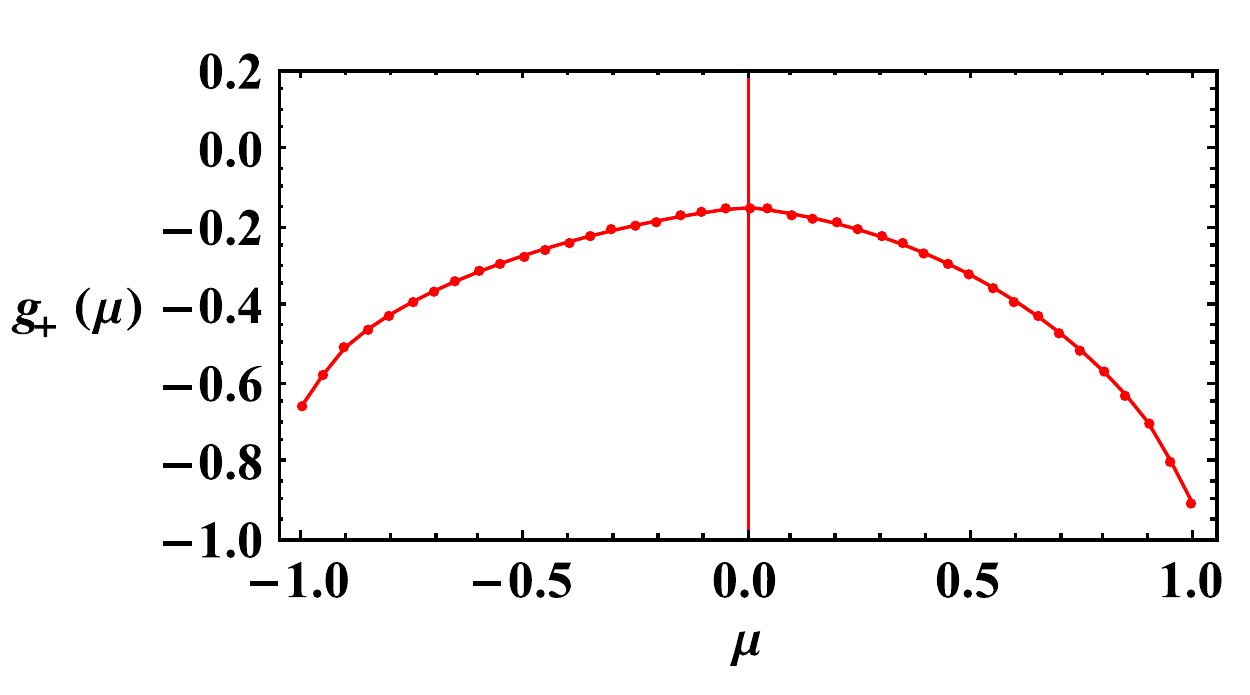}
\caption{
Top and bottom panels show $g_+(\mu)$ for certain non-spherical geodesics for $a=0.01$ and $a=0.65$ cases respectively, and the vertical lines are $\tilde{\mu}_0$. 
}
\label{fig:MuVsMu2}
\end{figure}

The initial and terminal Boyer-Lindquist $r$ are $(2.993 M, 3.381M)$ and $(3.01M, 3.554M)$ in the two slowly-spinning cases, and $(2.798M,3.701M)$ in the rapidly-spinning case. The corresponding $g_+(\mu)$ plots are shown in Fig.~\ref{fig:MuVsMu2}. Once again, we observe $\mu_0 =\tilde{\mu}_0$ as expected.

\begin{figure}[tb]
\includegraphics[width=0.95\columnwidth]{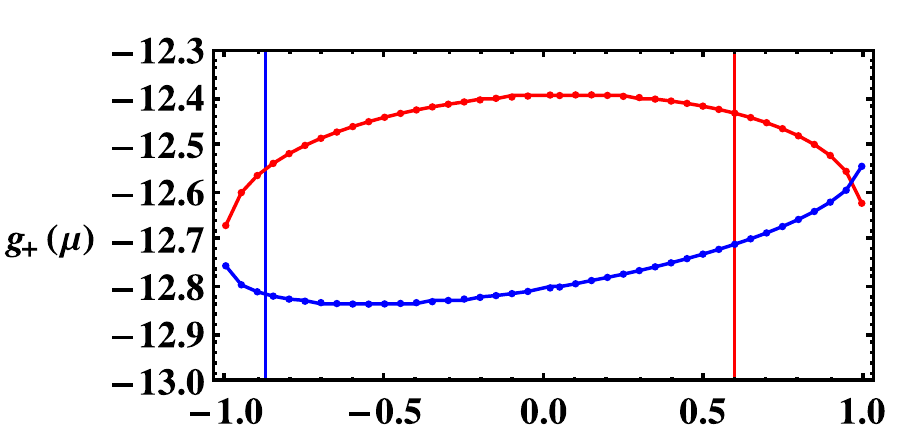}
\includegraphics[width=0.95\columnwidth]{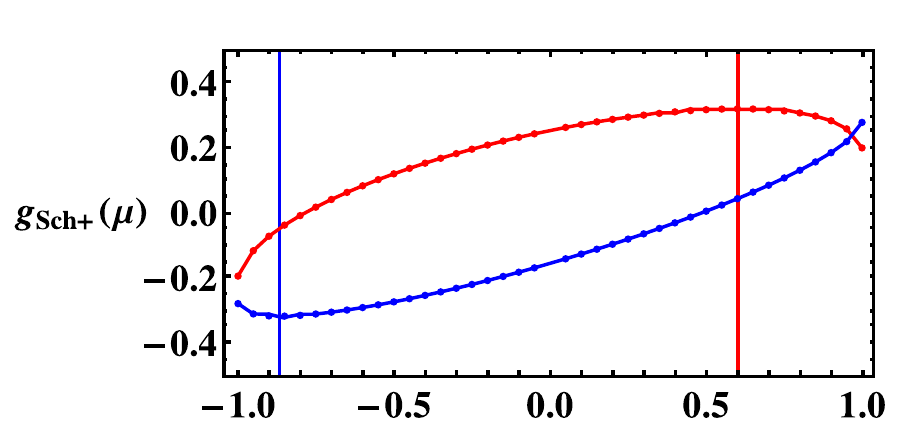}
\includegraphics[width=0.95\columnwidth]{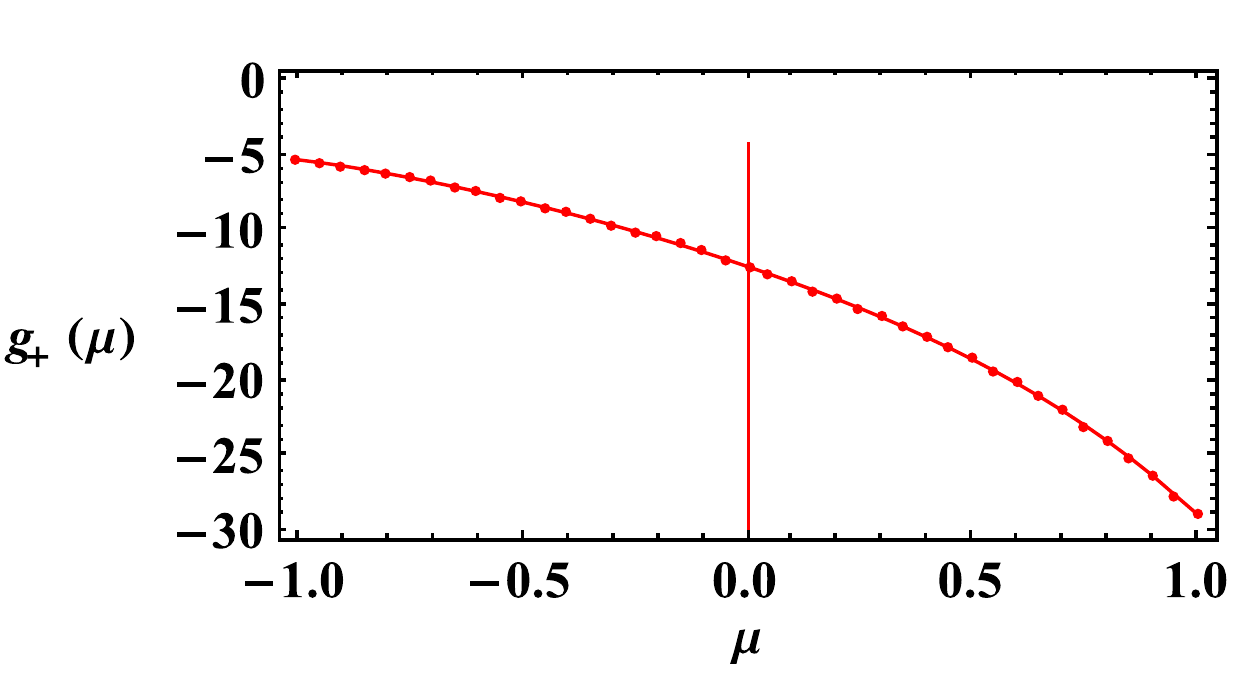}
\caption{
Top and bottom panels show $g_+(\mu)$ for geodesics with neither $x'$ nor $x$ on the photon sphere, for $a=0.01$ and $a=0.65$ cases respectively, and the vertical lines are $\tilde{\mu}_0$. The middle panel shows the $a=0.01$ case but including only those terms in $g^+$ that have $\mu$-dependence in the exact Schwarzschild limit.  
}
\label{fig:MuVsMu3}
\end{figure}

Lastly, we look at the case when no part of the geodesics is close to $r_p$. We expect that the equality between $\mu_0$ and $\tilde{\mu}_0$ to be broken for the rapidly-spinning case, and to a lesser extent for the slowly-spinning case. We launch the geodesics from an $x'$ on the equatorial plane at a radius of $r'=8M$, and with initial velocities such that $\tilde{\mu}_0$ has the same values as in the earlier scenarios. The $g_+(\mu)$ plots are given in Fig.~\ref{fig:MuVsMu3}. 
We observe a significant mismatch between $\mu_0$ and $\tilde{\mu}_0$ even with a very small spin of $0.01$. For comparison, and to rule out some numerical errors as the source of the mismatch (e.g. numerical error in the geodesic integrator), we also plot only the part of $g_+$ that has $\mu$-dependence in the Schwarzschild limit, which shows $\mu_0 =\tilde{\mu}_0$ as expected. 

\subsection{Coincidence of singular set with geodesics \label{sec:Residual}}

\begin{figure*}[tb]
\begin{minipage}[b]{0.49\textwidth}
\includegraphics[width=\textwidth]{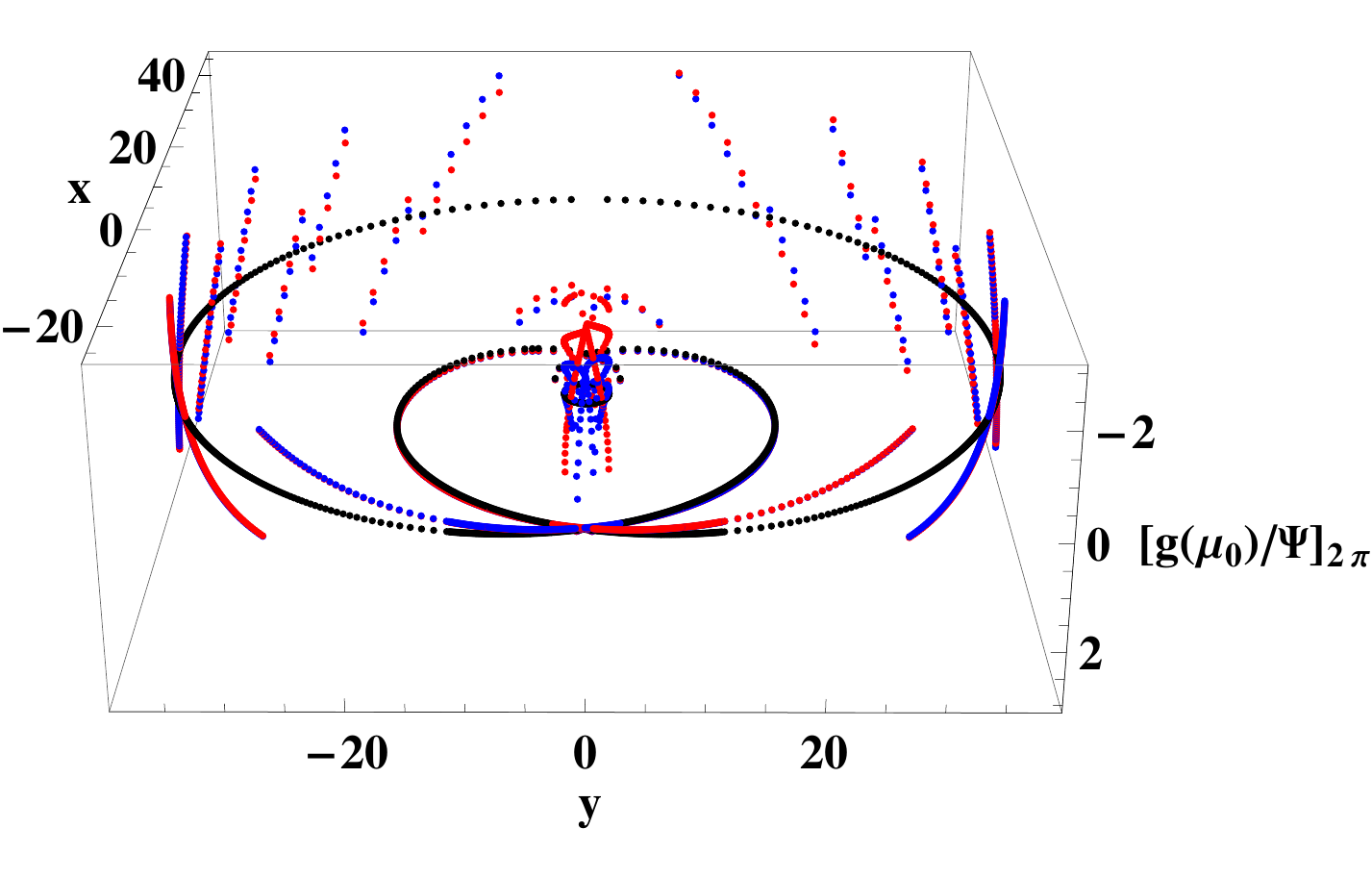}
\includegraphics[width=\textwidth]{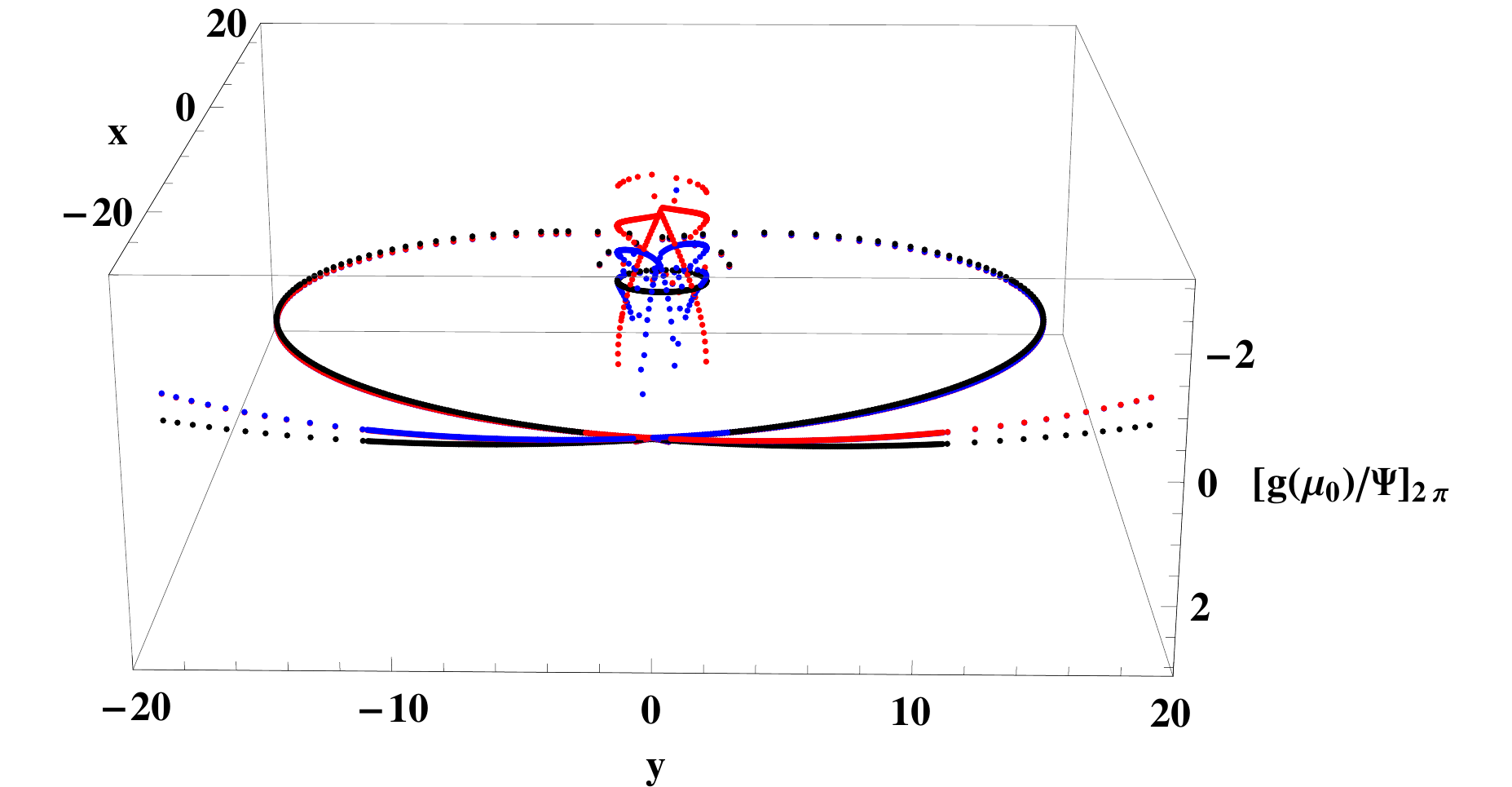}
\end{minipage}
\begin{minipage}[b]{0.49\textwidth}
\includegraphics[width=0.65\textwidth]{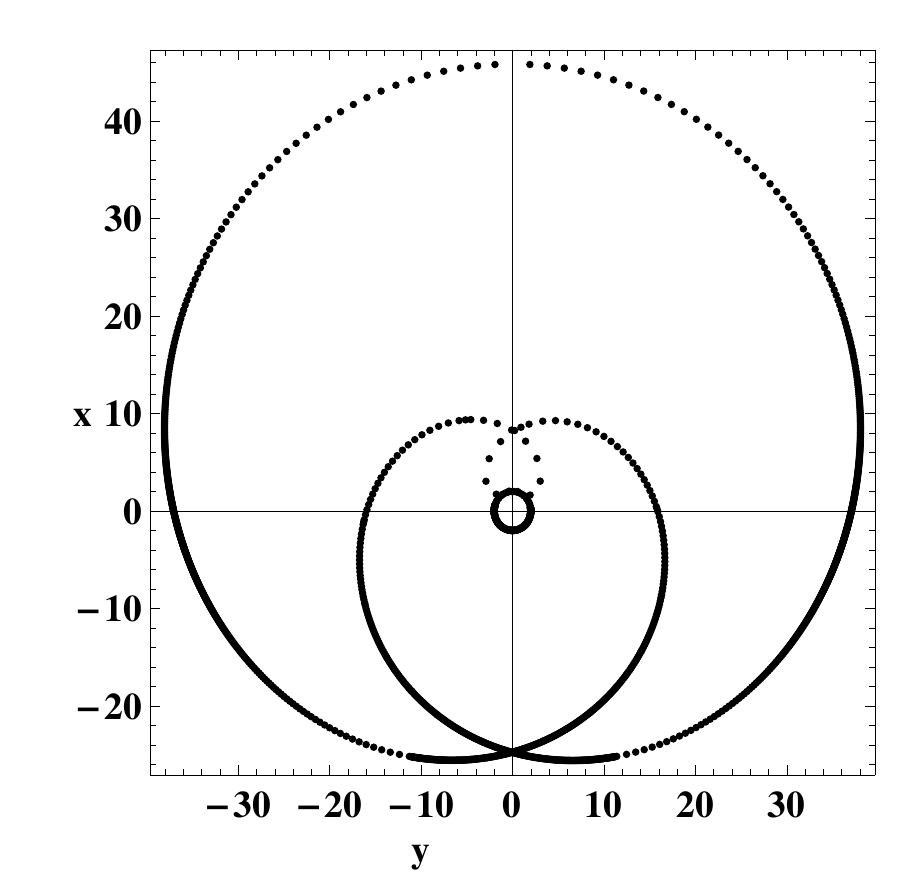}
\includegraphics[width=\textwidth]{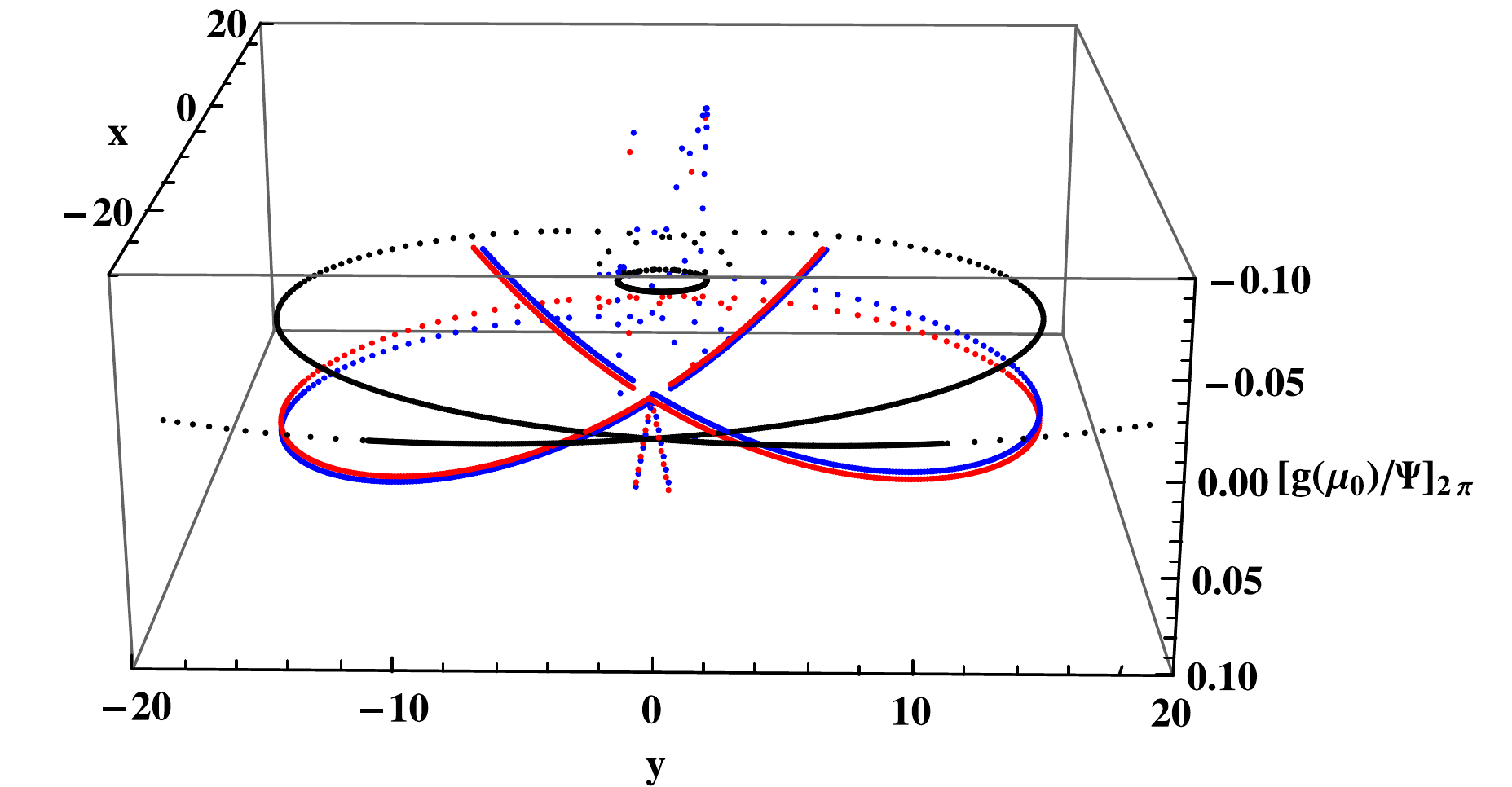}
\end{minipage}
\caption{ \fan{
For $a=0.01$, the residual $[g(\mu_0)]_{2\pi}$ ($[\Psi]_{2\pi}$) vs the Boyer-Lindquist Cartesian coordinates on equatorial plane $\Sigma$ are shown as red (blue) dots in the figure on the top left. The black dots are reference points that has the same spatial coordinates as the red and blue points but $[g(\mu_0)/\Psi]_{2\pi}\equiv 0$. 
Both $[g(\mu_0)]_{2\pi}$ and $[\Psi]_{2\pi}$ are shown for every geodesic, therefore whenever only red or blue dots are visible, the two quantities are simply coincident.
The bottom panels are the same as the top-left figure, but concentrating on the connecting segment and horizon, as well as zooming in on the $[g(\mu_0)/\Psi]_{2\pi}$ direction in the case of bottom right panel.
}}
\label{fig:SchwarzschildMatchDO3D}
\end{figure*}

In this section, we numerically study the singular structure of the QNM contribution to the Green function. 

We use a different measurement from that used in \cite{Dolan2011} to examine whether the singular set of the QNM part of the Green function, as given by 
\begin{align} 
\label{eq:PhCoKerr}
g(\mu_0) = 2\pi j \,,
\end{align}
for one of the geometric phases introduced in 
Sec.~\ref{sec:SumOverQNM},  
matches the null wave front emanating from the source point $x'$ (see Fig.~$1$ in \cite{Dolan2011} for an illustration of such a wave front). 
Namely, we do not solve Eq.~\eqref{eq:PhCoKerr} 
(which is numerically expensive)
to try and find the $x$ coordinates of the singular set and compare it to the wave front location.  
Instead we use the terminal coordinates of the geodesics constituting the wave front to calculate $g_{\pm}(\mu_0)$ and $g^{\rm P}_{\pm}(\mu_0)$, and subsequently their residuals $[g_{\pm}(\mu_0)]_{2\pi}$  and $[g^{\rm P}_{\pm}(\mu_0)]_{2\pi}$ modulo $2\pi$. If the residual vanishes for either of the $+$ or $-$ sign choices of $g$ or $g^{\rm P}$, then we know that there exists $j \in \mathbb{Z}$ such that Eq.~\eqref{eq:PhCoKerr} is satisfied, and the solution matches the geodesics. 
By adopting this alternative method, we bypass the problem that Eq.~\eqref{eq:PhCoKerr} does not have a solution when the approximations underlying it break down. This allows us to extend the study throughout the entire wave front.

In what follows, we will denote by $[g(\mu_0)]_{2\pi}$ whichever of $[g_+(\mu_0)]_{2\pi}$ and 
$[g_-(\mu_0)]_{2\pi}$ that has the smallest absolute value, and similarly for $g^{\rm P}$. 

\subsubsection{Singular structure of the slowly-spinning limit}
To demonstrate the numerical techniques adopted, we begin with the slowly-spinning limit.
According to the discussion in Sec.~\ref{sec:GFschw}, we expect our geometric phases from Sec.~\ref{sec:SumOverQNM} to limit to $\Psi^\pm$ from Eq.~\eqref{eq:PsiDO}. Therefore, the condition \eqref{eq:PhCoKerr} for the Green function to be singular reduces to 
\begin{align} 
\label{eq:PhCoSch}
\Psi^{\pm} = 2\pi j \,,
\end{align}
matching the expression in \cite{Dolan2011}. 
In this section, in addition to calculating the geometric phases, we also calculate $\Psi^\pm$ directly for comparison. 
To construct the null wave front, we launch a bundle of null geodesics in directions on the out-going celestial sphere that are confined to a plane $\Sigma$ containing the line $\ell_{\text{Ox'}}$ joining the coordinate origin $O$ to the initial launch point $x'$. 
The initial directions make angles between $0$ and $2\pi$ with $\ell_{\text{Ox'}}$. In the exact Schwarzschild limit, all other geodesics launched from the same $x'$ can be obtained from this bundle through symmetry arguments. 
For convenience, we adapt our coordinates such that $\Sigma$ is the equatorial plane.

Just as for Fig.~$5$ in \cite{Dolan2011}, we place $x'$ at a Boyer-Lindquist radius of $8M$ from the coordinate origin, and let the geodesics evolve for a Boyer-Lindquist time of $41.84M$. 
For our current case, where $\Sigma$ is the equatorial plane, we use $\mu^{\pm}_0 = \tilde{\mu}^{\pm}_0$, instead of trying to search for $\mu_0$ numerically near the boundaries of the range of valid values of $\mu$ (we consider a case later where we do search for $\mu_0$). To find $\tilde{\mu}^{\pm}_0$, we {use the the geometric correspondence equation for $\mathcal Q$,}
\begin{equation} 
\label{eq:QGeo}
\mathcal{Q} \approx L^2 \left[1- \mu^2 - \frac{a^2 \Omega_R^2}{2}\left(1-\mu^2 \right)\right]\,,
\end{equation}
which shows that when $\tilde{\mu}_0=\pm 1$, we have $\mathcal{Q}=0$. For our chosen launching point on the equatorial plane, this translates into $\Theta=0$ and subsequently $d\theta/d\lambda =0$. 
In other words, the choice $\tilde{\mu}^{\pm}_0 = \pm 1$ corresponds to launching geodesics in the equatorial plane (whatever the spin of the black hole). 
Furthermore, for these geodesics and $\mu^{\pm}_0$ values, $[g]_{2 \pi}$ and $[g^{\rm P}]_{2 \pi}$ become degenerate, and so we only calculate $[g]_{2 \pi}$.

In Fig.~\ref{fig:SchwarzschildMatchDO3D} we plot both the resulting $[g(\mu_0)]_{2\pi}$ and $[\Psi]_{2\pi}$ (the one among $[\Psi^+]_{2\pi}$ and $[\Psi^-]_{2\pi}$ with the smaller absolute value) in the Boyer-Lindquist Cartesian coordinates on $\Sigma$ as red and blue dots, respectively. 
The black dots are reference points that have the same spatial coordinates as the red and blue points but $[g(\mu_0)/\Psi]_{2\pi}\equiv 0$. They are included only as a visual aid. 
{The top left panel of Fig.~\ref{fig:SchwarzschildMatchDO3D} shows the residuals on the entire wave front, and the top right panel shows an overhead view of the wave front only (the residuals are not plotted).}
We recover the excellent match between the singular set of Green function and the wave front for that segment connecting the black hole horizon to the outer rim (this is the part of wave front examined in \cite{Dolan2011}). But the matching quality deteriorates significantly for the rim and the horizon. {The bottom left panel focuses on the region where the matching is good, and the bottom right panel zooms in to show the values of the residuals.}
We also observe a good match between $[g(\mu_0)]_{2\pi}$ and $[\Psi]_{2\pi}$, with their difference being smaller than the overall residual on the connecting segment (bottom right panel of Fig.~\ref{fig:SchwarzschildMatchDO3D}). This serves as a demonstration of the conclusion in Sec.~\ref{sec:GFschw}.

\begin{figure*}[tb]
\begin{minipage}[b]{0.49\textwidth}
\includegraphics[width=\textwidth]{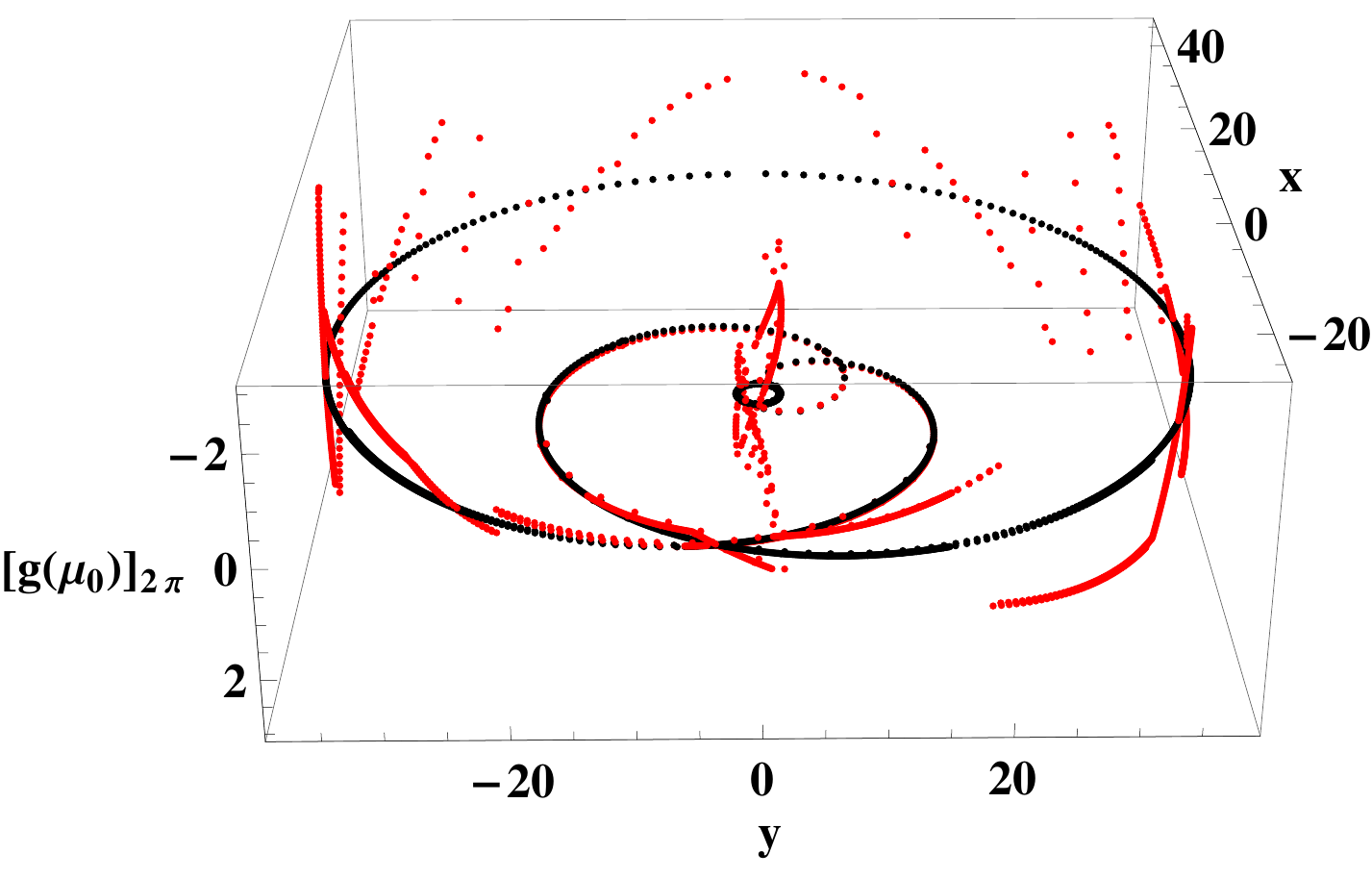}
\includegraphics[width=\textwidth]{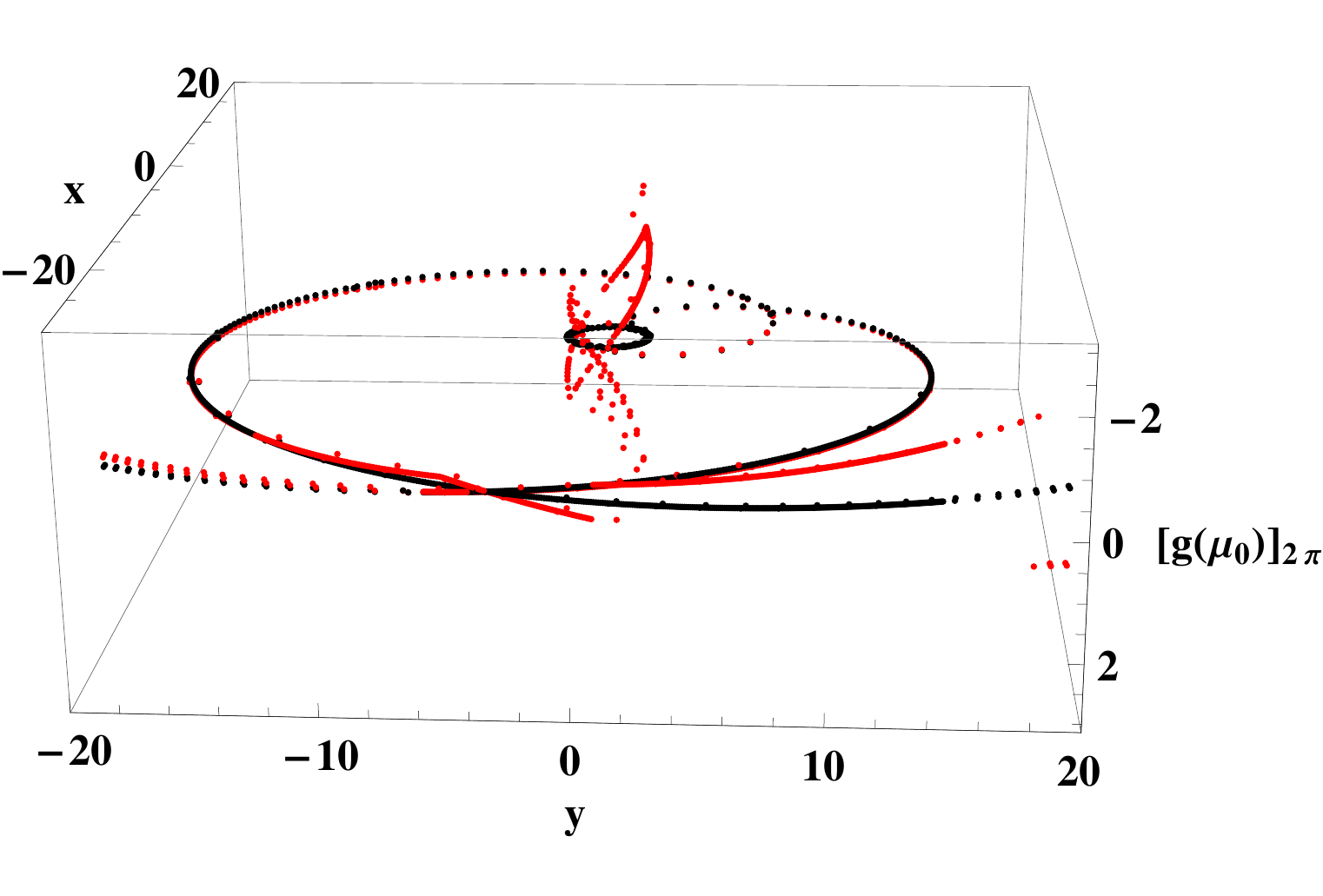}
\end{minipage}
\begin{minipage}[b]{0.49\textwidth}
\includegraphics[width=0.65\textwidth]{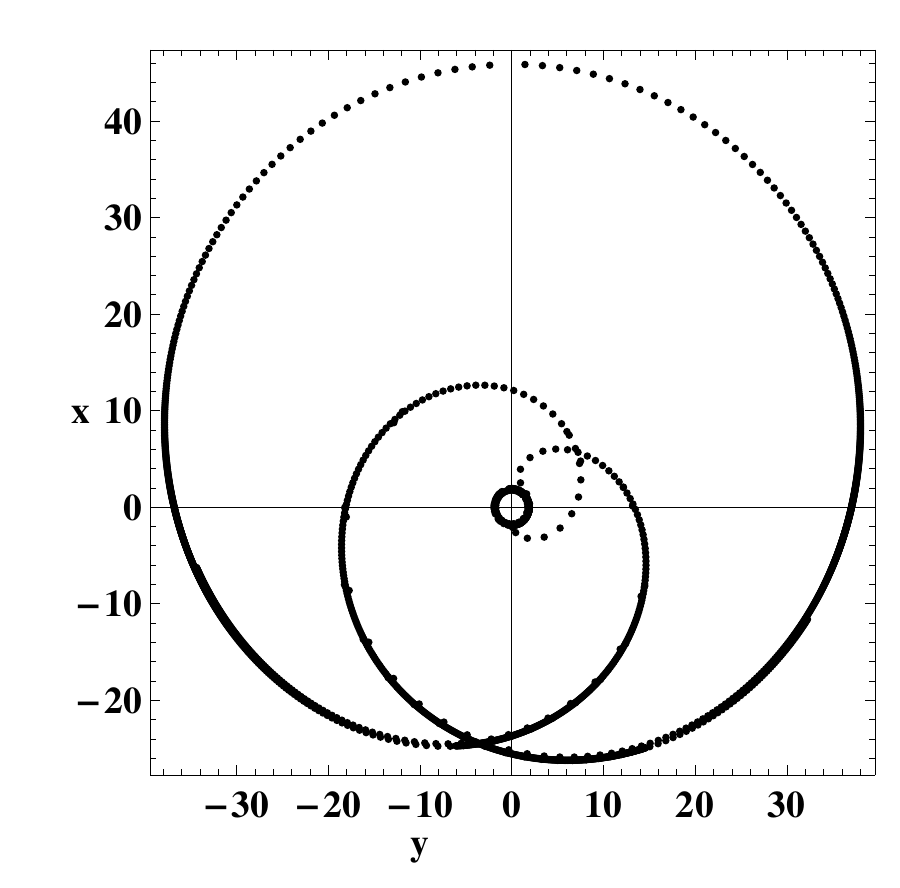}
\includegraphics[width=\textwidth]{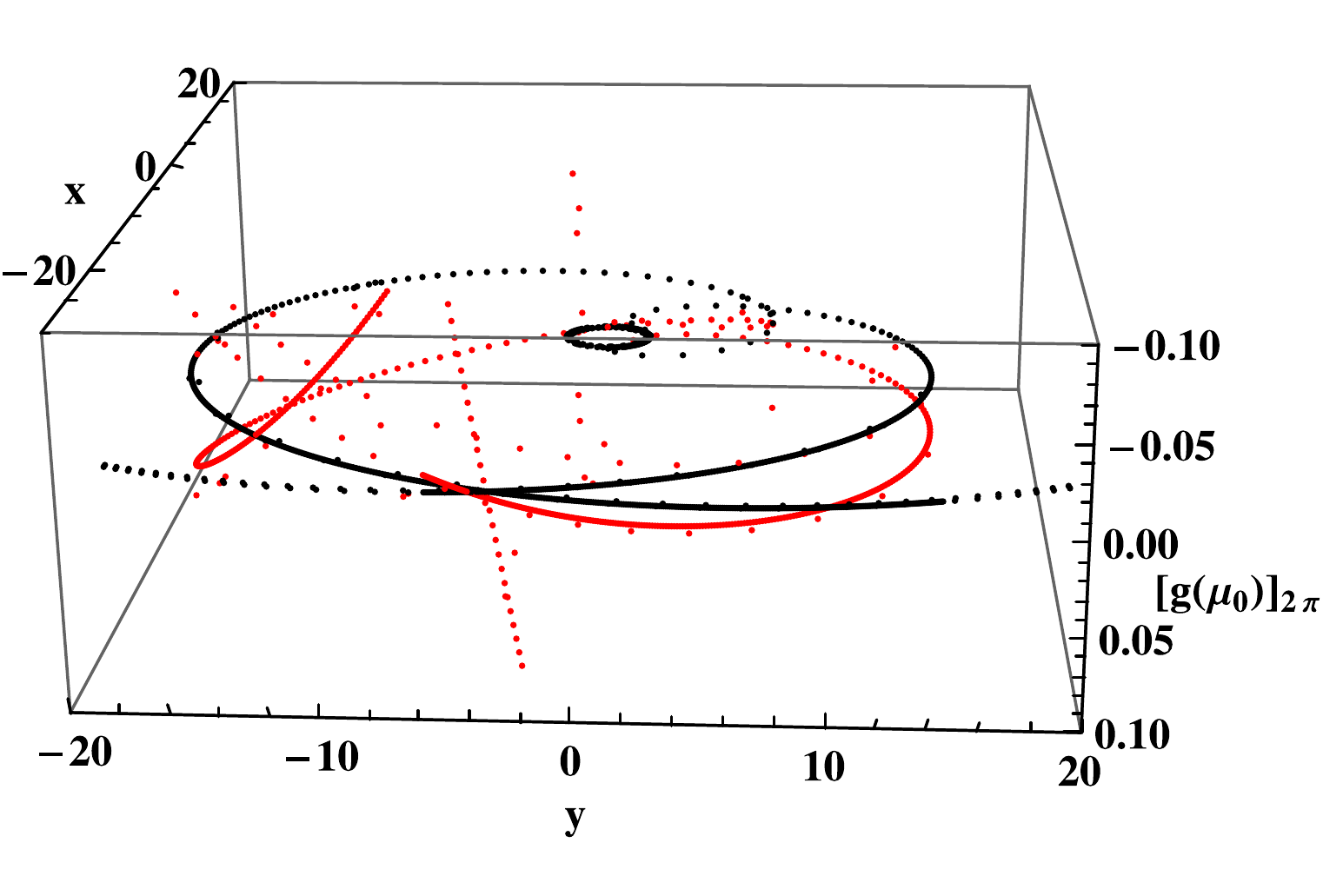}
\end{minipage}
\caption{ 
For $a=0.65$, the residual $[g(\mu_0)]_{2\pi}$ vs the Boyer-Lindquist Cartesian coordinates on the equatorial $\Sigma$ are shown as red dots in the figure on the top left. The black dots are reference points that has the same spatial coordinates as the red points but $[g(\mu_0)]_{2\pi}\equiv 0$. The figure on the top right is a portrait of the wave front on the $\Sigma$ plane. The bottom panels are the same as the top-left figure, but concentrating on the connecting segment and horizon, as well as zooming in on the $[g(\mu_0)]_{2\pi}$ direction in the case of the bottom right panel.  
}
\label{fig:KerrEquatorialPlane}
\end{figure*}

\begin{figure*}[tb]
\begin{minipage}[b]{0.49\textwidth}
\includegraphics[width=\textwidth]{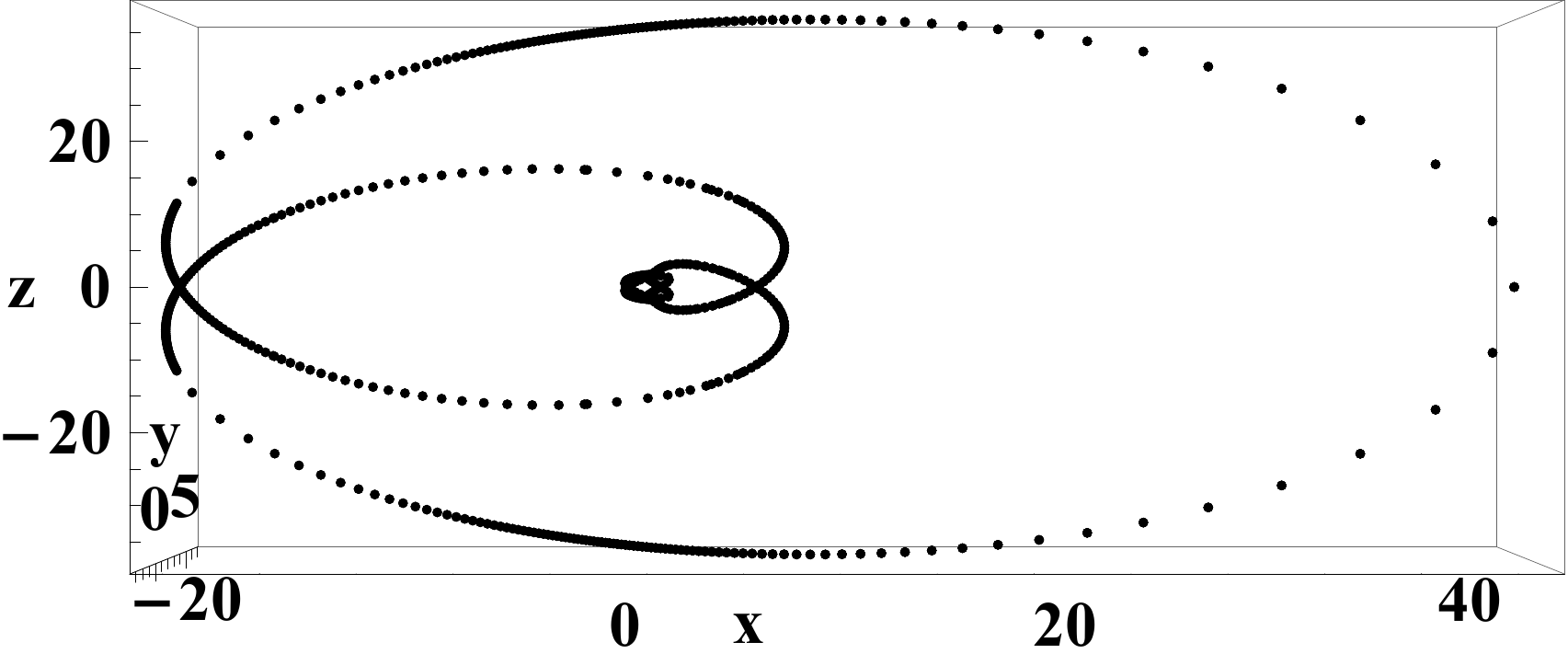}
\includegraphics[width=\textwidth]{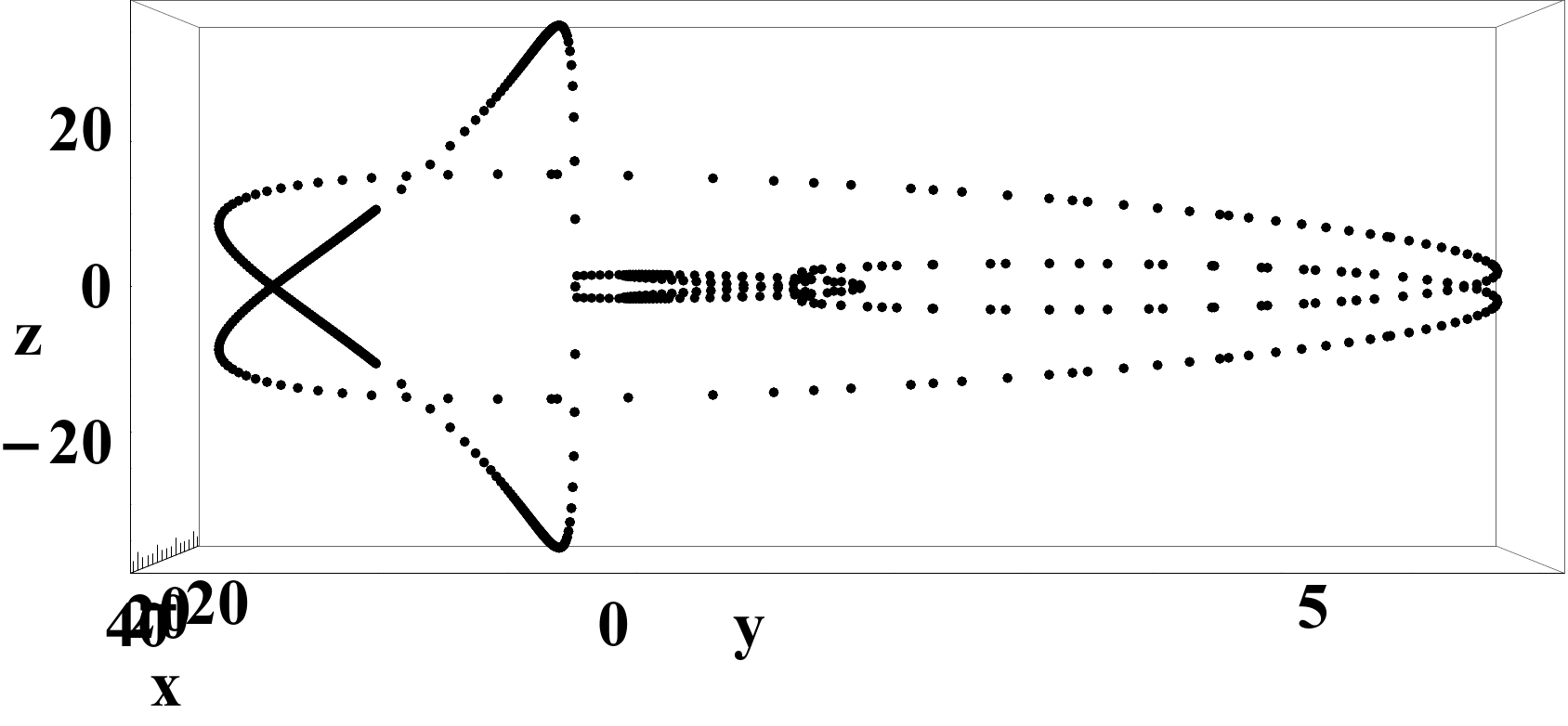}
\end{minipage}
\begin{minipage}[b]{0.49\textwidth}
\includegraphics[width=\textwidth]{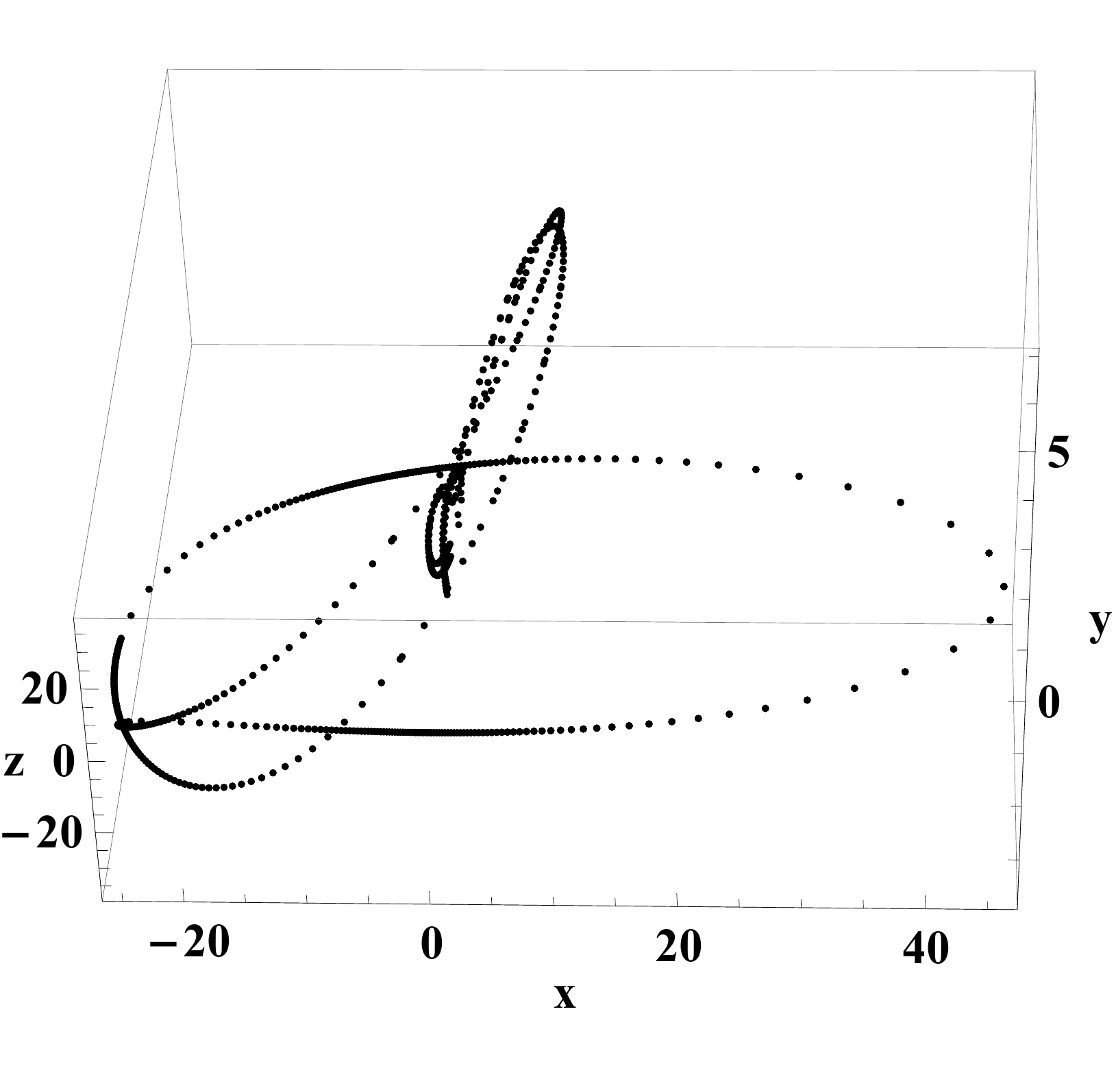}
\end{minipage}
\caption{ \fan{
For $a=0.65$, the wave front is more complicated when we launch the geodesics in the poloidal directions. Plotted above are the wave front viewed from different angles. 
}}
\label{fig:KerrPoloidalPlane}
\end{figure*}

\subsubsection{The singular structure for a generic Kerr black hole}

Next, we turn to a Kerr black hole of a generic spin $a=0.65$, which we call our rapidly-rotating case. 
We first consider the situation when the geodesics are confined to the equatorial plane, with the launching point $x'$ at $(8M,\pi/2,0)$. 
In this simpler case, we once again have $\tilde{\mu}^{\pm}_0 =\pm 1$ and we use these values for $\mu^{\pm}_0$ instead of numerically searching for them. 
We also only calculate $g$, as $g$ and $g^{\rm P}$ functions again share the same residuals. 
In addition, we do not need to restrict $\phi$ to within $[0,2\pi)$, because extra winding numbers simply introduce integer multiples of $2\pi$ to $g$ {when $\mu_0 = \pm 1$}. 
The residuals $[g(\mu_0)]_{2\pi}$ are plotted in Fig.~\ref{fig:KerrEquatorialPlane}, which demonstrates the same behaviour as in the Schwarzschild case, indicating that the relationship between the singular structure of the Green function and the null wave front as observed in 
\cite{Dolan2011} generalizes to the Kerr case, at least for the equatorial plane.

Finally, we show that this generalization remains valid when we move out of the equatorial plane for the $a=0.65$ Kerr black hole. To this end, we examine a different choice of the plane $\Sigma$ in which we initially fire our geodesics. Our choice of $\Sigma$ is the poloidal plane, meaning that initially the tangents to the geodesics have zero $\partial_{\phi}$ components. We do not otherwise select these geodesics to have particular characteristic parameters. We note that our choice of poloidal $\Sigma$ does not lead to geodesics with $\tilde{\mu}^{\pm} = 0$ [such geodesics equate to those launched in the 
poloidal plane only in the Schwarzschild limit, see discussion surrounding Eq.~\eqref{eq:OneParam1}].  
Instead, we now search for $\mu^{\pm}_0$ directly by solving $g_{\pm}'(\mu^{\pm}_0)=0$ or $g^{\rm P}_{\pm}{}'(\mu^{\pm}_0)=0$ for each geodesic. 
Since there is no longer a degeneracy between $g_{\pm}$ and $g^{\rm P}_{\pm}$, we calculate both and select the minimal residual amongst the four possibilities ($g$ versus $g^{\rm P}$ and $+$ versus $-$), which we denote as $[g/g^{\rm P}(\mu_0)]_{2\pi}$. We also restrict $\phi$ to be within $[0,2\pi)$, which was not necessary in the equatorial plane case.   

\begin{figure*}[tb]
\begin{minipage}[b]{0.49\textwidth}
\includegraphics[width=\textwidth]{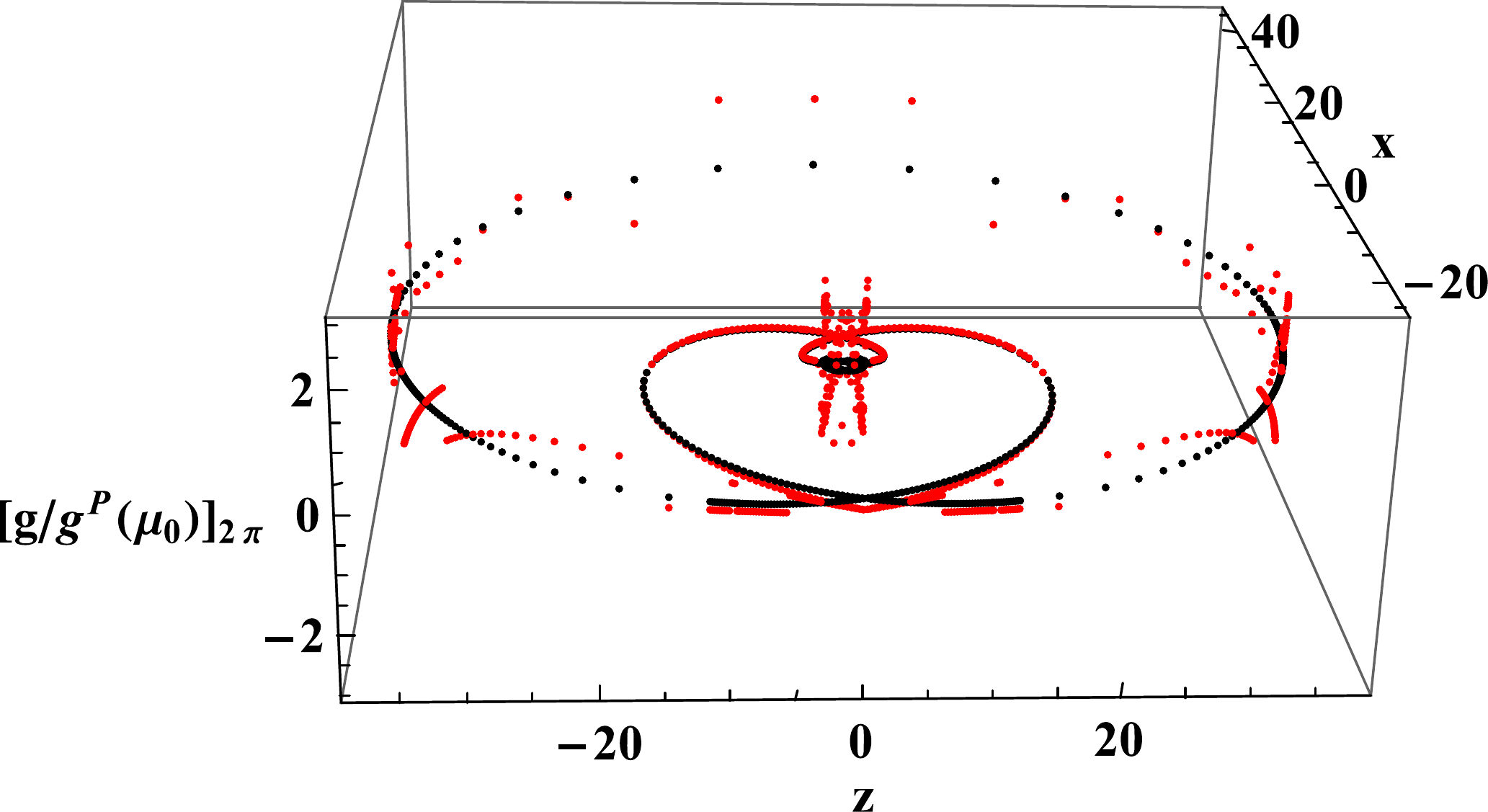}
\includegraphics[width=\textwidth]{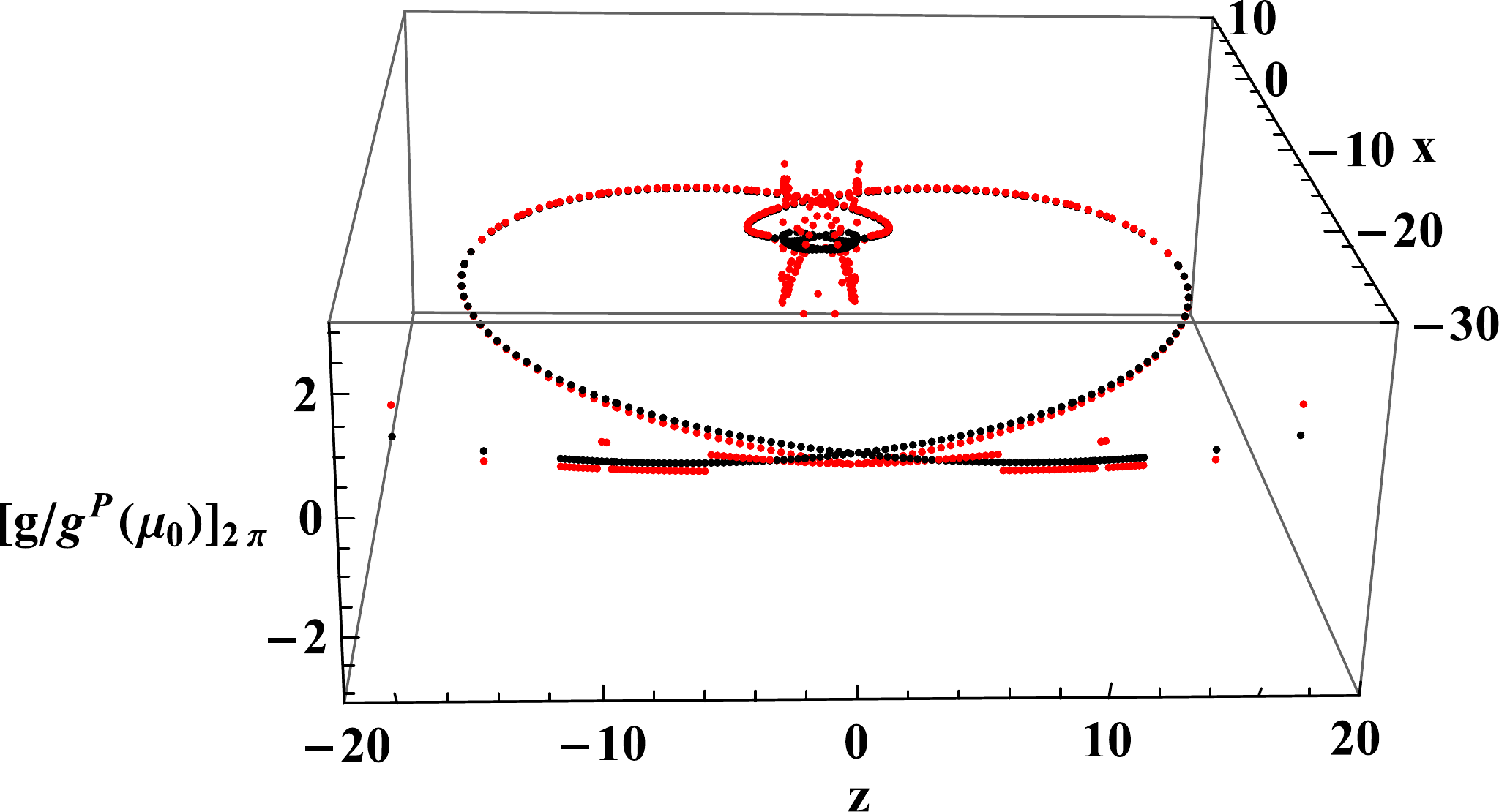}
\end{minipage}
\begin{minipage}[b]{0.49\textwidth}
\includegraphics[width=0.8\textwidth]{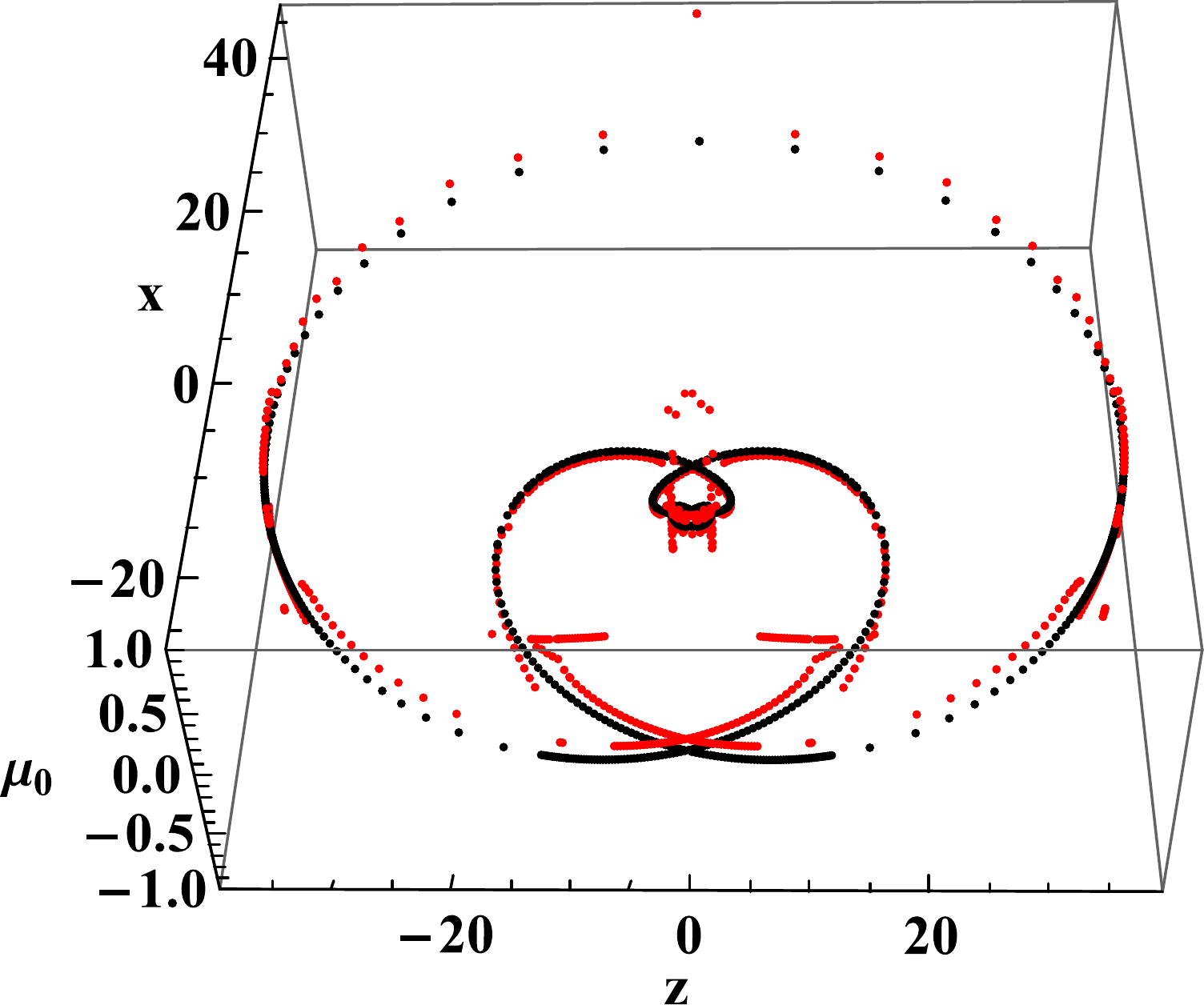}
\includegraphics[width=\textwidth]{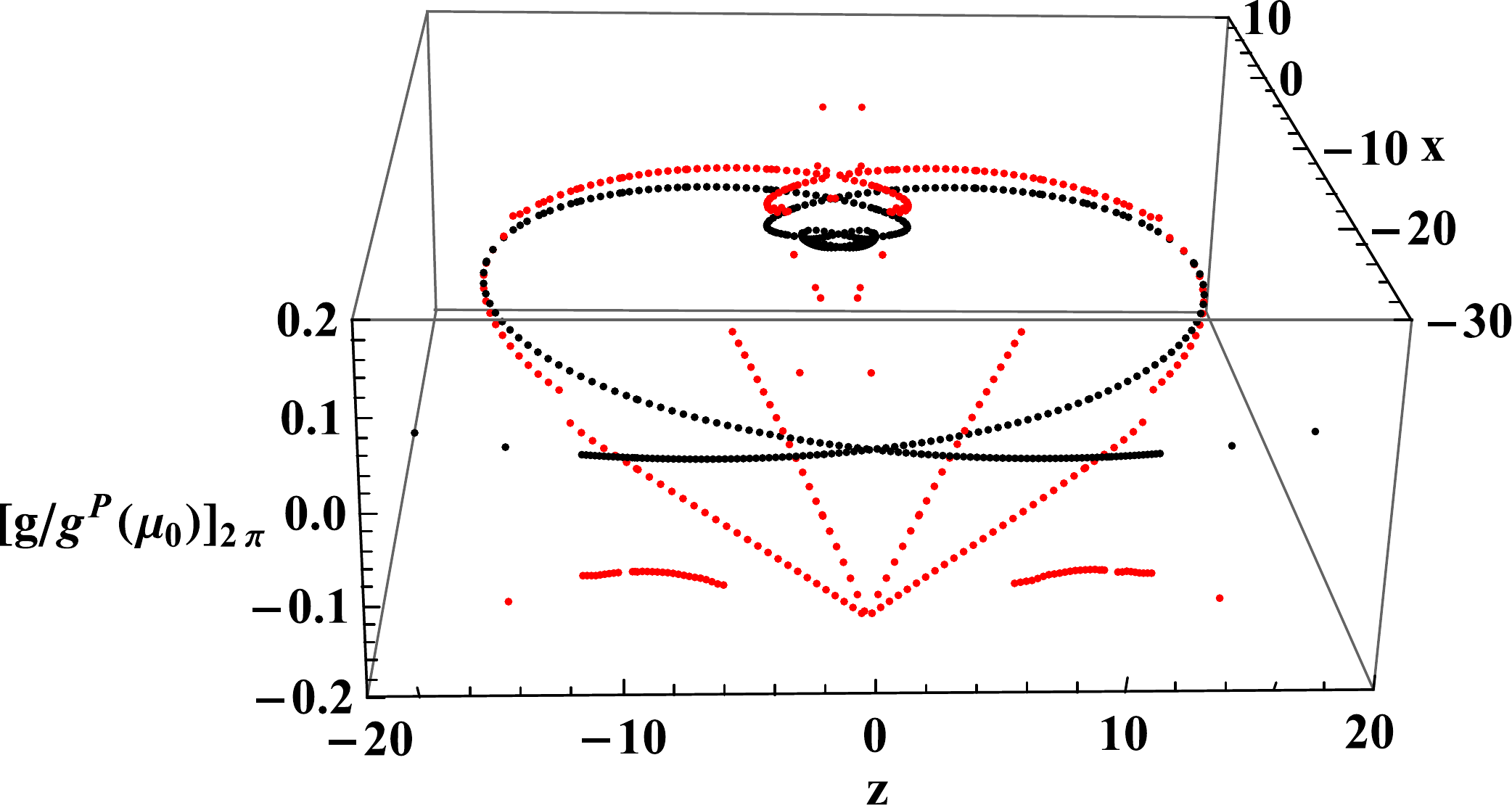}
\end{minipage}
\caption{ \fan{
For $a=0.65$ and $\Sigma$ being the poloidal plane, the residuals $[g/g^{\rm P}(\mu_0)]_{2\pi}$ vs the projection of the wave front onto $x-z$ plane are shown as red dots in the figure on the top left. The black dots are reference points that have the same spatial coordinates as the red points but $[g/g^{\rm P}(\mu_0)]_{2\pi}\equiv 0$. The figure on the top right shows the $\mu_0$ corresponding to the residual values in the top left panel. The bottom panels are the same as the top-left figure, but concentrating on the connecting segment and horizon, as well as zooming in on the $[g/g^{\rm P}(\mu_0)]_{2\pi}$ direction in the case of the bottom right panel.
}}
\label{fig:KerrPoloidalResidual}
\end{figure*}

The wave front for this initially poloidal $\Sigma$ plane case is not restricted to any particular plane, and is shown in Fig.~\ref{fig:KerrPoloidalPlane}. In order to achieve comparable accuracy for this wave front as in the equatorial plane case, a much smaller step size for the geodesic integrator is required. We therefore expect the uncertainty in the wave front location (limited by our computational resources) to constitute a significant proportion of the error in the residual. 
In order to present the residual $[g/g^{\rm P}(\mu_0)]_{2\pi}$ in a three dimensional figure, we replace the terminal points of the geodesics by their projection into the $x-z$ plane, and leave the third dimension for $[g/g^{\rm P}(\mu_0)]_{2\pi}$. Fig.~\ref{fig:KerrPoloidalResidual} shows of $[g/g^{\rm P}(\mu_0)]_{2\pi}$ (and corresponding $\mu_0$) plotted in this form, as in Fig.~\ref{fig:KerrEquatorialPlane}. Once again we observe a small residual for the ``connecting'' segment of the wave front, consistent with a generalization of the case of the Schwarzschild black hole.

 \section{Conclusion and future work}
\label{sec7}
 
In this study, we applied the WKB method to obtain an approximate expression for the QNM part of the Kerr spacetime's scalar Green function, which arises due to waves scattering off the strong-field region of the Kerr black hole. An immediate next step of this work would be to compare our analytical formula with numerical Green functions such as those presented in~\cite{Zenginoglu2012}. 

In addition, this study represents the first step in a future program whose aim is to extend the self-force calculation of \cite{casals2013b} to the Kerr spacetime. At the moment, less self-force results are available for Kerr black holes than Schwarzschild black holes, and any new method of attack on the problem is valuable. The development of the self-force approximation in Kerr is necessary for the modeling of sources for future space-based gravitational wave detectors. There is also growing interest in the possibility that self-force effects can accurately model binary black hole systems whose mass ratios would not normally be considered extreme~\cite{LeTiec:2011bk,LeTiec:2011dp,Tiec:2013twa}.

Besides the application in EMRI modeling, the Kerr Green function is useful for understanding any kind of wave propagation and emission in the strong-field region of the black hole. This includes electromagnetic (EM) radiation from stars, {compact objects, and accretion disks near SMBHs.} 
For example, it has been suggested that the phase front of EM waves can be distorted if the waves pass near the vicinity of a rotating black hole, and that this distortion can be characterized by the so called ``photon orbital angular momentum" carried by the wave~\cite{Harwitt}. By measuring the orbital angular momentum, one may be able to infer information such as the spin of the host black hole. {In order to generate an asymmetry in the orbital angular momentum expansion of a wave front, interference from the QNM part of the Green function is needed.} Our approximation for the Green function can be directly applied in this case to compute the signal emitted from a coherent source near a Kerr black hole, {scattered in the strong-field region,} and eventually observed by a distant observer. It enables us to build a map between the spacetime near the source and the EM observations at far distances. {We leave a description of this for a future study.}

\acknowledgements
We thank Marc Casals and Sam Dolan for helpful discussions on the Schwarzschild Green function, and An{\i}l Zengino\u{g}lu on the numerical Kerr Green function. We also would like to thank Marc Casals for his advice on an early draft of this paper. This research is funded by NSF Grants PHY-1068881 and PHY-1005655, CAREER Grants PHY-0956189 and PHY-1055103, NASA Grant No.NNX09AF97G, the Sherman Fairchild Foundation, the Brinson Foundation, and the David and Barbara Groce Startup Fund at Caltech. HY acknowledges supports from the Perimeter Institute of Theoretical Physics and the Institute for Quantum Computing. Research at Perimeter Institute is supported by the government of Canada and by the Province of Ontario though Ministry of Research and Innovation.
 
\appendix

\section{WKB quasinormal mode frequencies}
\label{sec:WKBFrequencies}

Analytic formulae for finding the QNM frequencies in Kerr were given in \cite{Yang2012a}. For completeness, we review the formulae here. In addition, note that the expression for $\Omega_R$ given in \cite{Yang2012a} becomes singular both at $a\to 0$ and $\mu \to 0$, which can cause difficulties in the evaluation of the various WKB quantities presented in this study. When the limits are taken carefully, the correct results for Schwarzschild and polar $(\mu =0)$ QNMs can be recovered, but for convenience we develop here an improved analytic formula for $\Omega_R$ which has manifestly the correct behavior for all values of $a$ and $\mu$.

The WKB method applied to the radial equation gives
\begin{align}
\label{eq:WKBfreq1}
\Omega_R = \frac{ \mu a (r_p - 1)}{(3 - r_p) r_p^2 - a^2(r_p+1)}\ .
\end{align}
As in the main text, $r_p$ is the WKB peak, which corresponds to the radius of the unstable null orbit associated with the high-frequency QNM.  Both the numerator and denominator of Eq~\eqref{eq:WKBfreq1} vanish when either $\mu =0$ (in which case the denominator is the cubic polynomial whose root is the position of the peak) or $a = 0$ (where $r_p = 3$). With the supplemental approximation~\eqref{eq:AlmApp} $r_p$ is found for generic $\mu$ and $a$ by solving the sixth-order polynomial
 \begin{align}
 \label{eq:PeakPoly}
0=& 2[ (3 - r_p) r_p^2 - a^2 (r_p+1) ]^2 - \mu^2 a^2 (4r_p^2  [r_p^2-3] \notag \\
& + a^2[3r_p^2 +2r_p+3 - \mu^2(r_p-1)^2 ])\,,
 \end{align}
 where we have separated Eq.~(2.36) of~\cite{Yang2012a} into $\mu$-dependent and $\mu$-independent parts. We see that the $\mu$-independent part is proportional to the square of the denominator in Eq.~\eqref{eq:WKBfreq1}, which confirms that this cubic vanishes at $r_p$ when $\mu =0$. Solving Eq.~\eqref{eq:PeakPoly} for this $\mu$-independent cubic allows us to rewrite Eq.~\eqref{eq:WKBfreq1} in a less compact but more useful form,
 \begin{widetext}
 \begin{align}
 \label{eq:WKBfreq2}
 \Omega_R = & \frac{\sqrt{2}(r_p-1)}{\sqrt{4 r_p^2  (r_p^2-3) +a^2[3r_p^2 +2r_p+3 - \mu^2(r_p-1)^2 ] }}\,.
 \end{align}
 It is straightforward to check that for $a =0$, Eq.~\eqref{eq:WKBfreq2} gives $\Omega_R = 1/\sqrt{27}$. The equation is also regular at $\mu =0$. It is not obvious that it approximates the value for $\Omega_R$ for the polar modes $(\mu =0)$ computed without the approximation~\eqref{eq:AlmApp}, which is~\cite{Dolan10}
 \begin{align}
 \label{eq:PolarWKB}
 \Omega_R = \frac{\pi \sqrt{\Delta(r_p)}}{2(r_p^2+a^2) {\rm EllipE}\left[\frac{a^2\Delta(r_p)}{(r_p^2+a^2)^2}\right]}\,.
 \end{align}
We have checked numerically that the two expressions are nearly identical when evaluated at the peak $r_p$. In fact Eq.~\eqref{eq:WKBfreq2} provides a simpler analytic expression for this case. The drawback of the Eq.~\eqref{eq:WKBfreq2} for $\Omega_R$ is that the numerator and denominator vanish in the extremal limit $a \to 1$ for those modes where $r_p \to 1$ (see \cite{Yang2012a,Yang2012b,Yang2013a} for a thorough discussion of the eikonal limit of nearly-extremal Kerr black holes).  For the study of nearly extremal Kerr black holes, Eq.~\eqref{eq:WKBfreq1} is the preferred form.
 
For completeness, we also give the analytic expression for $\Omega_I$ taken from~\cite{Yang2012a},
 \begin{align}
 \Omega_I & = \Delta(r_p)\frac{\sqrt{4(6r_p^2 \Omega_R^2 -1) + 2 a^2 \Omega_R^2 (3 - \mu^2)}}{2r_p^4\Omega_R - 4 a r_p \mu +a^2 r_p \Omega_R[r_p(3-\mu^2) + 2 (1+\mu^2)] + a^4 \Omega_R (1-\mu^2)} \,.
\end{align}
\end{widetext}

Again, for the Schwarzschild case, the equation immediately gives $\Omega_I = 1/\sqrt{27}$, and there are no issues with the case $\mu =0$ provided $\Omega_R$ is well-behaved.

In Sec.~\ref{sec6}, we use a more accurate expansion in orders of $a \omega_R/L$, taken from~\cite{casals} for the angular eigenfunction $A_{lm}$. Consequently, in that section we also use Eq.~\eqref{eq:WKBfreq1} for $\Omega_R$, although we still solve for the position of the peak using Eq.~\eqref{eq:PeakPoly}. We find that the additional accuracy in parts of our geometric phases that depend on $A_{lm}$ make no impact on the residuals of the geodesics, as compared to early tests where Eq.~\eqref{eq:AlmApp} was used.

\bibliography{green.bbl}

\begin{thebibliography}{44}
\expandafter\ifx\csname natexlab\endcsname\relax\def\natexlab#1{#1}\fi
\expandafter\ifx\csname bibnamefont\endcsname\relax
  \def\bibnamefont#1{#1}\fi
\expandafter\ifx\csname bibfnamefont\endcsname\relax
  \def\bibfnamefont#1{#1}\fi
\expandafter\ifx\csname citenamefont\endcsname\relax
  \def\citenamefont#1{#1}\fi
\expandafter\ifx\csname url\endcsname\relax
  \def\url#1{\texttt{#1}}\fi
\expandafter\ifx\csname urlprefix\endcsname\relax\def\urlprefix{URL }\fi
\providecommand{\bibinfo}[2]{#2}
\providecommand{\eprint}[2][]{\url{#2}}

\bibitem[{\citenamefont{Melia}(2007)}]{Melia}
\bibinfo{author}{\bibfnamefont{F.}~\bibnamefont{Melia}},
  \emph{\bibinfo{title}{The Galactic Supermassive Black Hole}}
  (\bibinfo{publisher}{Princeton University Press}, \bibinfo{year}{2007}),
  \bibinfo{edition}{1st} ed.

\bibitem[{\citenamefont{Ivanov et~al.}(2005)\citenamefont{Ivanov, Polnarev, and
  Saha}}]{Ivanov2005}
\bibinfo{author}{\bibfnamefont{P.~B.} \bibnamefont{Ivanov}},
  \bibinfo{author}{\bibfnamefont{A.}~\bibnamefont{Polnarev}}, \bibnamefont{and}
  \bibinfo{author}{\bibfnamefont{P.}~\bibnamefont{Saha}},
  \bibinfo{journal}{Mon.Not.Roy.Astron.Soc.} \textbf{\bibinfo{volume}{358}},
  \bibinfo{pages}{1361} (\bibinfo{year}{2005}), \eprint{astro-ph/0410610}.

\bibitem[{\citenamefont{Chen et~al.}(2009)\citenamefont{Chen, Madau, Sesana,
  and Liu}}]{chen}
\bibinfo{author}{\bibfnamefont{X.}~\bibnamefont{Chen}},
  \bibinfo{author}{\bibfnamefont{P.}~\bibnamefont{Madau}},
  \bibinfo{author}{\bibfnamefont{A.}~\bibnamefont{Sesana}}, \bibnamefont{and}
  \bibinfo{author}{\bibfnamefont{F.}~\bibnamefont{Liu}},
  \bibinfo{journal}{Astrophys. J.} \textbf{\bibinfo{volume}{697}},
  \bibinfo{pages}{L149} (\bibinfo{year}{2009}), \eprint{0904.4481}.

\bibitem[{eLI()}]{eLISA}
\urlprefix\url{http://www.elisascience.org}.

\bibitem[{\citenamefont{Mino et~al.}(1997)\citenamefont{Mino, Sasaki, and
  Tanaka}}]{Mino1996}
\bibinfo{author}{\bibfnamefont{Y.}~\bibnamefont{Mino}},
  \bibinfo{author}{\bibfnamefont{M.}~\bibnamefont{Sasaki}}, \bibnamefont{and}
  \bibinfo{author}{\bibfnamefont{T.}~\bibnamefont{Tanaka}},
  \bibinfo{journal}{Phys. Rev. D} \textbf{\bibinfo{volume}{55}},
  \bibinfo{pages}{3457} (\bibinfo{year}{1997}), \eprint{gr-qc/9606018}.

\bibitem[{\citenamefont{Barack}(2009)}]{Barack2009}
\bibinfo{author}{\bibfnamefont{L.}~\bibnamefont{Barack}},
  \bibinfo{journal}{Class.Quant.Grav.} \textbf{\bibinfo{volume}{26}},
  \bibinfo{pages}{213001} (\bibinfo{year}{2009}).

\bibitem[{\citenamefont{Poisson et~al.}(2011)\citenamefont{Poisson, Pound, and
  Bega}}]{Poisson2011}
\bibinfo{author}{\bibfnamefont{E.}~\bibnamefont{Poisson}},
  \bibinfo{author}{\bibfnamefont{A.}~\bibnamefont{Pound}}, \bibnamefont{and}
  \bibinfo{author}{\bibfnamefont{I.}~\bibnamefont{Bega}},
  \bibinfo{journal}{Living Review in Relativity} \textbf{\bibinfo{volume}{14}},
  \bibinfo{pages}{7} (\bibinfo{year}{2011}).

\bibitem[{\citenamefont{Dolan and Ottewill}(2011)}]{Dolan2011}
\bibinfo{author}{\bibfnamefont{S.~R.} \bibnamefont{Dolan}} \bibnamefont{and}
  \bibinfo{author}{\bibfnamefont{A.~C.} \bibnamefont{Ottewill}},
  \bibinfo{journal}{Phys. Rev. D} \textbf{\bibinfo{volume}{84}},
  \bibinfo{pages}{104002} (\bibinfo{year}{2011}), \eprint{1106.4318}.

\bibitem[{\citenamefont{Casals et~al.}(2009{\natexlab{a}})\citenamefont{Casals,
  Dolan, Ottewill, and Wardell}}]{Casals2009a}
\bibinfo{author}{\bibfnamefont{M.}~\bibnamefont{Casals}},
  \bibinfo{author}{\bibfnamefont{S.~R.} \bibnamefont{Dolan}},
  \bibinfo{author}{\bibfnamefont{A.~C.} \bibnamefont{Ottewill}},
  \bibnamefont{and} \bibinfo{author}{\bibfnamefont{B.}~\bibnamefont{Wardell}},
  \bibinfo{journal}{Phys. Rev. D} \textbf{\bibinfo{volume}{79}},
  \bibinfo{pages}{124043} (\bibinfo{year}{2009}{\natexlab{a}}),
  \eprint{0903.0395}.

\bibitem[{\citenamefont{Casals et~al.}(2009{\natexlab{b}})\citenamefont{Casals,
  Dolan, Ottewill, and Wardell}}]{Casals2009b}
\bibinfo{author}{\bibfnamefont{M.}~\bibnamefont{Casals}},
  \bibinfo{author}{\bibfnamefont{S.}~\bibnamefont{Dolan}},
  \bibinfo{author}{\bibfnamefont{A.~C.} \bibnamefont{Ottewill}},
  \bibnamefont{and} \bibinfo{author}{\bibfnamefont{B.}~\bibnamefont{Wardell}},
  \bibinfo{journal}{Phys. Rev. D} \textbf{\bibinfo{volume}{79}},
  \bibinfo{pages}{124044} (\bibinfo{year}{2009}{\natexlab{b}}),
  \eprint{0903.5319}.

\bibitem[{\citenamefont{Zenginoglu and Galley}(2012)}]{Zenginoglu2012}
\bibinfo{author}{\bibfnamefont{A.}~\bibnamefont{Zenginoglu}} \bibnamefont{and}
  \bibinfo{author}{\bibfnamefont{C.~R.} \bibnamefont{Galley}},
  \bibinfo{journal}{Phys. Rev. D} \textbf{\bibinfo{volume}{86}},
  \bibinfo{pages}{064030} (\bibinfo{year}{2012}), \eprint{1206.1109}.

\bibitem[{\citenamefont{Yang et~al.}(2012)\citenamefont{Yang, Nichols, Zhang,
  Zimmerman, Zhang et~al.}}]{Yang2012a}
\bibinfo{author}{\bibfnamefont{H.}~\bibnamefont{Yang}},
  \bibinfo{author}{\bibfnamefont{D.~A.} \bibnamefont{Nichols}},
  \bibinfo{author}{\bibfnamefont{F.}~\bibnamefont{Zhang}},
  \bibinfo{author}{\bibfnamefont{A.}~\bibnamefont{Zimmerman}},
  \bibinfo{author}{\bibfnamefont{Z.}~\bibnamefont{Zhang}},
  \bibnamefont{et~al.}, \bibinfo{journal}{Phys. Rev. D}
  \textbf{\bibinfo{volume}{86}}, \bibinfo{pages}{104006}
  (\bibinfo{year}{2012}), \eprint{1207.4253}.

\bibitem[{\citenamefont{Harte and Drivas}(2012)}]{Harte:2012uw}
\bibinfo{author}{\bibfnamefont{A.~I.} \bibnamefont{Harte}} \bibnamefont{and}
  \bibinfo{author}{\bibfnamefont{T.~D.} \bibnamefont{Drivas}},
  \bibinfo{journal}{Phys.Rev.D} \textbf{\bibinfo{volume}{85}},
  \bibinfo{pages}{124039} (\bibinfo{year}{2012}), \eprint{1202.0540}.

\bibitem[{\citenamefont{Casals and Nolan}(2012)}]{Casals2012}
\bibinfo{author}{\bibfnamefont{M.}~\bibnamefont{Casals}} \bibnamefont{and}
  \bibinfo{author}{\bibfnamefont{B.~C.} \bibnamefont{Nolan}},
  \bibinfo{journal}{Phys. Rev. D} \textbf{\bibinfo{volume}{86}},
  \bibinfo{pages}{024038} (\bibinfo{year}{2012}), \eprint{1204.0407}.

\bibitem[{\citenamefont{Casals et~al.}(2013)\citenamefont{Casals, Dolan,
  Ottewill, and Barry}}]{casals2013b}
\bibinfo{author}{\bibfnamefont{M.}~\bibnamefont{Casals}},
  \bibinfo{author}{\bibfnamefont{S.}~\bibnamefont{Dolan}},
  \bibinfo{author}{\bibfnamefont{A.}~\bibnamefont{Ottewill}}, \bibnamefont{and}
  \bibinfo{author}{\bibfnamefont{W.}~\bibnamefont{Barry}},
  \bibinfo{journal}{Phys. Rev. D} \textbf{\bibinfo{volume}{88}},
  \bibinfo{pages}{044022} (\bibinfo{year}{2013}), \eprint{gr-qc/1306.0884}.

\bibitem[{\citenamefont{Teukolsky}(1973)}]{Teukolsky1}
\bibinfo{author}{\bibfnamefont{S.~A.} \bibnamefont{Teukolsky}},
  \bibinfo{journal}{Astrophys.J.} \textbf{\bibinfo{volume}{185}},
  \bibinfo{pages}{635} (\bibinfo{year}{1973}).

\bibitem[{\citenamefont{Fackerell and Crossman}(1977)}]{Fackerell}
\bibinfo{author}{\bibfnamefont{E.~D.} \bibnamefont{Fackerell}}
  \bibnamefont{and} \bibinfo{author}{\bibfnamefont{R.~G.}
  \bibnamefont{Crossman}}, \bibinfo{journal}{J.~Math.~Phys.}
  \textbf{\bibinfo{volume}{18}}, \bibinfo{pages}{1849} (\bibinfo{year}{1977}).

\bibitem[{\citenamefont{Berti et~al.}(2006)\citenamefont{Berti, Cardoso, and
  Casals}}]{casals}
\bibinfo{author}{\bibfnamefont{E.}~\bibnamefont{Berti}},
  \bibinfo{author}{\bibfnamefont{V.}~\bibnamefont{Cardoso}}, \bibnamefont{and}
  \bibinfo{author}{\bibfnamefont{M.}~\bibnamefont{Casals}},
  \bibinfo{journal}{Phys. Rev. D} \textbf{\bibinfo{volume}{73}},
  \bibinfo{pages}{024013} (\bibinfo{year}{2006}), \eprint{gr-qc/0511111}.

\bibitem[{\citenamefont{Leaver}(1986)}]{leaver3}
\bibinfo{author}{\bibfnamefont{E.~W.} \bibnamefont{Leaver}},
  \bibinfo{journal}{Phys. Rev. D} \textbf{\bibinfo{volume}{34}},
  \bibinfo{pages}{384} (\bibinfo{year}{1986}).

\bibitem[{\citenamefont{Casals and Ottewill}(2012{\natexlab{a}})}]{casals2012a}
\bibinfo{author}{\bibfnamefont{M.}~\bibnamefont{Casals}} \bibnamefont{and}
  \bibinfo{author}{\bibfnamefont{A.}~\bibnamefont{Ottewill}},
  \bibinfo{journal}{Phys. Rev. D} \textbf{\bibinfo{volume}{86}},
  \bibinfo{pages}{024021} (\bibinfo{year}{2012}{\natexlab{a}}),
  \eprint{gr-qc/1112.2695}.

\bibitem[{\citenamefont{Casals and Ottewill}(2012{\natexlab{b}})}]{casals2012b}
\bibinfo{author}{\bibfnamefont{M.}~\bibnamefont{Casals}} \bibnamefont{and}
  \bibinfo{author}{\bibfnamefont{A.}~\bibnamefont{Ottewill}},
  \bibinfo{journal}{Phys. Rev. Lett} \textbf{\bibinfo{volume}{109}},
  \bibinfo{pages}{111101} (\bibinfo{year}{2012}{\natexlab{b}}),
  \eprint{gr-qc/1205.6592}.

\bibitem[{\citenamefont{Casals and Ottewill}(2013)}]{casals2013a}
\bibinfo{author}{\bibfnamefont{M.}~\bibnamefont{Casals}} \bibnamefont{and}
  \bibinfo{author}{\bibfnamefont{A.}~\bibnamefont{Ottewill}},
  \bibinfo{journal}{Phys. Rev. D} \textbf{\bibinfo{volume}{87}},
  \bibinfo{pages}{064010} (\bibinfo{year}{2013}), \eprint{gr-qc/1210.0519}.

\bibitem[{\citenamefont{Yang et~al.}(2013{\natexlab{a}})\citenamefont{Yang,
  Zhang, Zimmerman, Nichols, Berti et~al.}}]{Yang2012b}
\bibinfo{author}{\bibfnamefont{H.}~\bibnamefont{Yang}},
  \bibinfo{author}{\bibfnamefont{F.}~\bibnamefont{Zhang}},
  \bibinfo{author}{\bibfnamefont{A.}~\bibnamefont{Zimmerman}},
  \bibinfo{author}{\bibfnamefont{D.~A.} \bibnamefont{Nichols}},
  \bibinfo{author}{\bibfnamefont{E.}~\bibnamefont{Berti}},
  \bibnamefont{et~al.}, \bibinfo{journal}{Phys. Rev. D}
  \textbf{\bibinfo{volume}{87}}, \bibinfo{pages}{041502}
  (\bibinfo{year}{2013}{\natexlab{a}}), \eprint{1212.3271}.

\bibitem[{\citenamefont{Yang et~al.}(2013{\natexlab{b}})\citenamefont{Yang,
  Zimmerman, Zengino{\u g}lu, Zhang, Berti et~al.}}]{Yang2013a}
\bibinfo{author}{\bibfnamefont{H.}~\bibnamefont{Yang}},
  \bibinfo{author}{\bibfnamefont{A.}~\bibnamefont{Zimmerman}},
  \bibinfo{author}{\bibfnamefont{A.}~\bibnamefont{Zengino{\u g}lu}},
  \bibinfo{author}{\bibfnamefont{F.}~\bibnamefont{Zhang}},
  \bibinfo{author}{\bibfnamefont{E.}~\bibnamefont{Berti}},
  \bibnamefont{et~al.}, \bibinfo{journal}{Phys. Rev. D}
  \textbf{\bibinfo{volume}{88}}, \bibinfo{pages}{044047}
  (\bibinfo{year}{2013}{\natexlab{b}}), \eprint{1307.8086}.

\bibitem[{\citenamefont{Schutz and Will}(1985)}]{Schutz:1985km}
\bibinfo{author}{\bibfnamefont{B.~F.} \bibnamefont{Schutz}} \bibnamefont{and}
  \bibinfo{author}{\bibfnamefont{C.~M.} \bibnamefont{Will}},
  \bibinfo{journal}{Astrophys.J.} \textbf{\bibinfo{volume}{291}},
  \bibinfo{pages}{L33} (\bibinfo{year}{1985}).

\bibitem[{\citenamefont{Iyer and Will}(1987)}]{Iyer}
\bibinfo{author}{\bibfnamefont{S.}~\bibnamefont{Iyer}} \bibnamefont{and}
  \bibinfo{author}{\bibfnamefont{C.~M.} \bibnamefont{Will}},
  \bibinfo{journal}{Phys. Rev. D} \textbf{\bibinfo{volume}{35}},
  \bibinfo{pages}{3621} (\bibinfo{year}{1987}).

\bibitem[{\citenamefont{Dolan and Ottewill}(2009)}]{Dolan:2009nk}
\bibinfo{author}{\bibfnamefont{S.~R.} \bibnamefont{Dolan}} \bibnamefont{and}
  \bibinfo{author}{\bibfnamefont{A.~C.} \bibnamefont{Ottewill}},
  \bibinfo{journal}{Class.Quant.Grav.} \textbf{\bibinfo{volume}{26}},
  \bibinfo{pages}{225003} (\bibinfo{year}{2009}), \eprint{0908.0329}.

\bibitem[{\citenamefont{Dolan}(2010)}]{Dolan10}
\bibinfo{author}{\bibfnamefont{S.~R.} \bibnamefont{Dolan}},
  \bibinfo{journal}{Phys. Rev. D} \textbf{\bibinfo{volume}{82}},
  \bibinfo{pages}{104003} (\bibinfo{year}{2010}), \eprint{1007.5097}.

\bibitem[{\citenamefont{{Goebel}}(1972)}]{Goebel1972}
\bibinfo{author}{\bibfnamefont{C.~J.} \bibnamefont{{Goebel}}},
  \bibinfo{journal}{Astrophys. J.} \textbf{\bibinfo{volume}{172}},
  \bibinfo{pages}{L95} (\bibinfo{year}{1972}).

\bibitem[{\citenamefont{Ferrari and Mashhoon}(1984)}]{Ferrari}
\bibinfo{author}{\bibfnamefont{V.}~\bibnamefont{Ferrari}} \bibnamefont{and}
  \bibinfo{author}{\bibfnamefont{B.}~\bibnamefont{Mashhoon}},
  \bibinfo{journal}{Phys.Rev.D.} \textbf{\bibinfo{volume}{30}},
  \bibinfo{pages}{295} (\bibinfo{year}{1984}).

\bibitem[{\citenamefont{Cardoso et~al.}(2009)\citenamefont{Cardoso, Miranda,
  Berti, Witek, and Zanchin}}]{Cardoso:2008bp}
\bibinfo{author}{\bibfnamefont{V.}~\bibnamefont{Cardoso}},
  \bibinfo{author}{\bibfnamefont{A.~S.} \bibnamefont{Miranda}},
  \bibinfo{author}{\bibfnamefont{E.}~\bibnamefont{Berti}},
  \bibinfo{author}{\bibfnamefont{H.}~\bibnamefont{Witek}}, \bibnamefont{and}
  \bibinfo{author}{\bibfnamefont{V.~T.} \bibnamefont{Zanchin}},
  \bibinfo{journal}{Phys. Rev. D} \textbf{\bibinfo{volume}{79}},
  \bibinfo{pages}{064016} (\bibinfo{year}{2009}), \eprint{0812.1806}.

\bibitem[{\citenamefont{Mino}(2003)}]{Mino:2003yg}
\bibinfo{author}{\bibfnamefont{Y.}~\bibnamefont{Mino}},
  \bibinfo{journal}{Phys.Rev.} \textbf{\bibinfo{volume}{D67}},
  \bibinfo{pages}{084027} (\bibinfo{year}{2003}), \eprint{gr-qc/0302075}.

\bibitem[{\citenamefont{Olver et~al.}(2010)\citenamefont{Olver, Lozier,
  Boisvert, and Clark}}]{nist}
\bibinfo{author}{\bibfnamefont{F.~W.} \bibnamefont{Olver}},
  \bibinfo{author}{\bibfnamefont{D.~W.} \bibnamefont{Lozier}},
  \bibinfo{author}{\bibfnamefont{R.~F.} \bibnamefont{Boisvert}},
  \bibnamefont{and} \bibinfo{author}{\bibfnamefont{C.~W.} \bibnamefont{Clark}},
  \emph{\bibinfo{title}{NIST Handbook of Mathematical Functions}}
  (\bibinfo{publisher}{Cambridge University Press}, \bibinfo{address}{New York,
  NY, USA}, \bibinfo{year}{2010}), \bibinfo{edition}{1st} ed., ISBN
  \bibinfo{isbn}{0521140633, 9780521140638}.

\bibitem[{\citenamefont{Zhang et~al.}(2013)\citenamefont{Zhang, Berti, and
  Cardoso}}]{Zhang:2013ksa}
\bibinfo{author}{\bibfnamefont{Z.}~\bibnamefont{Zhang}},
  \bibinfo{author}{\bibfnamefont{E.}~\bibnamefont{Berti}}, \bibnamefont{and}
  \bibinfo{author}{\bibfnamefont{V.}~\bibnamefont{Cardoso}},
  \bibinfo{journal}{Phys. Rev. D} \textbf{\bibinfo{volume}{88}},
  \bibinfo{pages}{044018} (\bibinfo{year}{2013}), \eprint{1305.4306}.

\bibitem[{\citenamefont{Berti and Cardoso}(2006)}]{Berti:2006wq}
\bibinfo{author}{\bibfnamefont{E.}~\bibnamefont{Berti}} \bibnamefont{and}
  \bibinfo{author}{\bibfnamefont{V.}~\bibnamefont{Cardoso}},
  \bibinfo{journal}{Phys. Rev. D} \textbf{\bibinfo{volume}{74}},
  \bibinfo{pages}{104020} (\bibinfo{year}{2006}), \eprint{gr-qc/0605118}.

\bibitem[{Ber()}]{BertiSite}
\urlprefix\url{http://www.phy.olemiss.edu/~berti/qnms.html}.

\bibitem[{\citenamefont{Arfken and Weber}(2005)}]{ArfkenWeber}
\bibinfo{author}{\bibfnamefont{G.~B.} \bibnamefont{Arfken}} \bibnamefont{and}
  \bibinfo{author}{\bibfnamefont{H.~J.} \bibnamefont{Weber}},
  \emph{\bibinfo{title}{Mathematical Methods for Physicists}}
  (\bibinfo{publisher}{Elsevier Academic Press}, \bibinfo{year}{2005}),
  \bibinfo{edition}{6th} ed.

\bibitem[{\citenamefont{Kay et~al.}(1997)\citenamefont{Kay, Radzikowski, and
  Wald}}]{Kay:1996hj}
\bibinfo{author}{\bibfnamefont{B.~S.} \bibnamefont{Kay}},
  \bibinfo{author}{\bibfnamefont{M.~J.} \bibnamefont{Radzikowski}},
  \bibnamefont{and} \bibinfo{author}{\bibfnamefont{R.~M.} \bibnamefont{Wald}},
  \bibinfo{journal}{Commun.Math.Phys.} \textbf{\bibinfo{volume}{183}},
  \bibinfo{pages}{533} (\bibinfo{year}{1997}), \eprint{gr-qc/9603012}.

\bibitem[{\citenamefont{{Misner} et~al.}(1973)\citenamefont{{Misner}, {Thorne},
  and {Wheeler}}}]{MTW}
\bibinfo{author}{\bibfnamefont{C.~W.} \bibnamefont{{Misner}}},
  \bibinfo{author}{\bibfnamefont{K.~S.} \bibnamefont{{Thorne}}},
  \bibnamefont{and} \bibinfo{author}{\bibfnamefont{J.~A.}
  \bibnamefont{{Wheeler}}}, \emph{\bibinfo{title}{{Gravitation}}}
  (\bibinfo{publisher}{San Francisco: W.H.~Freeman and Co.},
  \bibinfo{year}{1973}).

\bibitem[{\citenamefont{Carter}(1968)}]{Carter:1968rr}
\bibinfo{author}{\bibfnamefont{B.}~\bibnamefont{Carter}},
  \bibinfo{journal}{Phys. Rev.} \textbf{\bibinfo{volume}{174}},
  \bibinfo{pages}{1559} (\bibinfo{year}{1968}).

\bibitem[{\citenamefont{Le~Tiec et~al.}(2011)\citenamefont{Le~Tiec, Mroue,
  Barack, Buonanno, Pfeiffer et~al.}}]{LeTiec:2011bk}
\bibinfo{author}{\bibfnamefont{A.}~\bibnamefont{Le~Tiec}},
  \bibinfo{author}{\bibfnamefont{A.~H.} \bibnamefont{Mroue}},
  \bibinfo{author}{\bibfnamefont{L.}~\bibnamefont{Barack}},
  \bibinfo{author}{\bibfnamefont{A.}~\bibnamefont{Buonanno}},
  \bibinfo{author}{\bibfnamefont{H.~P.} \bibnamefont{Pfeiffer}},
  \bibnamefont{et~al.}, \bibinfo{journal}{Phys.Rev.Lett.}
  \textbf{\bibinfo{volume}{107}}, \bibinfo{pages}{141101}
  (\bibinfo{year}{2011}), \eprint{1106.3278}.

\bibitem[{\citenamefont{Le~Tiec et~al.}(2012)\citenamefont{Le~Tiec, Barausse,
  and Buonanno}}]{LeTiec:2011dp}
\bibinfo{author}{\bibfnamefont{A.}~\bibnamefont{Le~Tiec}},
  \bibinfo{author}{\bibfnamefont{E.}~\bibnamefont{Barausse}}, \bibnamefont{and}
  \bibinfo{author}{\bibfnamefont{A.}~\bibnamefont{Buonanno}},
  \bibinfo{journal}{Phys.Rev.Lett.} \textbf{\bibinfo{volume}{108}},
  \bibinfo{pages}{131103} (\bibinfo{year}{2012}), \eprint{1111.5609}.

\bibitem[{\citenamefont{Le~Tiec et~al.}(2013)\citenamefont{Le~Tiec, Buonanno,
  Mrou\'e, Pfeiffer, Hemberger, Lovelace, Kidder, Scheel, Szil\'agyi, Taylor
  et~al.}}]{Tiec:2013twa}
\bibinfo{author}{\bibfnamefont{A.}~\bibnamefont{Le~Tiec}},
  \bibinfo{author}{\bibfnamefont{A.}~\bibnamefont{Buonanno}},
  \bibinfo{author}{\bibfnamefont{A.~H.} \bibnamefont{Mrou\'e}},
  \bibinfo{author}{\bibfnamefont{H.~P.} \bibnamefont{Pfeiffer}},
  \bibinfo{author}{\bibfnamefont{D.~A.} \bibnamefont{Hemberger}},
  \bibinfo{author}{\bibfnamefont{G.}~\bibnamefont{Lovelace}},
  \bibinfo{author}{\bibfnamefont{L.~E.} \bibnamefont{Kidder}},
  \bibinfo{author}{\bibfnamefont{M.~A.} \bibnamefont{Scheel}},
  \bibinfo{author}{\bibfnamefont{B.}~\bibnamefont{Szil\'agyi}},
  \bibinfo{author}{\bibfnamefont{N.~W.} \bibnamefont{Taylor}},
  \bibnamefont{et~al.}, \bibinfo{journal}{Phys. Rev. D}
  \textbf{\bibinfo{volume}{88}}, \bibinfo{pages}{124027}
  (\bibinfo{year}{2013}).

\bibitem[{\citenamefont{Harwitt}(2003)}]{Harwitt}
\bibinfo{author}{\bibfnamefont{M.}~\bibnamefont{Harwitt}},
  \bibinfo{journal}{Astrophys. J.} \textbf{\bibinfo{volume}{597}},
  \bibinfo{pages}{1266} (\bibinfo{year}{2003}).

\end{thebibliography}

\end{document}